\documentstyle[psfig,floats,aps,preprint]{revtex}
\begin{document}
\draft

\title{Theory of spiral wave dynamics in weakly excitable media:
 asymptotic reduction to a kinematic model and applications}

\author{Vincent Hakim$^{(1)}$ and Alain Karma$^{(2)}$}

\address{$^{(1)}$ Laboratoire de Physique Statistique\dag,
Ecole Normale Sup{\'e}rieure, 24 rue Lhomond, 75231 Paris Cedex 05,
France\\
\dag associ{\'e} au CNRS et aux Universit{\'e}s Paris VI et VII\\
$^{(2)}$ Department of Physics and Center for Interdisciplinary
Research on Complex Systems,\\ Northeastern University, Boston,
Massachusetts 02115}

\date{\today}
\maketitle
\begin{abstract}
In a weakly excitable 
medium, characterized by a large
threshold stimulus, the free end of 
an isolated broken plane wave (wave tip) can either 
rotate (steadily or unsteadily) around a large excitable core,
thereby producing a spiral pattern, or retract causing
the wave to vanish at boundaries. 
An asymptotic analysis of spiral motion and retraction 
is carried out in this weakly excitable large core regime
starting from the free-boundary limit of the reaction-diffusion
models, valid when the excited region is delimited by a thin interface.
The wave description is shown to naturally
split between the tip region and a far region that
are smoothly matched on an intermediate scale.
This separation allows us to
rigorously derive an equation of motion 
for the wave tip, with the large scale motion
of the spiral wavefront slaved to the tip. 
This kinematic description
provides both a physical picture and exact
predictions for a wide range of 
wave behavior, including: (i)
steady rotation (frequency and core radius), (ii) 
exact treatment of the meandering instability in the
free-boundary limit with the prediction that the
frequency of unstable motion is half 
the primary steady frequency
(iii) drift under external actions 
(external field with application to axisymmetric scroll ring
motion in three-dimensions, and
spatial or/and time-dependent variation of excitability),
and (iv) the dynamics of multi-armed spiral waves with the new
prediction that steadily rotating waves with two or more arms are 
linearly unstable. 
Numerical simulations of FitzHug-Nagumo kinetics are used
to test 
several aspects of our results.
In addition, we discuss the
semi-quantitative extension of this theory 
to finite cores and pinpoint mathematical subtleties related to the
thin interface limit of singly diffusive reaction-diffusion models.
\end{abstract}

\section{Introduction}

Spiral waves are characteristic 
structures of excitable media 
\cite{Win0,Zyk} that have been
observed in systems as different as catalytic surface 
oxidation \cite{cat}, the 
Belousov-Zhabotinsky chemical reaction \cite{Win1,Plesetal,Skietal,belfles}
aggregating colonies of slime mold \cite{dic},
and heart tissue where they are suspected to play
an essential role in cardiac arrhythmia and fibrillation \cite{heart}.
Spiral waves are prone to a variety 
of instabilities, the best studied
of which is meander, and they can be made 
to drift and be controlled in 
diverse ways, for instance by varying the medium excitability in space 
or/and time, or by adding an external field. 

\begin{figure}
\centerline{
\psfig{file=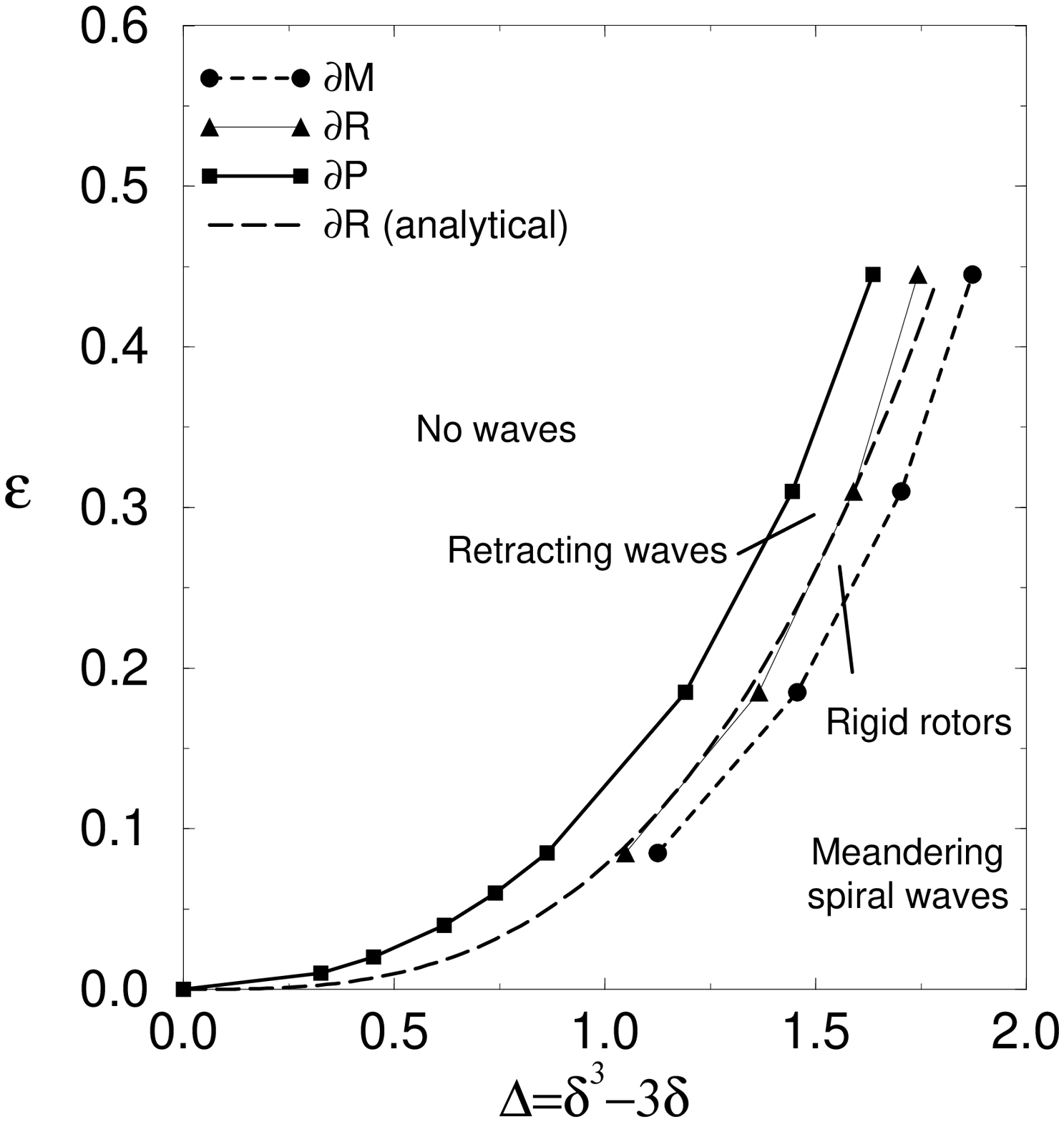,width=.8\textwidth}}
\caption{Plots of the propagation ($\partial P$), rotor ($\partial R$), 
and meander ($\partial M$) boundaries in the parameter space
$\epsilon$ (the ratio of the fast
activator to the slower controller time scale)
and $\Delta$ (the medium 'excitability' defined as $v_s-v_0$)
for the numerically simulated FitzHugh-Nagumo 
kinetics ($f(u,v)=3 u-u^3-v,\ g(u,v)=u-\delta$). 
Our analysis predicts
that the three boundaries approach
smoothly the origin without crossing 
as $\epsilon \rightarrow 0$, with $\Delta_p \sim
-\epsilon^{1/2}{\rm ln}\, \epsilon$ for $\partial P$, $\Delta_c \sim
\epsilon^{1/3}$ for $\partial R$, and $\Delta_m-\Delta_c \sim
-\epsilon^{5/9}/({\rm ln}\, \epsilon)^{2/3}$ for $\partial M$. The
prediction $\Delta_c=(2^{1/2}4\epsilon/B_c)^{1/3}$
with $B_c=0.535$ for the $\partial R$ boundary is in good agreement 
with the simulations.}
\label{diag}
\label{flowergarden}
\end{figure}

\begin{figure}
\centerline{
  \psfig{file=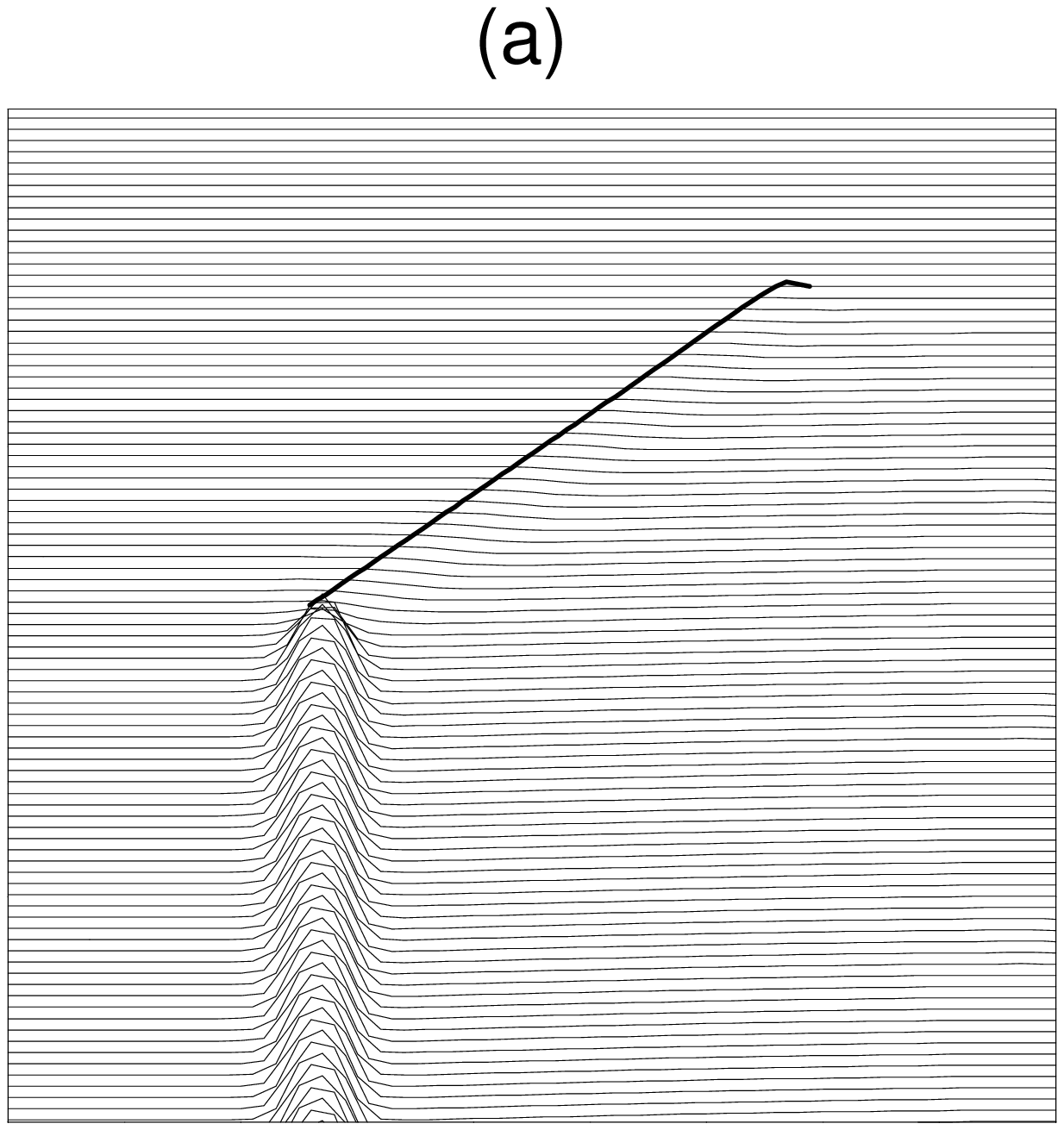,width=.45\textwidth}
\hfill
  \psfig{file=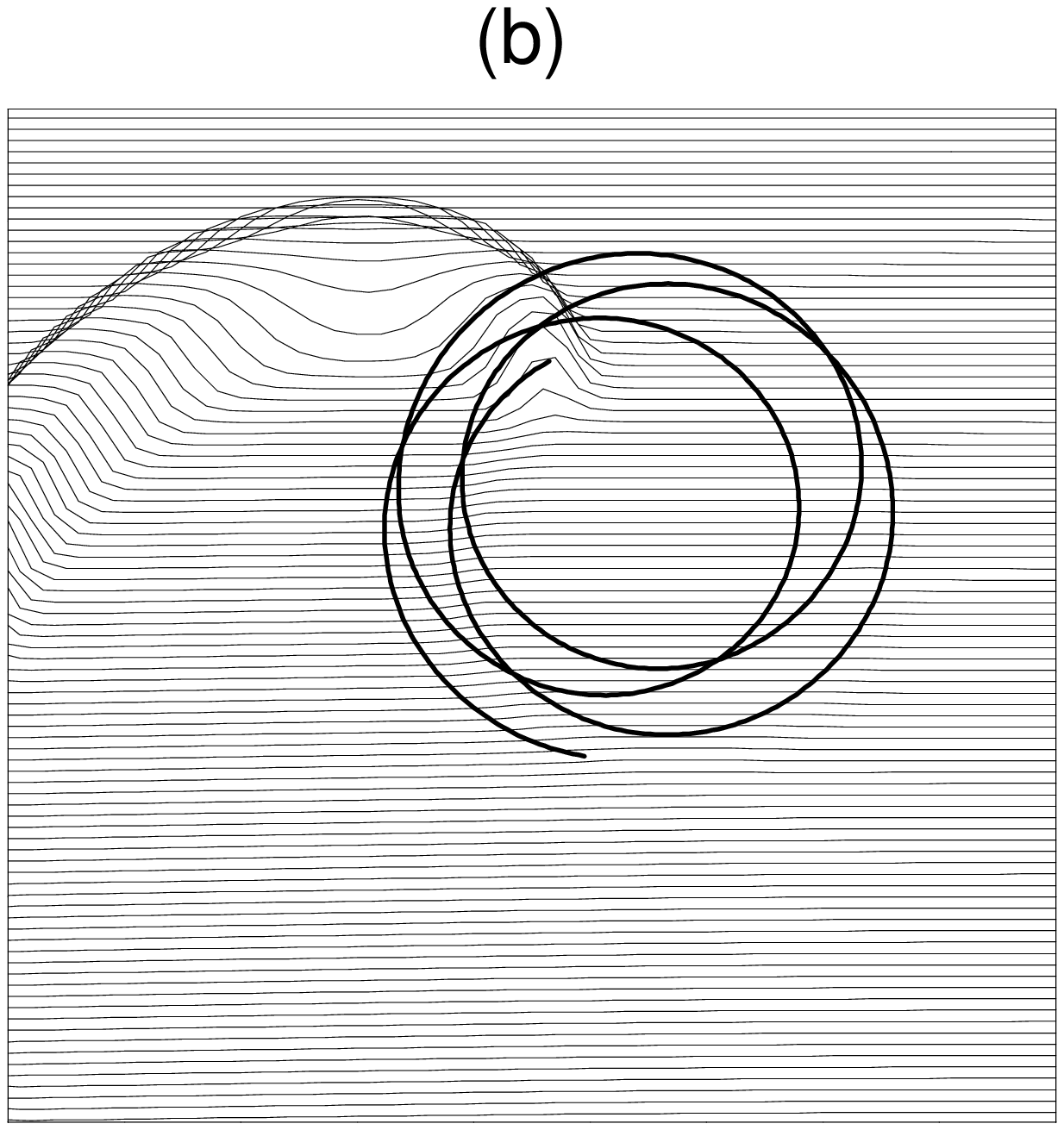,width=.45\textwidth}}
\caption{Surface plots of $u$
and wave tip trajectories (thick solid line) illustrating in
(a)
 a
retracting wave for $\delta=-1.4$ and $\epsilon=0.27$,
in between the $\partial P$ and $\partial R$ boundaries,
 and in (b)
 a large core meandering spiral wave for
$\delta=-1.4$ and $\epsilon=0.18$,
close to the $\partial M$ boundary.
}
\label{examples}
\end{figure}

Much of the observed experimental
phenomenology has been reproduced 
by using simplified two-variable activator-controller types of
description, like the classic FitzHugh-Nagumo (FN) model
\cite{fitzhugh} and mild variations of it.
Extensive surveys \cite{zykfg,Win2} of the possible types 
of wave motion in such models have been performed
in a reduced parameter space where the only two parameters
left to vary are the medium excitability 
$\Delta$, defined in section II in such a
way that the isolated
pulse speed is proportional to $\Delta$ for weak excitability,
and the ratio $\epsilon$ between the time 
scale of the activator and controller kinetics,
which controls the abruptness of the wave front (i.e.
the thickness of the interface delimiting
the excited region).
Different regimes have been identified that are
summarized in Fig.~\ref{flowergarden} for simple FN kinetics \cite{notefig1}.
In the
whole region above the {\it propagation} boundary ($\partial P$),
the medium excitability is too weak for any plane 
wave to propagate persistently. In the narrower region comprised
between $\partial P$ and the {\it rotor} boundary ($\partial R$),
the medium excitability is sufficient for a plane wave to propagate
but not for a spiral wave to form. In this region, the end
of a broken wave-front, referred hereafter as the `wave tip', 
simply retracts steadily (Fig.~\ref{examples}a), such that 
this finger-shaped wave must shrink in length and eventually 
vanish at boundaries in a finite system.
In the even narrower region comprised between $\partial R$ and
the {\it meander} boundary ($\partial M$), 
the excitability is sufficient for large core spirals to
form and the wave tip now rotates steadily at a frequency
$\omega_1$ around a circular core of radius $R_0$.
Right on the $\partial R$ boundary, a half plane
wave, referred hereafter as the ``critical finger'',
propagates without changing its shape. It can be equivalently interpreted
as a retracting finger with vanishing retracting velocity or as a spiral
wave of infinite core radius. 
As one keeps increasing the excitability,
the radius of the spiral core decreases and below
the $\partial M$ boundary the spiral tip traces a classic ``flower-like''
meander pattern (Fig.~\ref{examples}b). 
It has been shown that meander originates from a 
supercritical Hopf bifurcation at $\partial M$
which adds a second frequency $\omega_2$ to
the basic spiral rotation \cite{Hopf,hopfak}. 
The meander patterns exhibit first
inward petals as $\omega_2<\omega_1$. Outward petals appear as 
$\omega_2$ becomes greater than $\omega_1$.
Further away from $\partial M$, the spiral tip
motion becomes more complex and possibly chaotic passed the still poorly
characterized $\partial C$ boundary \cite{Win2} (not shown in
Fig.~\ref{flowergarden}). Given that a Hopf bifurcation takes place
on $\partial M$, symmetry arguments fix its resonant coupling to the
translation modes when $\omega_2=\omega_1$ and thus determine the bifurcation
structure of the tip motion near the codimension 2 
point $\omega_2=\omega_1$ on $\partial M$ \cite{Cod2,Lietal}.

In contrast to this rather detailed knowledge, the
precise mechanisms that govern spiral formation 
and motion remain less well understood from both
physical and predictive viewpoints.
A simple picture remains missing to
answer even basic questions, such as why does meander occur
and why is this instability oscillatory (i.e. a Hopf bifurcation) 
beyond numerical observation ? 
From a predictive viewpoint, we still 
lack a quantitative analytical understanding of what controls 
the $\partial M$ boundary or the frequency ratio $\omega_2/\omega_1$
in the parameter space of reaction-diffusion models. 
Similar uncertainties are to a large extent also
present for other phenomena like 
spiral drift under external action.

A kinematical model
of spiral dynamics, aimed at the 
weakly excitable large core limit,
has been proposed some years ago
on a purely phenomenological basis \cite{mikzyk}. 
It has been helpful to
rationalize experimental facts but it
has not been derived from the
underlying reaction-diffusion equations. Thus it remains
limited in its predictions, e.g. it
falls short of predicting the $\partial M$ boundary
and the ratio $\omega_2/\omega_1$. Moreover, at a more
conceptual level, the general validity of the boundary
condition assumed for the free end of the wave-front in this 
kinematic theory has remains somewhat unclear.

The first goal of this paper 
is to present a rigorous asymptotic derivation
of a kinematic theory of spiral 
wave motion in the weakly excitable and free-boundary 
limit (lower left hand corner of Fig.~1) onto which we focus. 
As we shall see, the structure of this
theory differs from the one
proposed phenomenologically in Ref. \cite{mikzyk}.
The second goal of this paper is to demonstrate, through selected 
applications of this theory, that it is able to provide
a physical and quantitative understanding
of a wide range of wave phenomena such as, 
meander, drift under various
external actions considered in
previous studies \cite{agl,ag,stein,apmun,bel,mit,krins}, 
and multi-arm spiral wave motion.
Highlights of our results include 
an asymptotically exact treatment of the meander 
instability for $\epsilon\ll 1$, which gives the
precise location of the $\partial M$ boundary
and shows that the instability 
arises from a supercritical Hopf 
bifurcation with $\omega_2/\omega_1= 1/2$ in
this limit, the finding that multi-arm spiral
waves with two or more arms are always linearly unstable, 
in contrast to a previous numerical study \cite{vas},
and predictions of the spiral drift speed
and drift angle in an external field.
These results are generally found to be in good quantitative
agreement with our simulations of FN kinetics.

The starting point of our analysis is
the standard free-boundary limit of 
reaction-diffusion models \cite{fbp} described in section II, 
which is valid when the excited region is
surrounded by a thin interface of width $\epsilon\ll 1$.
In this limit, the fast activator variable is eliminated
in favor of an eikonal equation that gives the
normal velocity of this boundary.
This velocity generally depends on the
local radius curvature of this interface, assumed large
compared to $\epsilon$, as well as the local 
value of the slow controller variable at the
interface. 

This free-boundary problem is 
non-trivial to solve because it requires to 
treat both the dynamics of the wave-front, which is the part 
of the boundary where the excited region propagates
into the recovery region of the medium, 
and the wave-back where the reverse process
occurs. Far from the tip, the front and back 
behave essentially identically, such that 
a `single front' description is rigorously
possible. In the tip region, however, the front
and back must be matched at the tip (i.e. point of zero normal
velocity along the boundary), which is a difficult task.
For this reason a single-front description with
a somewhat arbitrary tip condition was first used historically to relate
the steady rotation frequency and core radius of spirals \cite{keenty}.
The kinematic theory of Ref. \cite{mikzyk} is an attempt to extend this
picture to an unsteady situation. Subsequent solutions
of the complete free-boundary problem, with a rigorous matching of
front and back that provides a unique and independent determination of the
spiral frequency and wavelength, 
focused on two limits. One of these limits (see \cite{klr} and
earlier references therein)
is obtained mathematically by
assuming $\epsilon\ll 1$ while keeping $\Delta$ fixed
of order unity, which corresponds physically to
a highly excitable medium.
The wavelength and frequency obey in this 
limit \cite{ak92} certain scaling
laws with $\epsilon$ first proposed by Fife \cite{fife}. The wave-front
and wave-back, however, are matched onto singular 
core solutions (of size $\epsilon$ with only activator diffusion)
that have later been shown to be generically unstable \cite{klr}; this result
actually seems to concord with the numerical observation of complex 
meander (and thus unstable motion) in this limit \cite{Win2}.
Thus these solutions do not provide a proper
starting point for a kinematic theory aimed to describe
the onset of meander. 
A better starting point, on which we
focus here, is the second limit originating from
Ref.~ \cite{ps1} where one constructs smooth core solutions to
the free-boundary problem. It was shown in \cite{ak1}
that this can in fact only be consistently 
done for a weakly excitable medium 
when the radius of curvature of the boundary at the 
tip remains much larger than the
front and back interface width, i.e. 
by assuming simultaneously $\epsilon\ll 1$
and $\Delta \sim \epsilon^{1/3}\ll 1$. 
This allowed a rigorous derivation
of the line $\partial R$ in the weak excitability limit \cite{ak1}
in good quantitative agreement with numerical simulations
of FitzHugh-Nagumo kinetics (lower left hand corner of Fig.~1),
as well as a semi-analytic derivation of the selected 
core radius/frequency of spiral waves and retracting
wave speed in the same limit \cite{ak2}. 

The present kinematic theory is derived
by first refining analytically the description of steady
retraction and rotation in this weakly excitable limit
(section IV), and then extending it to an unsteady regime (section V)
for the non-trivial case of self interacting spirals,
i.e. where the wave tip motion is influenced
by the average controller concentration left by the
previous passage of the wave-front. This allows us to
derive an equation of motion for the wave tip that is then used
to analyze meander in a linear and nonlinear regime
(section VI). Results of these two 
sections have been 
summarized in a previous short
publication \cite{hk1}.
Further applications of the kinematic
theory are then contained in subsequent sections that
include spiral drift under various external actions 
without self-interaction (section VII)
and interacting multi-arm spiral 
waves (section VIII). 
Finally, corrections to the large core results are discussed
in Section \ref{smallcore} and several points 
are further analyzed in four appendices.
In addition, for clarity of exposition,
we have found best to first give a simple  physical
picture  of the kinematic
theory and summarize the main results of its application in section III.
This section is purposely aimed to discuss this theory
in terms of experimentally measurable quantities 
as well as to provide a guide to
the rest of the paper.

\section{Reaction-diffusion model and free-boundary limit}
We consider the 
classic activator ($u$) controller ($v$)
two-variable reaction-diffusion
model of excitable media \cite{fitzhugh}
\begin{eqnarray}
\partial_t u&=&D_u\nabla^2u+f(u,v)/\tau_u\label{eqd1}
\\
\partial_t v&=&D_v\nabla^2 v+ g(u,v)/\tau_v
\label{eqd2}
\end{eqnarray}
with a linearly stable rest state ($u_0,v_0$). 
We focus in this paper 
on the singly diffusive case $D_v=0$, although we shall also
briefly consider the slow controller 
diffusion limit $\gamma\equiv D_v/D_u\ll 1$ in section \ref{seclinstab}.
The $u$-nullcline ($f(u,v)=0$) is assumed to have the standard S-shape
in the $(u,v)$ plane. A simple choice of FN kinetics that we use 
for the numerical simulations is $f(u,v)=3u-u^3-v$, $g(u,v)= u-\delta$ with
the rest state $u_0=\delta,\ v_0=3\delta-\delta^3$. 
It is convenient to rewrite Eqs. (\ref{eqd1}) and (\ref{eqd2})
in a standard dimensionless form by measuring time and length in units
of $\tau_v$ and $(D_u\tau_v^2/\tau_u)^{1/2}$, respectively, which
yields for the singly diffusive case
\begin{eqnarray}
\partial_t u&=&\epsilon\,\nabla^2u+f(u,v)/\epsilon\label{eqrd1}
\\
\partial_t v&=& g(u,v)
\label{eqrd2}
\end{eqnarray}
where $\epsilon\equiv \tau_u/\tau_v$. We study this dimensionless
form of the equations in the rest of this paper, except in the
next section where we summarize the essential 
ingredients of the kinematic theory
in dimensional units.
For small $\epsilon$, the excited region ($u\simeq\sqrt{3}$ with the previous
choice) of a propagating wave
is separated by a sharp boundary from the unexcited or recovering medium
($u\simeq-\sqrt{3}$ with the previous choice). The
wave description can thus be reduced to determining
the motion of its boundary (i.e. a free boundary problem)
\cite{Zyk,keenty,ps1}:
\begin{eqnarray}
c_n&=&c(v)-\epsilon\,\kappa\label{eikonal},\\
\partial_t v&=&g(u^\pm (v),v)~~~~~{\rm in}
~~{\cal D}^\pm ,\label{veq}
\end{eqnarray}
where $c_n$ is the normal velocity of
the interface separating the excited
and recovery regions of the medium denoted by
(${\cal D}^+$) and (${\cal D}^-$), respectively, $\kappa$ is
the local curvature of this interface, and $u^\pm (v)$ denotes
the right-most ($+$) and left-most ($-$) branch of the
$u$-nullcline ($f(u,v)=0$). The function $c(v)$ is entirely determined
by Eq.~(\ref{eqrd1}) with $v$ fixed. 
We measure the excitability of the medium, i.e. the threshold
stimulus necessary to cause a response,
by the parameter $\Delta\equiv v_s-v_0$, where $v_s$ is the
stall value of $v$ at which $c(v_s)=0$. The
isolated pulse speed $c_0\equiv c(v_0)$
is then a monotonously increasing function of $\Delta$
with $c_0=\alpha\Delta$ for $\Delta\ll 1\ (\alpha=1/\sqrt{2}$ for our numerical
choice). For values of
of $v$ near $v_s$, Eq.~(\ref{veq}) can
be simplified even further to
\begin{eqnarray}
\partial_t v&=&1/\tau_e
~~~~~{\rm in}
~~{\cal D}^+
\label{exkin}\\
\partial_t v&=&-(v-v_0)/\tau_R
~~~~~{\rm in}
~~{\cal D}^-
\label{inkin}
\end{eqnarray}
where the activator time scale $\tau_e=1/g(u^+(v_s),v_s)$ controls
the pulse duration and the recovery time 
\begin{equation}
\tau_R= \frac{\partial_u f}
{(\partial_u g\partial_v f-\partial_v g\partial_u f)}|_{
u=u^-(v_s),v=v_s}
\end{equation}
is the time scale over which the controller variable
returns to its rest state after an excitation
(for our numerical choice, $\tau_e=1/(2\sqrt{3}), \tau_R=6 $).

\section{Physical picture of kinematic theory and main results}

\subsection{Retraction and rotation}

\begin{figure}
\centerline{
\psfig{file=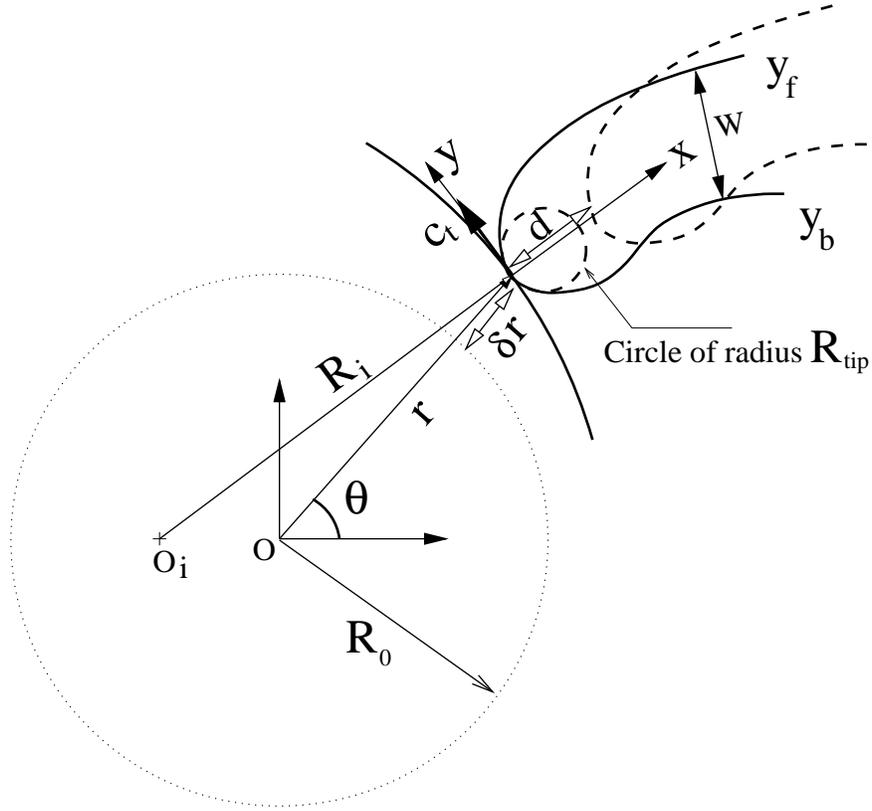,width=.7\textwidth}}
\bigskip
\caption{Sketch of the spiral 
tip region and unsteady tip trajectory (solid lines).
($r,\theta$) denotes the polar coordinates 
of the wave tip with respect to the fixed steady-state
center of rotation $O$. 
$R_i$ is the
instantaneous radius of curvature of the unsteady tip trajectory
about the instantaneous center of
rotation $O_i$, with $O_i=O$ and $R_i=R_0$ for steady rotation,
and $R_{tip}$ is the radius of curvature of the
boundary between the excited and recovery
regions of the medium at the tip. $c_t$ denotes the
instantaneous tangential velocity of 
the wave tip along this trajectory, 
with $c_t=\omega_1R_0$ for steady-state rotation.
The coordinates $\delta r=r-R_0\equiv \epsilon q/c_0$ 
and $\psi=\theta-\omega_1 t$
measure the radial and angular departure from
steady-state rotation, respectively. 
The cartesian coordinate system $(x,y)$
that moves with the wave tip is also shown
with the $y$-axis parallel to $c_t$.
$y_f(x)$ and $y_b(x)$
denote the instantaneous wave-front and wave-back boundaries. 
Finally, $d=q(\theta)-q(\theta-2\pi)$
measures the radial displacement of the wave tip
after one $2\pi$ rotation.}
\label{coordinates}
\label{figcoord}
\end{figure}

In a typical chemical 
or biological excitable media, many
parameters (chemical or ionic concentrations, 
temperature, light etc.), control the excitability 
of the medium. However, independently
of the complexity of the medium, it is 
generally possible to construct
a single dimensionless parameter \cite{ak1,ak2}
\begin{equation}
B=\frac{2 R_{tip}}{W}=\frac{2 D_u}{c_0W},\label{Bdef}
\end{equation}
which determines whether the tip 
rotates, and thus forms a spiral wave,
or whether it simply retracts. 
This parameter is expressed as the ratio of two lengthscales that
characterize the tip region of a broken plane wave that is
shown schematically in  Fig.~\ref{figcoord}.
The first is the radius of curvature $R_{tip}$ of the 
wave boundary at the tip. In the limit ($\epsilon\ll 1$) 
where this boundary is thin, $R_{tip}$ is obtained
by applying the eikonal equation at the tip, 
which yields $c_n=0=c_0-D_u/R_{tip}$,
and thus $R_{tip}=D_u/c_0$ where $c_0$ is the plane
wave speed. The second lengthscale $W$ is the constant width of
the excited region away from the tip. As argued in \cite{ak1,ak2},
the wave boundary in the tip region can only be smooth
if $R_{tip}\sim W$, such that the $\partial R$ boundary
must correspond to a fixed value of $B=B_c$ 
of order unity; an explicit calculation (Refs. \cite{ak1,ak2} and
section IV.A here) yields $B_c=0.535$. For $B>B_c$, the excitability
of the medium is not sufficient to overcome diffusion
in the most highly curved part of the tip region, which retracts.
In contrast for $B<B_c$, the excitability is sufficient
to overcome retraction, and the increase of $c_n$ 
away from the tip induces rotation.

Increasing (decreasing) $B$ corresponds to decreasing (increasing)
the excitability of the medium while moving 
normal to the $\partial R$ boundary
in the multi-dimensional parameter space that
characterizes a given excitable medium. $B$ is therefore the
most natural parameter to characterize this
excitability close to this boundary, and is used
throughout this paper. From an experimental
standpoint, $c_0$ and $W$ are in principle 
measurable quantities and $D_u$ can be either measured 
or estimated, such that one could attempt to
quantify $B$ directly. Of course, in practice, 
the definition of $W$, and thus $B$, becomes less
precise when $\epsilon$ is not small and it
is simpler to use an experimental control
parameter $P_{ex}$ (such as a concentration) that can be
varied to cross the $\partial R$ boundary.
Therefore we will briefly describe 
in subsection \ref{subsecC}  a more direct way
to obtain a quantitative relationship between $B$
and $P_{ex}$ close to the $\partial R$ boundary
without the need to use Eq. (\ref{Bdef}). 

\subsection{Rigid wave tip and slaved wave-front}
                       
Close to the $\partial R$ boundary, we will show that
a spiral wave in its tip region, behaves as
a `rigid body' whose motion
can be characterized by giving two {\it instantaneous}
quantities: the tip tangential velocity
$c_t(t)$, and its rotation rate $\omega_i(t)=c_t(t)/R_i(t)$ where
$R_i(t)$ is the radius of curvature of the
tip trajectory as depicted in Fig. \ref{figcoord}.
Two key ingredients 
make this  kinematic description possible.
The first is that, near $\partial R$, the wave shape  
is close to a critical finger
(i.e. the broken plane wave that simply translates without
retraction or rotation for $B=B_c$)
on a length scale $\ell\sim D_u/(c_0\sqrt{1-c_t/c_0})$ (as explained
in section IV.C)
that is large compared to the
scale of the tip itself $R_{tip}=D_u/c_0$.
The second is
that the relations that govern $c_t$ and $R_i$
are established on time scales that are 
both much shorter than the steady rotation period, $T_0=2\pi/\omega_1$,
as discussed at the start of section V and in quantitative
details in appendix B.
This separation of time scales, which makes our
adiabatic treatment possible, becomes exact in the large
core limit $B\rightarrow B_c$, but, importantly, 
it does not depend on $\epsilon$ being small. Therefore
the present kinematic description should rigorously extend beyond the
free-boundary limit but we shall assume that
$\epsilon\ll 1$ in this paper to compute $c_t$.

A key difference between this description 
and the one proposed in \cite{mikzyk} should already be
apparent. Namely here, the dynamics is driven entirely by the 
rigid tip region (which is just a point on the core scale). 
In contrast, in the kinematic model of \cite{mikzyk}, 
the tip motion is determined by the wave-front
dynamics via a boundary condition imposed at the tip.
In the present context, the dynamics of the spiral wave-front outside
the tip region {\it need not} be 
invoked to calculate the tip motion.

\subsection{Tangential velocity}
\label{subsecC}
The tip tangential velocity is 
determined by the controller concentration 
(equivalently the spatial variation
of excitability in which the wave-front propagates)
in the tip neighborhood. To compute this velocity
we exploit the fact that, close to the $\partial R$ boundary,
the equations of motion can be linearized 
around the critical finger.
This allows us to obtain a solvability
condition (i.e. a general condition for the
existence of a solution to these linearized equations)
that uniquely relates the tangential velocity $c_t(\{v\})$ to an 
arbitrary spatial distribution of $v$; this distribution is 
only constrained to deviate slightly from 
the rest state $v_0$ in order for the wave tip 
to remain close to the critical finger.
This solvability
condition
is first used in section IV.B in the simplest
steady-state situation where the tip propagates into a
uniform controller concentration.
The result of interest is
\begin{equation}
c_t/c_0=1\,+\,(B-B_c)/K\label{e1}
\end{equation}
where $K\simeq 0.63$ is a numerical constant.
This result implies that on the weak 
excitability side of the $\partial R$ boundary ($B>B_c$),
steady retracting waves form with $c_t/c_0=\sqrt{c_r^2+c_0^2}/c_0>1$, 
where $c_r$ is the tip retracting speed.
On the other side ($B<B_c$), spiral forms with
$c_t/c_0<1$ and a second relation discussed in the next subsection
is needed in this case to determine the
rotation rate. The calculation of
the tangential velocity is extended to the case of
steady self-interacting spirals that propagate in
a not fully recovered medium
in section IV.D, to unsteady self-interacting
spirals in section V.B, and to
spirals in an external field in section VII.B. In the
latter case, $c_t(\{v\},E_\parallel,E_\perp)$ depends both on the controller
concentration and the components $E_\parallel$ and $E_\perp$ of the external
field respectively parallel and orthogonal to $c_t$.

A relation between $B$ and an arbitrary experimental
control parameter $P_{ex}$ can be obtained 
close to the $\partial R$
boundary by simply measuring the slope $S$ of the curve
$c_t/c_0$ vs $P_{ex}$, which should be the same on both sides of
$\partial R$ and presumably simpler to obtain
on the retracting side. It then follows at once from
Eq.~(\ref{e1}) that
\begin{equation}
B-B_c=K\,S(P_{ex}-P_{ex,c})
\label{experiment}
\end{equation}
close to $\partial R$, where $P_{ex,c}$ is the value of $P_{ex}$ 
where $\partial R$ is crossed.
This relationship can be used to
relate quantitatively the results of the
rest of this paper to experiments, keeping in mind
that these results are only accurate asymptotically
close to $\partial R$ and for small $\epsilon$.

\subsection{Rotation rate}

The motion of the
wave tip region, although rigid, must generally be consistent
with the motion of the rest of the wave-front 
away from the tip. On the spiral side of $\partial R$,
the tip region must necessarily rotate to accommodate the
fact that its tip end translates at a slower speed than the
plane waves radiated outward from the core 
(on the other side of $\partial R$, $c_t>c_0$ simply 
implies retraction of the tip).
In section IV.C, we show that the tip
and the far regions
can be matched on the gently curved intermediate scale $\ell$,
yielding a rotation rate $c_t/R_i$, with
\begin{equation}
R_i\,=\,\frac{D_u}{c_{0}}\,
\left[\frac{b}{\left(1-c_t/c_0\right)}\right]^{3/2}\label{e2}
\end{equation}
where $b\simeq 2.946$ is a constant that
is obtained by matching the curved tip and far regions. 
It should be noted that
this constant differs from the constant $b'$ \cite{note32} obtained
by arbitrarily imposing a radial departure of
the wavefront from the steady-state circular core trajectory as
in Ref. \cite{mikzyk}. The 3/2 exponent, however, is the same
both here and in Ref. \cite{mikzyk} since it 
does not depend on details of
the matching on the tip scale.

Eq.~(\ref{e2}) holds both for
steady rotation (in the context of which it is
derived in section IV.C) and for unsteady rotation
owing to the aforementioned adiabatic approximation.
For steady rotation, the core radius $R_0$ 
is simply obtained by substituting the expression for $c_t$ 
from Eq.~(\ref{e1}) into Eq.~(\ref{e2}), which yields,
\begin{equation}
R_0\,=\,\frac{D_u}{c_{0}}\,
\left[\frac{bK}{(B_c-B)}\right]^{3/2}\label{e3}
\end{equation}
The generalization
to self-interacting spirals is given in section IV.D.

\subsection{Parameterization of the wave tip trajectory}

Knowing how to compute $c_t$
and $R_i$ gives in principle a complete kinematic 
theory of the wave tip motion, since this uniquely
predicts the Euclidean trajectory of the tip in time.
However to characterize analytically 
the tip dynamics in unsteady 
situations (such as drift, meander, etc)
it is convenient to measure 
the instantaneous tip position by the standard polar 
coordinates $(r,\theta)$ with respect to 
a fixed origin at the center of steady rotation
(Fig.~\ref{coordinates}),
and to relate the tip motion in these coordinates
to $c_t$ and $R_i$. This part of our
analysis is carried out in section V.A and yields 
a simple forced harmonic oscillator equation 
\begin{equation}
\frac{d^2\delta r}{dt^2}\,+\,\omega_1^2\,\delta r\,=\,
\omega_1^2 \delta R_i\label{osci}
\end{equation}
where $\delta r(t)\equiv r(t)-R_0$ is the radial displacement
of the wave tip from its radius $R_0$ of steady rotation
and $\delta R_i(t)\equiv R_i(t)-R_0$. Eq.~(\ref{osci})
is valid for a small radial displacement ($|\delta r|/R_0\ll 1$)
and is accompanied by an independent equation for the angular displacement
$\psi(t)=\theta(t)-\omega_1t$ from steady rotation. For a
small radial displacement, however, the two equations are not
coupled such that $\delta r$ can be computed independently.
Without forcing, the solution of Eq.~(\ref{osci}), 
namely an harmonic motion at frequency $\omega_1$, is a simple
superposition of the two translation modes: it
gives the tip displacement of a steady
spiral which is slightly translated with respect to the reference
unperturbed spiral. 

\subsection{Main results}

In summary, the application of the
present kinematic theory contains three steps: (i) using
a solvability condition to calculate $c_t$ in terms
of the local controller concentration, external field, etc.,
with a resulting expression that depends on the situation
considered, (ii) using Eq.~(\ref{e2}) to express $R_i$ in terms of $c_t$,
and (iii) solving Eq.~(\ref{osci}) to obtain the radial
displacement of the tip for a given forcing, which also
obviously depends on the situation considered. We now
summarize the result of this procedure for 
the selected applications examined in this paper (in a different
order than in subsequent sections).

The simplest example (section VII.A) is to compute the 
the tip motion induced by a small periodic
spatially uniform variation
of excitability 
$B(t)= B_0\,+\delta B\sin(\omega_1 t+\phi)$.
Following the above steps, and using the fact
that the perturbation is small,
we obtain at once
\begin{equation}
\frac{d^2\delta r}{dt^2}\,+\,\omega_1^2\,\delta r\,=\,
\omega_1^2 (dR_0/dB)\delta B
\sin(\omega_1 t+\phi)\label{osci1}
\end{equation}
where the function $R_0(B)$ is defined
by Eq.~(\ref{e3}), and $dR_0/dB$ is to be
evaluated at $B=B_0$. This resonant forced
harmonic oscillator equation has a 
growing sinusoidal solution with
an amplitude that increases linearly in time,
and which thus corresponds to a spiral drift 
at a speed $c_d=\omega_1(dR_0/dB)\delta B/2$, 
or $c_d=\omega_1\,dR_0/dP_{ex}\delta P_{ex}/2$
in an experiment. The action of 
an electric field is considered in section \ref{efield}
and produces a similar type of periodic forcing 
that leads to spiral drift. In agreement with
previous studies \cite{ag,stein,bel,mit,krins}
the spiral is found to drift at an angle with the external field.
This result also determines the curvature-induced
motion of a scroll filament. There the main prediction
is that rings expand in the large core limit 
(i.e. the filament tension is negative) in agreement with
previous numerical observations in this limit 
(see \cite{Biktension} and earlier references therein).

In the case of meander
the tip tangential velocity and hence
the forcing on the right-hand-side of Eq.~(\ref{osci})
depends on the radial displacement $\delta r(t)-\delta r(t-T_0)$
of the tip  after one complete rotation (Fig.~\ref{coordinates}) 
due to the self-interaction
of the wave-front with its own recovery tail. 
If this displacement is positive, the average controller 
concentration will be slightly more elevated in the tip region 
(i.e. the medium will be slightly less excitable in this region) than if
it is negative, which then affects $c_t(\{v\})$ and thus
$R_i$ and the forcing of the tip. This effect leads 
to a differential equation
with delay of the form
\begin{equation}
\frac{d^2q}{dt^2}\,+\,\omega_1^2\,q\,=\,
\omega_1^2\, m\, F\left(q(t)-
q(t-T_0)\right)
\label{osci2}
\end{equation}
where we have defined the dimensionless
radial displacement $q(t)=\delta r(t)/R_{tip}=c_0\delta r(t)/D_u$,
the parameter 
$m=3B_c(dR_0/dB)e^{-T_0/\tau_R}$,
and $F$ is a $\tanh$-shaped function
which we compute in section V.B. The saturation of $F$
at large radial displacement is due to the fact that the controller
concentration only varies appreciably on the scale of $R_{tip}$.
A linear stability analysis of this equation in
section VI.A yields that the onset of meander
occurs when $m$
exceeds a threshold $3/[8F'(0)]$ that depends in a
singular way on the diffusivity 
ratio $D_v/D_u$ and $\epsilon$.
Namely, the function $F$ 
is non-analytic at $0$ in the pure sharp boundary 
limit (\ref{eikonal},\ref{veq}) of the singly diffusive model,
which sheds some light on difficulties 
that were previously encountered when attempting to perform
a linear stability analysis in this limit \cite{ps2}.
However,
for a finite interface width $\sqrt{D_u \tau_u}$, small
compared to the spiral tip radius $D_u/c_0$,
the slope at the origin is 
finite, with $F'(0)\sim -{\rm ln}(c_0 \sqrt{\tau_u/D_u})$,
such that there is a finite meander threshold that will be
typically of order unity in experiments or simulations.
In addition, this analysis predicts that
$\omega_2/\omega_1=1/2$ at onset 
in the large core limit and a simple physical interpretation of
both the existence of a threshold and oscillatory motion
is given at the end of section VI.A. Slightly away from the large core
limit, the discussion 
in section IX leads to a modified differential equation
with delay that shows that $\omega_2/\omega_1$ increases
above $1/2$ as the core radius $R_0$ is decreased, in 
semi-quantitative agreement with numerical simulations. This
actually provides a simple picture of the onset of quasiperiodicity
(i.e. how $\omega_2/\omega_1$ becomes irrational) as one moves
away from the large core limit.

Finally, for spirals with $N$ arms, we obtain a system of $N$
coupled differential equations with delay and with the interaction between
the arms controlled by the parameter 
$m_N=3B_c(dR_0/dB)e^{-T_0/(N\tau_R)}$.
An exact linear stability analysis shows that, 
unlike for meander, there is no finite threshold of instability.
Moreover, this instability develops on a time scale proportional
to $1/m_2^2$ for $N=2$ and $1/m_N$ for $N>2$, such that the
time necessary to observe it grows exponentially 
as the $\partial R$ boundary is approached.

\section{Steady states}
\label{sstates}

We start by analyzing steady wave patterns of the free boundary problem
(\ref{eikonal}-\ref{inkin}). As shown in ref.\cite{ps1,ak1},
when excitability is decreased, the spiral core radius and spiral
period diverge
on the line $\partial R$ with
$\Delta=\Delta_c(\epsilon)$ which 
marks the lower excitability limit of spiral
wave propagation in the ($\epsilon,\Delta$) plane.
As described in section I, on the line $\partial R$, 
spirals degenerate into critical fingers that translate at
$c_0$, the plane wave speed. For $\Delta<\Delta_c(\epsilon)$,
the steady waves are retracting fingers. 
Laws for the tip retraction speed and
spiral tip divergence were obtained in ref.\cite{ak2}
from numerical computations in the neighborhood of the line 
$\Delta_c(\epsilon)$ for $\epsilon\ll 1$. 
Here, we begin by recalling the result of \cite{ak1} about
the line of existence of critical fingers. We then proceed and study steady
patterns in the neighborhood of $\Delta_c(\epsilon)$ by perturbation around
the critical fingers. On the retracting wave side, we determine the tangential
speed of the tip as a function of $\Delta-\Delta_c(\epsilon)$ by a solvability
condition \cite{pier}. This is the simplest example 
of the method that we will use in
more complicated situations to determine the tip tangential speed.
For $\Delta>\Delta_c(\epsilon)$, the critical finger winds up around
its tip and becomes a steady spiral rotating around a circular
core $R_0$ at a constant tangential speed $c_t$. We first consider the
case where $R_0$ is large enough so that the spiral front interface
can be assumed to propagate in the medium rest state (i.e. the disturbance
of the medium induced by the tip previous passage can be neglected).
We obtain analytically the 
divergence of the spiral radius by adding to the previous determination of
$c_t$, 
an analysis of the
Burton-Cabrera-Frank (BCF) 
equation \cite{bcf} in the large radius limit using matched asymptotics,
thus confirming the laws obtained in ref.\cite{ak2}. Finally, we determine
the modification of the steady spiral parameters induced by
the perturbation of the
medium characteristics due to previous passages of the spiral.

\subsection{The line $\Delta_c(\epsilon)$ of critical fingers}
\label{seccritfing}
We now examine the critical fingers that propagate in a shape
preserving way at the pulse speed $c_0$ on the
boundary $\Delta_c(\epsilon)$ in the parameter space
$(\Delta,\epsilon)$.
For a small medium excitability,
the scaling of the line $\Delta_c(\epsilon)$ is easily determined by
comparing two lengthscales \cite{ak1} as reviewed in section III.A.
Firstly, the condition that the normal velocity vanishes
at the wave tip requires that the tip radius of curvature is equal to
$\epsilon/c_0$. This gives the order of magnitude of the distance between the
wave front and back interfaces. Secondly, the front and back interfaces should
move at the same velocity. The value of the controller field $v$ should
therefore increase from $v_0=v_s-\Delta$ on the front interface to
$v_s+\Delta$ on the back interface in a time $(\epsilon/c_0)/c_0$. This gives
$\tau_e \epsilon/c_0^2\sim \Delta$ and, remembering that $c_0=\alpha \Delta$,
the scaling $\Delta_c(\epsilon)\sim (\epsilon/\alpha^2 \tau_e)^{1/3}$.

\begin{figure}
\centerline{
\psfig{file=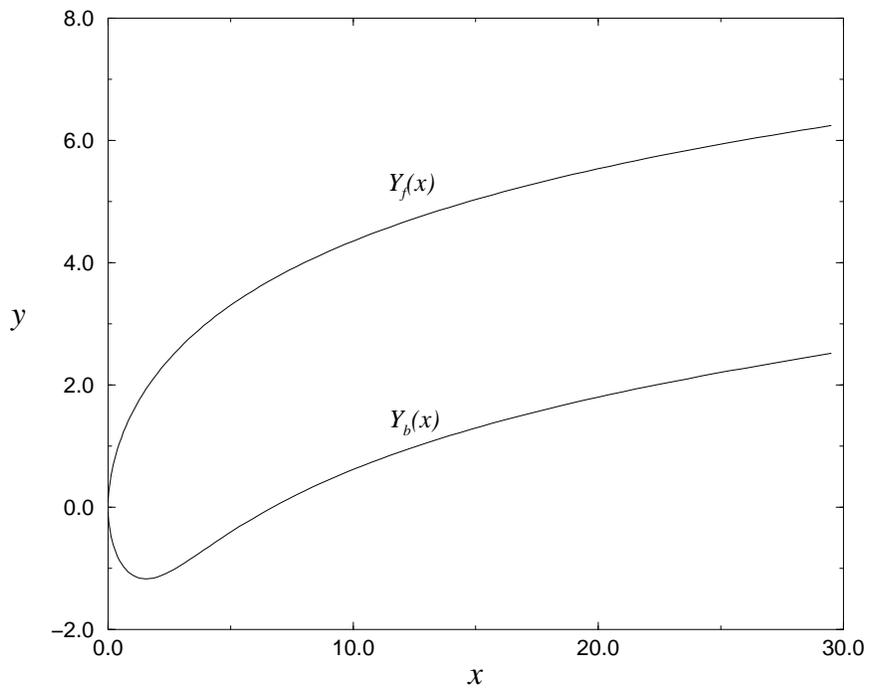,angle=-90,width=.8\textwidth}}
\caption{The critical finger: the solution of Eq.~(\ref{fingfront}) and
(\ref{fingback}) for $B=B_c$.
}
\label{critfing}
\end{figure}

A more detailed analysis is required to
determine the constant in this asymptotic relation and the critical finger
shape that we will subsequently need. 
We follow ref.\cite{ak1} and search for a steady state finger shape translating
at $c_0$, the isolated pulse speed. It is convenient to work in the frame of
the finger with the origin at the finger tip (see Fig.\ref{critfing})
 and to use as length unit
$\epsilon/c_0$, the finger tip radius. On the front interface $Y_f(x)$, the
 value of the controller field is equal to the rest state value $v_0$. At
point $x$ on the back interface, the controller field value has increased
to $v_0+\epsilon (Y_f(x)-Y_b(x))/c_0^2\tau_e$ from Eq.~(\ref{exkin}).
Eqs.(\ref{eikonal}) therefore become :
\begin{eqnarray}
\frac{d^2 Y_f}{dx^2}&=&\left[1+ (\frac{dY_f}{dx})^2\right]-\left[1
+ (\frac{dY_f}{dx})^2\right]^{3/2}
\label{fingfront}\\
\frac{d^2 Y_b}{dx^2}&=&\left[1+ (\frac{dY_b}{dx})^2\right]
+\left[1- B\left[Y_f(x)-Y_b(x)\right]\right]\left[1
+ (\frac{dY_b}{dx})^2\right]^{3/2}
\label{fingback}
\end{eqnarray}
where $Y_f(x)$ denotes the front interface of the finger and $Y_b(x)$ its
back interface. These equations depends on the single parameter
$B=\epsilon/(\alpha^2\tau_e\Delta^3)$ \cite{noteB}.
 The searched solutions should
satisfy at the tip the boundary conditions $Y_f(0)=Y_b(0)=0$,$\ dY_f/dx (0)=-dY_b/dx(0)=
+\infty$,  and asymptotically $dY_f/dx (+\infty)=dY_b/dx(+\infty)=0$. 

The solution of (\ref{fingfront}) does not require any supplementary condition.
At $x=0$, it tends to zero as $Y_f(x)\sim \sqrt{2x}$ in agreement with the
chosen tip radius of length unity. At $x=+\infty$, it diverges logarithmically
$Y_f(x)\sim 2\ln(x)$. In fact, $Y_f(x)$  
 can be obtained analytically,
\begin{eqnarray}
x&=& 2\arctan(v)+\frac{2}{v-1}-\pi
\nonumber\\
Y_f&=&\ln\left[\frac{v^2+1}{(v-1)^2}\right]
\label{csolfront}
\end{eqnarray}
with $1\le v < +\infty$.

On the contrary, Eq.(\ref{fingback}) can be solved with the appropriate boundary
 conditions only for a particular value of the parameter $B$. The two boundary
conditions at $x=0$ entirely determines the solution of (\ref{fingback})
once the front interface is determined. The solution $Y_b(x)$ should
approach $[Y_f(x)-2/B_c]$ as $x\rightarrow +\infty$ to satisfy the
boundary condition at infinity. A linearization of
(\ref{fingback}) around this asymptotic behavior gives a convergent mode and 
a divergent mode growing like $\exp(\sqrt{B_c} x)$. So, the solution
obeys the right boundary condition at $x=+\infty$ only for
the special values of $B$ which cancel the prefactor of the 
diverging mode. This is numerically found to happen for
$B_c=0.5353\cdots$ which defines the line of existence
$\Delta_c(\epsilon)$ of the critical fingers in the $(\epsilon,\Delta)$ plane.
In the following, we refer to the solution of Eq.~(\ref{fingfront}) and
(\ref{fingback}) with $B=B_c$ as the "critical finger shape". It is plotted in
Fig~.\ref{critfing}.

{\bf Remark.} On can note that at the level of Eq.~(\ref{fingfront}) and
(\ref{fingback}) the interface is continuous as well as its first two
derivatives. However, its third derivative is discontinuous at the finger
tip ($x=0,y=0$) since one has $Y_f(x)=\sqrt{2x}+x/3+\cdots$ while
$Y_b(x)=-\sqrt{2x}+x (1-2 B_c)/3+ \cdots$ (and $B_c\neq 0$). This weak
non-analyticity  can be cured by introducing
a small boundary layer near the tip as discussed in Appendix \ref{apptipbl}.

\subsection{Retracting fingers}
\begin{figure}
\centerline{
\psfig{file=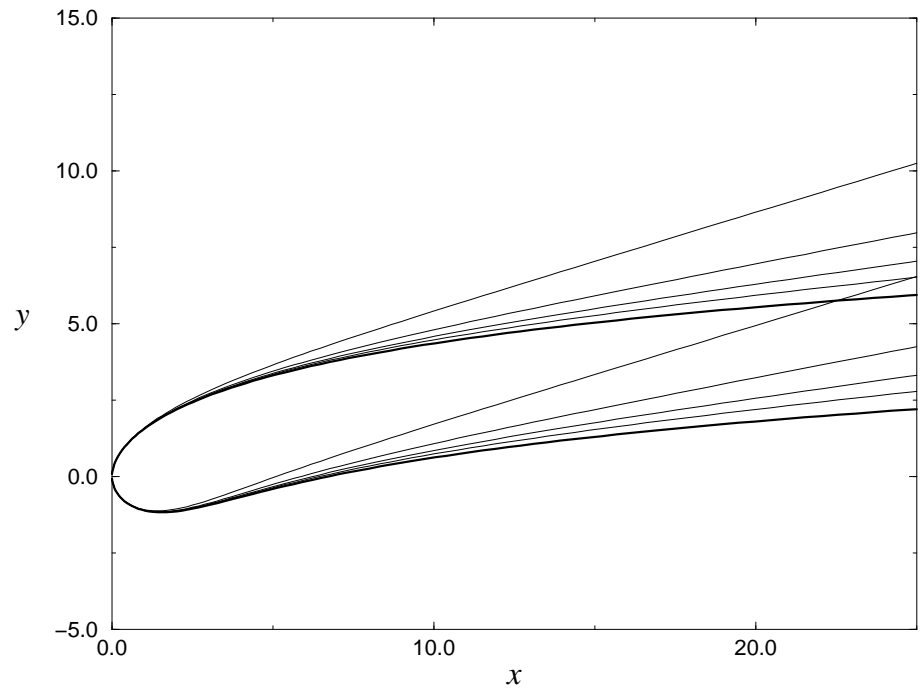,angle=-90,width=.8\textwidth}}
\caption{ Several retracting fingers ($U=1.05,1.02,1.01,1.005$
and corresponding values of $B=.5669,.5479,.5416,.53848$) compared
to the critical finger (bold line).
}
\label{retfing}
\end{figure}

We consider a medium characterized by a parameter $B=
\epsilon/(\alpha^2\tau_e\Delta^3)$ higher than $B_c$, that
is, not excitable enough to allow for the existence of spirals. We look for
steady state shapes propagating at $c_t$. We use as before $\epsilon/c_0$
as unit length where $c_0=\alpha\Delta$ is the velocity of planar front
in the considered medium. Eqs.(\ref{eikonal}) determining the
front $y_f$ and back $y_b$ interfaces become,
\begin{eqnarray}
\frac{d^2 y_f}{dx^2}&=&\frac{c_t}{c_0}
\left[1+ (\frac{dy_f}{dx})^2\right]-\left[1
+ (\frac{dy_f}{dx})^2\right]^{3/2}
\label{rfingfront}\\
\frac{d^2 y_b}{dx^2}&=& \frac{c_t}{c_0}\left[1+ (\frac{dy_b}{dx})^2\right]
+\left[1- B\frac{c_0}{c_t}\left[y_f(x)-y_b(x)\right]\right]\left[1
+ (\frac{dy_b}{dx})^2\right]^{3/2}
\label{rfingback}
\end{eqnarray}
The solutions should satisfy at the tip the same boundary conditions as
the critical fingers, 
$y_f(0)=y_b(0)=0$,$\ dy_f/dx (0)=-dy_
b/dx(0)=
+\infty$. Asymptotically, they should obey $dy_f/dx (+\infty)=
dy_b/dx(+\infty)=\sqrt{(c_t/c_0)^2-1}$.
As for the critical finger, the solution of Eq. (\ref{rfingfront}) for the
front interface can be obtained for any value of the ratio $U=c_t/c_0 >1$
and is given by
\begin{eqnarray}
x&=& \frac{2}{U}\arctan(v)-\frac{1}{U\sqrt{U^2-1}}\ln\left(
\frac{v-U-\sqrt{U^2-1}}{
v-U+\sqrt{U^2-1}}\right)-\frac{\pi}{U}
\nonumber\\
y_f&=&\frac{1}{U}\ln\left[\frac{v^2+1}{v^2+1-2Uv}\right]
\label{rsolfront}
\end{eqnarray}
with $U+\sqrt{U^2-1}<v<+\infty$.
On the contrary, Eq.(\ref{rfingback}) for the back interface can be solved
with the correct boundary condition only if  $B$
is chosen appropriately for each value of $U$. Several
obtained shapes are shown in Fig~.\ref{retfing}.
In
ref.\cite{ak2}, the solution of (\ref{rfingback}) was computed in such a way
for several
values of $U$ close to one and it was found that $c_t/c_0$ extrapolates
linearly to one when $B\rightarrow B_c$,
\begin{equation}
\frac{c_t}{c_0} =1 + \frac{B-B_c}{K}
\label{cvsb}
\end{equation}
with the constant $K\simeq.63$.

\begin{figure}
\centerline{
\psfig{file=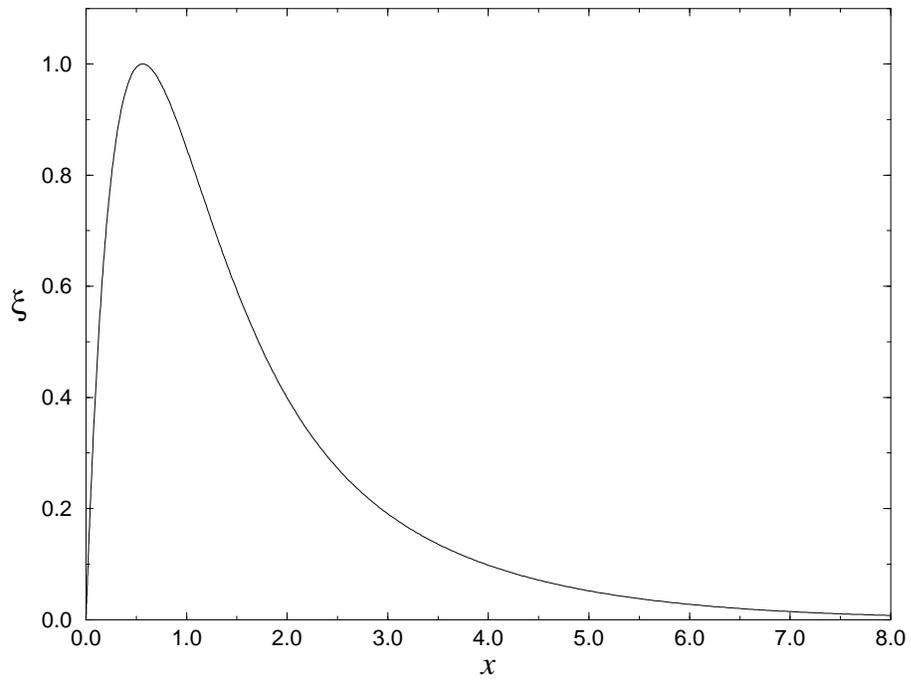,angle=-90,width=.8\textwidth}}
\caption{Graph of the zero-mode $\xi(x)$ of the operator ${\cal L}_b^{\dag}$
normalized by imposing that the maximum value of $\xi(x)$ is one. 
}
\label{figadj}
\end{figure}

We show how the result (\ref{cvsb}) can be derived by analyzing perturbatively 
Eqs.(\ref{rfingfront},
\ref{rfingback}) around the critical finger \cite{pier}.
For $|c_t/c_0-1|\ll 1$, 
the front interface
of the retracting finger $y_f$ is close to the front interface of the critical
finger $Y_f$
on distances of the finger tip small
compared to
$(c_t/c_0-1)^{-1/2}$. In this region, we linearize Eq.(\ref{rfingfront},
\ref{rfingback}) around 
the critical finger shape as  $y_f= Y_f +\delta y_f, y_b=Y_b+\delta y_b$. The
corrections $\delta y_f,\delta y_b$ obey the inhomogeneous linear equations
\begin{eqnarray}
{\cal L}_f(\delta y_f)&=& \frac{\delta c_t}{c_0} \left[1+\left(\frac{dY_f}{dx}
\right)^2\right]
\label{linrfront}\\
{\cal L}_b(\delta y_b)&=& \frac{\delta c_t}{c_0} \left[1+\left(\frac{dY_b}{dx}
\right)^2\right]
\nonumber\\
&+&\left[[B_c\,\frac{\delta c_t}{c_0}-\delta B]
[Y_f(x)-Y_b(x)]- B_c\,
\delta y_f \right]
\left[1+\left(\frac{dY_b}{dx}
\right)^2\right]^{3/2}
\label{linrback}
\end{eqnarray}
We have introduced $\delta B=B-B_c,\,\delta c_t=c_t-c_0$ and the linear
operators ${\cal L}_f,{\cal L}_b$ which are given by 
\begin{eqnarray}
{\cal L}_f &=& \frac{d^2}{dx^2}+\left[-2 +3\left[1+\left(\frac{dY_f}{
dx}\right)^2\right]^{1/2}\right] \frac{dY_f}{dx} \frac{d}{dx}
\label{oplinfront}\\
{\cal L}_b &=& \frac{d^2}{dx^2} -a(x)\frac{d}{dx}-b(x)
\label{oplinback}
\end{eqnarray}
with,
\begin{eqnarray}
a(x)&=&
\left[2+3\left[1-B_c\left(Y_f\left(x\right)-Y_b(x)\right)
\right]\left[1+\left(\frac{dY_b}{dx}\right)^2\right]
^{1/2}\right] \frac{dY_b}{dx}\nonumber\\
b(x)&=& B_c \left(1+(\frac{dY_b}{dx})^2\right)^{3/2}
\label{eqab}
\end{eqnarray}
The boundary values at the tip are $\delta y_f(0)=\delta y_b(0)=0$. For
the derivatives, one obtains using the asymptotic behavior $Y_f(x)\sim
-Y_b(x)\sim\sqrt{2x}$ near $x=0$,
\begin{eqnarray}
\frac{d \delta y_f}{dx}|_{x=0}&=&\frac{1}{3}\frac{\delta c_t}{c_0}\\
\frac{d \delta y_b}{dx}|_{x=0}&=&\frac{1}{3}\left[\frac{\delta c_t}{c_0}
(1+2 B_c) -2 \delta B\right]
\end{eqnarray}
As before, Eq.(\ref{linrfront}) can be integrated and one obtains
$\delta y_f=\eta_1 \delta c_t/c_0$ where $\eta_1$ is the solution of
\begin{equation}
{\cal L}_f (\eta_1)= 1+\left(\frac{dY_f}{dx}\right)^2
\label{eta1} 
\end{equation}
such that
$\eta_1(0)=0,\ \eta_1'(0)=1/3$. When $x\rightarrow+\infty$, $\eta_1$
grows like $x^2/6$. The situation is different for Eq.(\ref{linrback}).
For large $x$, ${\cal L}_b$ reduces to
$d^2/dx^2-B_c$. So, in general $\delta y_b$ grows exponentially like
$\exp(\sqrt{B_c}\, x)$ on distances of order one much smaller than the region
where the linearized equation (\ref{linrback}) is valid. It is only when
$\delta c_t/c_0$ is related in a particular way to $\delta B$
that the exponential growth
is absent and that $\delta y_b$ can grow algebraically like $\delta y_f$, 
as it should. In order to determine the relation between $\delta c_t/c_0$ and
$\delta B$ that should be imposed, we find it convenient to introduce the
zero mode $\xi(x)$ of the adjoint ${\cal L}_b^{\dag}$ which vanishes
(exponentially)
at infinity,
\begin{equation}
{\cal L}_b^{\dag}(\xi)=\frac{d^2\xi}{dx^2}+\frac{d}{dx}(a(x)\xi)-b(x)\xi=0,\ 
\xi(+\infty)=0
\label{adj}
\end{equation}
where the functions $a(x)$ and $b(x)$ are defined by Eq.(\ref{eqab}).
$\xi$ is uniquely defined up to a global normalization.
A local analysis  shows that $\xi$ automatically vanishes at $x=0$ and that it
tends linearly to zero when $x\rightarrow 0$. For definiteness, we normalize
$\xi(x)$ so that its maximum value is equal to 1.
A graph of $\xi$ is shown on
Fig.~\ref{figadj}. 
We now multiply both sides of Eq.(\ref{linrback}) by $\xi(x)$ and integrate
over $x$ from $x_1$ to $x_2$. Integration by parts gives for the l.h.s.,
\begin{equation}
\int_{x_1}^{x_2} \!dx\,\xi(x)\,{\cal L}_b(\delta y_b)=
\left[\xi \frac{d\delta y_b}{dx}-\delta y_b\frac{d\xi}{dx}-a(x)\xi(x)
\delta y_b\right]_{x_1}^{x_2}+\int_{x_1}^{x_2}\! dx\,\delta y_b(x)
{\cal L}_b^{\dag}
(\xi)
\label{intpart}
\end{equation}
The integral on the l.h.s of (\ref{intpart}) vanishes since ${\cal L}_b^{\dag}$
annihilates $\xi$ (\ref{adj}). Moreover, when $x_1\rightarrow 0$ and $x_2
\rightarrow+\infty$ the boundary terms also vanish when $\delta y_b$ satisfies
the correct boundary condition. Terms at $x=+\infty$ vanish when $\delta y_b$
grows algebraically since $\xi(x)$ vanishes exponentially. There is no
contribution at zero since $\delta y_b$ and $\xi(x)$ vanish linearly  which
compensate for the singular behavior of
$a(x)=-3/(2x)+\cdots$. Therefore,
the r.h.s. of (\ref{linrback}) has to satisfy the solvability condition,
\begin{equation}
\frac{\delta c_t}{c_0} [I_1 + B_c (-I_2+I_3)]-\delta B\, I_3=0
\label{eq23}
\end{equation}
where the constants $I_1,I_2,I_3$ are given by the following integrals
which have been numerically evaluated,
\begin{eqnarray}
I_1=\int_0^{+\infty}\!\!dx\,\xi(x)\left[1+\left(\frac{dY_b}{dx}\right)^2\right]
\simeq 2.771
\nonumber
\\
I_2=\int_0^{+\infty}\!\!dx\,\xi(x)\eta_1(x)
\left[1+\left(\frac{dY_b}{dx}\right)^2\right]^{3/2}\simeq
3.814
\nonumber\\
I_3=\int_0^{+\infty}\!\!dx\,\xi(x)\, \left[Y_f(x)-Y_b(x)\right]
\left[1+\left(\frac{dY_b}{dx}\right)^2\right]^{3/2}\simeq 7.708
\label{i123}
\end{eqnarray}
Eq.(\ref{eq23}) shows that the tangential velocity
 of the retracting finger tip depends linearly on the
departure of $B$ from $B_c$ as stated in Eq.~(\ref{cvsb}).
The proportionality constant $K$ is  in excellent
agreement with the value obtained by numerical extrapolation in ref.\cite{ak2}.
\begin{equation}
K=B_c+(I_1-B_c\,I_2)/I_3\simeq.630
\label{eqK}
\end{equation}
Note that  the values (\ref{i123}) of the integrals depend on the 
normalization of $\xi$ but that the expression of
the physical constant $K$ appear as a ratio of such integrals and is thus
independent of this (arbitrarily chosen) normalization. It is also important
to remark that Eq.~(\ref{eq23}) shows 
that the tangential tip velocity is an appropriate quantity for a perturbative
calculation around $\partial R$ since it has a smooth behavior
when $\partial R$ is crossed. This should be contrasted
with the retracting tip velocity which
decreases like the square root of the distance to $\partial R$ on the 
retraction side and does not appear to have a simple continuation on the
spiral side of $\partial R$.

\subsection{Steadily rotating spirals}
\label{srspi}
For $B=\epsilon/(\alpha^2\tau_e\Delta^3)<B_c$,  steady spiral
waves exist. Their tip rotates around a circular core $R_0$ at a 
constant tangential tip velocity $c_t=\omega_1 R_0$. When $B\rightarrow B_c$,
$R_0$ diverges, $c_t\rightarrow c_0$ and the tip of the spiral becomes closer
and closer to a critical finger. In this subsection, we determine $R_0$ and
$\omega_1$ as a function of $B$ ("the excitability of the medium"). We
begin by considering spirals of radius large enough so that one can neglect
the disturbance of the medium due to the spiral previous passage. In this case, 
the front interface can be assumed to propagate in the medium rest state.
The spiral shape is analyzed by decomposing it into three overlapping regions
where different approximations can be performed. Close to the tip, on distances
of order $R_{tip}=\epsilon/c_0$, the curvature
of the tip trajectory can be neglected and a transposition of the analysis of
the previous subsection shows that the tangential velocity is linearly related
to $\delta B= B-B_c$ by Eq.~(\ref{cvsb}), namely $\delta c_t=\delta B/K$
(both sides being negative now). Far from the tip, it is
the effect
of the interface curvature on the normal velocity (Eq.~(\ref{eikonal})) which 
can be neglected. The normal velocity can be taken constant,
equal to $c_0$ and the spiral shape is then simply determined.
These two approximate descriptions match at a distance
of order $\ell$ from the spiral tip in an intermediate region
where the interface is almost normal to the tip circle of rotation and the
interface curvature is small. The intermediate scale $\ell$ appears as the
balance between two effects. On one hand, the tip tangential velocity is
smaller than $c_0$ by about $\omega_1 \ell$ for purely kinematical reasons so
that $|\delta c_t| \sim c_0 \ell/R_0$. On the other hand, $\ell$
is the distance where curvature effects become small enough to
be comparable to this velocity drop.
At a distance $\ell$ from its tip,
the critical finger curvature is of the order $2 R_{tip}/\ell^2$. This provides
the alternative estimate $|\delta c_t| \sim \epsilon R_{tip}/\ell^2$. Comparing
both expressions and remembering that $R_{tip}=\epsilon/c_0$ 
gives $\ell\sim (R_{tip}^2 R_0)^{1/3}$ and $|\delta c_t/c_0|
\sim (R_{tip}/R_0)^{2/3}$. The detailed analysis reported below replaces this
simple order of magnitude estimate by the precise asymptotic relation,
\begin{equation}
\frac{c_0-c_t}{c_0}\simeq b \left(\frac{R_{tip}}{R_0}\right)^{2/3}
\label{rvsc}
\end{equation}
where the numerical constant $b$ is obtained from the first zero $a_1$ of Airy
function $Ai$ \cite{abr}, $b=-2^{1/3} a_1\simeq 2.946$
\cite{note32}. Comparing (\ref{rvsc})
with (\ref{cvsb}) determines the frequencies and core radii of steady spirals
near the line $\Delta_c(\epsilon)$,
\begin{equation}
R_0 = R_{tip} \left(\frac{b K}{B_c-B}\right)^{3/2},\ 
\omega_1 = \frac{c_0}{R_{tip}} \left(\frac{B_c-B}{b K}\right)^{3/2}
\label{asyRom}
\end{equation}
\subsubsection{Front interface}
\label{matchas}
We first consider the front
interface and assume that the spiral propagates in the medium rest
state (this is of course justified only if the spiral period is long enough and
it requires to be sufficiently close to the line $\partial R$).
The equation
for the front spiral interface is then identical to the classic equation for
the growth of screw dislocations on crystal surfaces \cite{bcf},
For a steady rotation at frequency 
$\omega_1$ in a counterclockwise direction, Eq.(\ref{eikonal}) gives
 for the front spiral interface in polar coordinates
$(r,\theta)$,
\begin{equation}
r\,\omega_1= c_0 \left(1+\left(r\frac{d\theta_f}{dr}\right)^2\right)^{1/2}
+ \epsilon\,\left( \frac{d\theta_f}{dr} +\frac{\frac{d}{dr} (r 
\frac{d\theta_f}{dr})
}{1+\left(r\frac {d\theta_f}{dr}\right)^2}\right)
\label{bcf}
\end{equation}
with the boundary condition at infinity 
$d\theta_f/dr \rightarrow -\omega_1/c_0$.
Rescaling coordinates makes it clear that Eq.(\ref{bcf}) depends on
the single dimensionless parameter $\Omega=\omega_1 \epsilon/c_0^2$. For
$0<\Omega<.331$, it is found that $d\theta_f/dr\rightarrow +\infty$ at $r=R_0$
when Eq.~(\ref{bcf}) is integrated from $r=+\infty$. $R_0$ is the location of
the spiral tip and is found to increase from $0$ to $+\infty$ when
$\Omega$ decreases from $.331$ to $0$. The limit $\Omega\rightarrow .331$
has been considered in previous works \cite{fife,klr}.
 We focus here on the other
limit $\Omega\rightarrow 0$ where the excited region width ($\sim \epsilon/c_0$)
becomes much smaller than the core radius ($\sim\omega_1/c_0$). In this limit,
the front spiral interface can be separated into three distinct regions:\\
- Far from
the spiral tip (the outer region), the interface scale of variation is $c_0/\omega_1$. Introducing
the rescaled coordinates $r=  z c_0/\omega_1$ shows that the terms involving the
interface curvature are multiplied by the small parameter $\Omega$. Neglecting
them,
Eq.~(\ref{bcf}) becomes 
\begin{equation}
z\frac{d \theta_{out}}{dz}= -\sqrt{z^2-1}
\label{out}
\end{equation}
which is of course integrable.
This first approximation breaks down near $z=1$, where  the solution
of Eq.~(\ref{out}) has a fast variation on the $z$-scale and the formally
negligible terms are important.\\ 
-Close to the spiral tip, Eq.~(\ref{bcf}) can be
simplified in a different way. One can introduce the radial distance $x$
 from the
tip circle of rotation measured in unit of the tip radius such that $r=R_0 +
\epsilon/c_0 \ x$ and the tangential displacement $y\epsilon/c_0=R_0 \theta_f$.
At lowest order in $\epsilon/(c_0 R_0)$, 
Eq.~(\ref{bcf}) becomes identical
in these variables to Eq.~(\ref{fingfront}) for the front
interface of a critical finger.\\
- These two different descriptions do not 
directly match. The transition occurs in an intermediate region where 
the interface curvature is small and the interface tangent almost radial.
We thus assume (and check {\em a posteriori}) that $dy/dx$ is small and
expand the square root and denominator in Eq.~(\ref{bcf}). This gives
\begin{equation}
\frac{c_t-c_0}{c_0}+ \frac{\epsilon}{c_0 R_0} x = \frac{1}{2} \left(
\frac{dy}{dx}\right)^2 + \frac{d^2 y}{dx^2}
\label{eqint}
\end{equation}
where the tangential tip velocity $c_t=R_0 \omega$ has been introduced.
The different terms of Eq.~(\ref{eqint}) are of comparable magnitude for
$x\sim (R_0 c_0/\epsilon)^{1/3},\, dy/dx\sim (\epsilon/R_0 c_0)^{1/3}\,$ and
$c_t/c_0-1\sim (\epsilon/R_0 c_0)^{2/3}$. In the limit $R_{tip}=\epsilon/c_0\ll 
R_0$, this justifies the expansion leading to Eq.~(\ref{eqint}) and the neglect
of higher order terms. Introducing the rescaled variable $\xi= x\, (\epsilon
/(2c_0 R_0))^{1/3}$,
 Eq.~(\ref{eqint}) becomes
\begin{equation}
\frac{1}{2} \frac{d^2y}{d\xi^2}+\frac{1}{4}\left(\frac{dy}{d\xi}\right)^2=
\xi + a_1
\label{riccati}
\end{equation}
where we have defined
\begin{equation}
a_1= 2^{-1/3} \left(\frac{c_0 R_0}{\epsilon}\right)^{2/3} \, \frac{c_t-c_0}{c_0}
\label{a1}
\end{equation}
 The Riccati equation (\ref{riccati})  can be 
transformed into the linear Airy equation. Matching with
(\ref{out}) imposes that $dy/d\xi <0$ when $\xi\rightarrow+\infty$.
This imposes that the Airy function decreases at infinity and
be proportional to $Ai$ \cite{abr}. It 
gives
$dy/d\xi= 2\, A_i'(\xi+a_1)/A_i(\xi+a_1)$. Using the asymptotic behavior
\cite{abr} $ 
Ai(\xi)\sim 1/2 \pi^{-1/2} \xi^{-1/4} \exp(-2/3\, \xi^{3/2})$, one indeed checks
that
the obtained 
large $\xi$ behavior $dy/d\xi\sim -2\sqrt{\xi}$ coincides with the behavior
of (\ref{out}) near $z=1$.  Matching with the tip region requires that the small
$\xi$ behavior of $dy/d\xi$ coincides with the
asymptotic behavior of (\ref{fingfront}) when $x\rightarrow+\infty$, namely
$dy/dx\sim 2/x^2$. This requires $a_1$ to be a zero of $Ai$.  Since
$y(\xi)$ should be well-defined for all real positive $\xi$ it is necessarily
the first one $a_1=-2.3381\cdots$\cite{abr}.
Comparing with the definition (\ref{a1}) of $a_1$ directly leads to the relation
(\ref{rvsc}).

The  
relation $\omega_1(R_0)$, numerically determined in \cite{pan},
 was approximately
obtained in \cite{mikzyk,keenty} 
by assuming a radial departure of the front
interface imposed on a fictitious inner radius $R_0$. This boundary condition
is equivalent to requiring that  $\theta$ be maximum at $R_0$. It
is worth noting that, in the present limit, it  
would 
simply amount to replace the exact value of the constant 
$a_1$ by the location of the maximum
of $Ai$ namely $a'_1=-1.01879\cdots$\cite{abr}. Correspondingly, this would
replace the exact value $b\simeq 2.946$ in Eq.(\ref{rvsc}) by $b'\simeq 1.283$.

\subsubsection{The back interface}

The equation for the back interface reads in polar coordinates,
\begin{equation}
r\,\omega_1= \left(-c_0+\frac{\alpha}{\omega_1 \tau_e}\left(\theta_f-\theta_b
\right)\right) 
\left(1+\left(r\frac{d\theta_b}{dr}\right)^2\right)^{1/2}
+ \epsilon\,\left( \frac{d\theta_b}{dr} +\frac{\frac{d}{dr} (r
\frac{d\theta_b}{dr})
}{1+\left(r\frac {d\theta_b}{dr}\right)^2}\right)
\label{bcfback}
\end{equation}
As for the front interface, we proceed by separately analyzing three regions. We
consider first the tip region which plays here the dominant role.
 Introducing as 
before the coordinates $x$ and $y$ such that $r=R_0+\epsilon/c_0 x,
\ y\epsilon/c_0=R_0\theta_b$, Eq.~(\ref{bcfback}) becomes at lowest order in
$\epsilon/c_0 R_0$ identical to Eq.~(\ref{rfingback})
 describing the back interface of
retracting fingers (except that now $c_t<c_0$ and $B<B_c$. As in this previous 
case, requiring that the back interface does 
not diverge exponentially from the 
front interface relates $c_t/c_0$ to $B$. For $B$ close to $B_c$, one can 
linearize around the critical finger and follow the previous analysis
(\ref{linrfront}-\ref{eqK}) which leads to (\ref{cvsb}). The comparison of
 (\ref{cvsb}) and (\ref{rvsc})
gives the expression (\ref{asyRom}) for the spiral core radius and frequency of 
rotation as a function of $B$.

As one moves away from the tip, the back $Y_b$ of the critical finger
relaxes exponentially toward $Y_f-2/B_c$ on the scale of the finger tip width.
The equations describing the back and
front interfaces 
are thus essentially identical in the intermediate and far regions and the 
analysis of subsection \ref{matchas} applies as well to the back equation.

\subsection{Steady self-interacting spirals}
The analysis of the previous section applies when the spiral period 
$T_0=2\pi/\omega_1$ is long
enough compared to the recovery time constant $\tau_R$ 
so that the front interface can be assumed to propagate in the
medium rest state. This applies for
$\Delta$ sufficiently close to $\Delta_c(\epsilon)$ but as the medium 
excitability increases the spiral radius decreases and the front interface
begins to feel the medium disturbance due to the spiral previous passages. This
eventually leads to spiral meander as we show in the next 
section. As a preliminary step,
we analyze here the influence of this medium modification on the steady spiral
parameters $R_0$ and $\omega_1$.

The concentration of the controller $v$ on the front and
 back interfaces follow from Eq.~(\ref{exkin}) and (\ref{inkin}). For a
spiral rotating steadily at frequency $\omega_1$ in a counterclockwise
direction, they are given by,
\begin{eqnarray}
v_f(r)&=& v_0 + \delta v_f(r)\\
v_b(r)&=& v_f(r)+\frac{\theta_f(r)-\theta_b(r)}{\omega_1 \tau_e}
\end{eqnarray}
with
\begin{equation}
\delta v_f(r)=\frac{\theta_f(r)-\theta_b(r)}{\omega_1\tau_e}
\frac{\exp-\left(\frac{2\pi+\theta_b(r)-\theta_f(r)}{\omega_1\tau_R}\right)}{
1-\exp-\left(\frac{2\pi+\theta_b(r)-\theta_f(r)}{\omega_1\tau_R}\right)}
\label{dvf}
\end{equation}
Near the line $\Delta_c(\epsilon)$, $\omega_1$ tends to zero,  $\delta v_f(r)$
 becomes negligible and the concentration of $v$ on the front interface can
be taken equal to $v_0$ as done in the previous  subsection. This approximation
is justified as long as $\delta v_f$ induces a change in the front 
velocity which is
negligible compared to the difference $c_t-c_0$
between the tip velocity and $c_0$. That is, for $\exp(-2\pi R_0/c_0\tau_R)\ll 
\delta c_t/c_0$ or using the estimate (\ref{asyRom}),
$\exp(-2\pi R_{tip}/c_0 (bK/B_c-B)^{3/2})\ll B_c-B$. Therefore, one can neglect
the perturbation of the medium as long as $B_c-B\ll \epsilon^{2/9}$ (up to
logarithmic corrections) or equivalently $\Delta-\Delta_c\ll \epsilon^{5/9}$.
The results (\ref{asyRom}) are modified when $\delta v_f(r)$ becomes comparable
to $\delta c_t$. The transition regime where $\delta v_f(r)$ is still small
can be analyzed along the lines of the previous subsections
\begin{eqnarray}
r\,\omega_1&=& c(v_0+\delta v_f(r))
 \left(1+\left(r\frac{d\theta_f}{dr}\right)^2\right)^{1/2}
+ \epsilon\,\left( \frac{d\theta_f}{dr} +\frac{\frac{d}{dr} (r
\frac{d\theta_f}{dr})
}{1+\left(r\frac {d\theta_f}{dr}\right)^2}\right)
\label{frontinter}
\\
r\,\omega_1&=& -c(v_b(r))
\left(1+\left(r\frac{d\theta_b}{dr}\right)^2\right)^{1/2}
+ \epsilon\,\left( \frac{d\theta_b}{dr} +\frac{\frac{d}{dr} (r
\frac{d\theta_b}{dr})
}{1+\left(r\frac {d\theta_b}{dr}\right)^2}\right)
\end{eqnarray}
In the tip region, it is useful to introduce as previously the coordinates
$x$ and $y$, with $r=R_0 +\epsilon/c_0 x, y \epsilon/c_0=R_0 \theta$. At 
lowest order in $\epsilon/(c_0 R_0)$, Eq.~(\ref{frontinter}) becomes
\begin{equation}
\frac{d^2 y_f}{dx^2}=\frac{c_t}{c_0}
\left[1+ (\frac{dy_f}{dx})^2\right]-\left[1-\frac{\alpha}{c_0}
\delta v_f(r)\right]\left[1
+ (\frac{dy_f}{dx})^2\right]^{3/2}
\label{eq41}
\end{equation}
We are interested in the parameter region where $\alpha \delta v_f(r)/c_0$ is
of the same order as $\delta c_t/c_0$. As found above, this happens when
the spiral period is large but only logarithmically in $\epsilon$. This allows
to expand the exponential in (\ref{dvf}) and to obtain the expression of
the medium perturbation as a function of the critical finger shape,
\begin{equation}
\frac{\alpha}{c_0}\delta v_f(r)= B_c\, [Y_f(x)-Y_b(x)]\,
 \exp(-\frac{2\pi R_0}{
c_0 \tau_R})
\label{dvflo}
\end{equation}

We expand the spiral front
 in the tip region around the critical finger shape as
$y_f(x)=Y_f(x)+\delta y_f(x)$. The correction $\delta y_f(x)$ obeys the 
equation
\begin{equation}
{\cal L}_f(\delta y_f)= \frac{\delta c_t}{c_0} \left[1+\left(\frac{dY_f}{dx}
\right)^2\right] +B_c e^{-\frac{2\pi R_0}{
c_0 \tau_R}}(Y_f(x)-Y_b(x)) \left[1+\left(\frac{dY_f}{dx}\right)^2\right]^{3/2}
\end{equation}
where the linear operator ${\cal L}_f$ is defined by (\ref{oplinfront}). 
$\delta y_f$ can be expressed as
\begin{equation}
\delta y_f = \frac{\delta c_t}{c_0}\, \eta_1 + B_c e^{-
\frac{2\pi R_0}{c_0\tau_R}}\,
 \eta_{v,0}
\end{equation}
where $\eta_1$ is defined in (\ref{eta1}).  $\eta_{v,0}$ obeys
\begin{equation}
{\cal L}_f (\eta_{v,0})= (Y_f(x)-Y_b(x)) 
\left[1+\left(\frac{dY_f}{dx}\right)^2\right]^{3/2}
\label{etav0}
\end{equation}
with the boundary conditions $\eta_{v,0}(0)=0,\frac{d\eta_{v,0}}{dx} (0)=2/3$. 

In the same way, we obtain, in the tip region,
 the lowest order correction $\delta y_b$ to the
back interface of the critical finger $Y_b(x)$
\begin{eqnarray}
{\cal L}_b(\delta y_b)&=& \frac{\delta c_t}{c_0} \left[1+\left(\frac{dY_b}{dx}
\right)^2\right]
\nonumber\\
&+&
\left[[B_c\,\frac{\delta c_t}{c_0}-\delta B-B_c e^{-(2\pi R_0/c_0\tau_R)}]
[Y_f(x)-Y_b(x)]- B_c\,
\delta y_f \right]
\left[1+\left(\frac{dY_b}{dx}
\right)^2\right]^{3/2}
\label{linintback}
\end{eqnarray}
where the linear operator ${\cal L}_b$ is defined by Eq.~(\ref{oplinback}).
Eq.~(\ref{linintback}) is similar to Eq.~(\ref{linrback})
 and can be analyzed in the same way. The solvability condition
 that should be obeyed
in order for $\delta y_b$ not to diverge exponentially as $x\rightarrow +\infty$
is found by integrating both members of (\ref{linintback}) with the zero mode
$\xi(x)$ of the adjoint of ${\cal L}_b$. This gives the following generalization
of Eq.~(\ref{cvsb})
\begin{equation}
K \frac{\delta c_t}{c_0} =\delta B + B_c\  J\ 
 \exp(-\frac{2\pi R_0}{c_0 \tau_R})
\label{eq45}
\end{equation}
where $K\simeq .630$ (Eq.~(\ref{eqK}))and the constant $J$ is
\begin{equation}
J=1+B_c \frac{I_{v,0}}{I_3}\simeq=1.872
\label{eqJ}
\end{equation}
$I_3$ is defined by (\ref{i123}) and $I_{v,0}$ is given
in terms of $\xi(x)$ (\ref{adj}) and $\eta_{v,0}$ (\ref{etav0}) by
\begin{equation}
I_{v,0}=\int_0^{+\infty}\!\!dx\, \xi(x)\,\eta_{v,0}
 \left[1+\left(\frac{dY_b}{dx}
\right)^2\right]^{3/2}= 12.553
\label{cvsbint}
\end{equation}

To complete the analysis, it remains to match the tip region to the outer part
of the spiral. As one moves away from the tip, the finger width $Y_f(x)-Y_b(x)$
relaxes exponentially toward its asymptotic value $2/B_c$ on the scale of the
tip region. Therefore in the intermediate and far region, $\alpha \delta v_f(r)
/c_0$ (\ref{dvflo}) is equal to lowest order to $2\exp(-2\pi R_0/c_0\tau_R)$ and the matching equation
becomes instead of (\ref{eqint})
\begin{equation}
\frac{c_t-c_0}{c_0}+2 \exp(-\frac{2\pi R_0}{c_0\tau_R})
+ \frac{\epsilon}{c_0 R_0} x = \frac{1}{2} \left(
\frac{dy}{dx}\right)^2 + \frac{d^2 y}{dx^2}
\label{eqint2}
\end{equation}
Matching with the tip region gives in a similar way
\begin{equation}
2^{-1/3} \left(\frac{c_0 R_0}{\epsilon}\right)^{2/3} \left[ \frac{c_t-c_0}{c_0}+2 \exp(-\frac{2\pi R_0}{c_0\tau_R})\right]
=a_1\simeq-2.338\cdots
\label{rvscint}
\end{equation}
The difference between Eq.~(\ref{rvscint}) and (\ref{rvsc}) is simply that
$c_t$ is not compared to the velocity of a single planar pulse but to the
velocity of a train of pulses of wavelength $2\pi R_0$ 
(the asymptotic wavelength
of the spiral to lowest order). Comparing (\ref{eq45}) and (\ref{rvscint})
determines the radius $R_0$ of a steadily rotating spiral as a function of the
medium characteristics $B$,
\begin{equation}
B_c-B= K b \left(\frac{\epsilon}{c_0 R_0}\right)^{2/3}+ (B_c J +2 K) 
\exp(-\frac{2\pi R_0}{c_0\tau_R})
\label{rint}
\end{equation}
The medium disturbance due to the spiral previous passage has two distinct 
effects which are comparable to lowest order:\\
-the medium "excitability" is reduced in the tip region which modify the
tip velocity (Eq.~(\ref{eq45})),\\
-the tip velocity should be compared to the velocity of a periodic train
of planar waves which is slower than the velocity of a single planar pulse.

\section{Derivation of kinematic theory}
\label{secdyn}

We consider now the spiral dynamics in the vicinity of the line $\Delta_c
(\epsilon)$ (for $\epsilon\ll 1$). In this limit, several
simplifying features made the previous analysis of
the steady states possible. These still hold when one is interested in
an unsteady motion taking place on a time scale comparable to the steady
rotation period which is long compared to the time
scales of the internal modes of the wave tip:\\
- i) the dominant effect which shapes  both the steady spirals
 and retracting fingers tips 
is the curvature dependence of the normal
velocity. As a consequence, the shape of a wave tip is
close to a critical finger 
up to a distance $\ell$ from the tip where
the curvature effects have become small enough to be comparable to the
velocity difference between the tip and planar front velocity,
namely when $c_0-c_t\sim \epsilon R_{tip}/\ell^2$ (where we have
evaluated the curvature $-d^2Y_f/dx^2\sim R_{tip}/\ell^2$ 
at $x\sim \ell$ using the asymptotic behavior
$Y_f/R_{tip}\sim {\rm ln}(x/R_{tip})$ for $x/R_{tip}\gg 1$).
This yields the relation $\ell \sim R_{tip}/\sqrt{1-c_t/c_0}$ that
remains also true in the unsteady case.
The motion of this 'solid' shape can be determined from the knowledge
of its instantaneous tangential velocity $c_t$
and of its instantaneous rotation rate $\omega$ obtained by extending  our
previous analysis of the steady states,
\\
-ii) the tangential velocity $c_t$ depends on
the 'average' concentration of the controller $v$ in the vicinity of the tip.
The precise definition of the average is obtained by using a solvability
condition which generalizes Eq.(\ref{eq23}) and (\ref{eq45}).\\
-iii)  a tangential velocity $c_t$ smaller
than the asymptotic normal velocity $c_0$
 of the wave gives rise to a rotation of
the solid tip at a rate $\omega$ which can be estimated as in the steady case.
As said above,
the wave tip 
has a solid character (i.e.  is close to a critical finger shape)
up to a  distance 
 $\ell \sim R_{tip}/\sqrt{1-c_t/c_0}$
 from the tip. Since
kinematics requires that $\omega\ell\sim c_0-c_t$, one obtains for the rotation
rate $\omega\sim
c_0/R_{tip} (1-c_0/c_t)^{3/2}$. As shown in Appendix \ref{appdyn}, 
this relation is established on the time scale
$\sim R_{tip}/(c_0 -c_t)$ much shorter than the steady rotation 
period. Therefore, on this latter slow time scale, the slowly varying rotation
rate is linked in an adiabatic manner to the slowly varying tangential velocity
by the same relation Eq. (\ref{rvsc}) or (\ref{rvscint}) 
which relates the steady state frequency to the
tip velocity.

We begin our analysis by considering the kinematics of the wave tip motion.
We then compute the tangential velocity of the tip as a function of the
concentration of the controller $v$ in the medium left by previous passages of
the wave. As a result, we obtain an ordinary differential equation
with delay which describes the motion of the wave tip. An analysis of
this equation at the linear and the weakly
non-linear levels determines the
characteristics of the meandering instability near threshold in the weak
excitability limit.
\subsection{Parameterization of the wave tip motion}
\label{kin}
We use polar coordinates $(r,\theta)$
 with the origin at the center of the circular steady
spiral
core.
The wave tip motion is determined by its tangential velocity $c_t(\{v\})$,
a functional of the (space and time dependent) 
controller concentration  which will be computed in the
next subsection, and by the rotation rate of the shape (or, equivalently
by the radius of curvature of the tip trajectory). We use a complex notation
$z(t)=r(t)\exp(i\theta)$
 to denote the tip position. Then, the tip velocity is $|\dot{z}|$
and the shape rotation rate $Im(\ddot{z} \dot{\bar z}/|\dot{z}|^2)$ where
 time differentiation is denoted by a dot and $\bar z$ is the complex conjugate of $z$.
The tip motion is thus determined by the two equations
\begin{eqnarray}
|\dot{z}|&=&c_t(\{v\})
\label{eqct}\\
Im(\ddot{z} \dot{\bar z}/|\dot{z}|^2)&=&c_t(\{v\})/R_i[c_t(\{v\})]
\label{eqri}
\end{eqnarray}
where at this stage $c_t(\{v\})$ can be thought of as a given function of
time. The instantaneous radius of rotation $R_i$ is a function
of $c_t(\{v\})$ given to
lowest order in the interaction parameter by Eq.~(\ref{rvscint}),
\begin{equation}
 \left(\frac{c_0 R_i}{\epsilon}\right)^{2/3} \left[ \frac{c_t-c_0}{c_0}+2 \exp(-\frac{2\pi R_0}{c_0\tau_R})\right]
=-b
\label{rivscint}
\end{equation}
We will actually find that the meander threshold occurs before
a significant modification of the steady state radius by the interaction so
that Eq.~(\ref{rivscint}) can be replaced by the simpler Eq.~(\ref{rvsc})
\begin{equation}
R_i[c_t]\simeq\frac{\epsilon}{c_0}\left(\frac{b\, c_0}{c_0-c_t}\right)^{3/2}
\label{rivsc}
\end{equation}
 
We consider the motion
of  a spiral tip which is
displaced from its steady state position
 $z=(R_0+\epsilon q(t)/c_0) e^{i\omega_1 t + \psi(t)}$
(see Fig.~\ref{figcoord}). We restrict ourselves 
to displacements
of the tip which are comparable to the tip radius of curvature
$\epsilon/c_0$ ( i.e. $q(t)\sim 1$) and therefore 
small compared to the radius of the steady core $R_0$.
As a consequence of the tip displacement, the controller concentration and thus
$c_t(\{v\})$ and $R_i$ depart slightly from their steady-state values,
$c_t(\{v\})=c_t^0+\delta c_q(\{v\}), R_i=R_0 +\delta R_i$. 

 We assume (and will check {\em a posteriori})
that the time scale of
the unsteady motion is of the order of the steady state period 
$T_0=2\pi/\omega_1$. 
We expand Eq.~(\ref{eqct}) and (\ref{eqri})
in the small
parameter $\epsilon/(c_0 R_0)$
 and keep
only the dominant terms,$\  
|\dot{z}|=\omega_1 R_0 (1+\dot{\psi}/\omega_1+ q\epsilon/(c_0R_0)+\cdots),\ 
Im(\ddot{z} \dot{\bar z}/|\dot{z}|^2)=\omega_1 (1+\dot{\psi}/\omega_1-
\epsilon \ddot{q}/(c_0\omega_1^2R_0)+\cdots)$.
Eq.~(\ref{eqct})
gives therefore at lowest order in $\epsilon/(c_0 R_0)$
\begin{equation}
\dot{\psi}=\frac{\delta c_q}{R_0}-\frac{\epsilon}{c_0 R_0}\, \omega_1 q
\label{eqct2}\\
\end{equation}
which shows that $\dot{\psi}/\omega_1\sim \epsilon/(c_0 R_0)$.
Using this scaling, 
Eq.(\ref{eqri}) becomes at lowest order 
\begin{equation}
- \frac{\epsilon}{c_0 R_0} \frac{\ddot{q}}{\omega_1} +\dot{\psi}
=\frac{\delta c_q}{R_0}- c_t^0
\frac{\delta R_i}{R_0^2}
\label{eqri2}
\end{equation}
We obtain the  equation for the radial motion of the tip
by substituting in (\ref{eqri2}) 
the expression (\ref{eqct2}) 
of $\dot{\psi}$,
\begin{equation}
\ddot{q}+\omega_1^2 q= \omega_1^2 \delta R_i c_0/\epsilon
\label{eqm0}
\end{equation}
Finally, it is convenient to use the tip angular position
$\theta=\omega_1 t +\psi(t)$ instead of time. To lowest order in 
$q/R$, this simply gives
\begin{equation}
\frac{d^2q}{d\theta^2}+ q=  \frac{c_0 R'_i[c_t^0]}{\epsilon} \delta c_q
\label{eqm1}
\end{equation}
with
\begin{equation}
\frac{c_0 R'_i[c_t^0]}{\epsilon}= \frac{3}{2b\, c_0}
\left(\frac{c_0 R_0}{\epsilon}
\right)^{5/3} 
\label{dri}
\end{equation}
from a differentiation of Eq.~(\ref{rivscint}).
In order to have a closed equation for $q$, it remains to express
$\delta c_q$ in terms of the previous 
positions of the wave. 
We now proceed to this task. 

\subsection{Computation of the tangential tip velocity for
self-interacting spirals}

\begin{figure}
\centerline{
\psfig{file=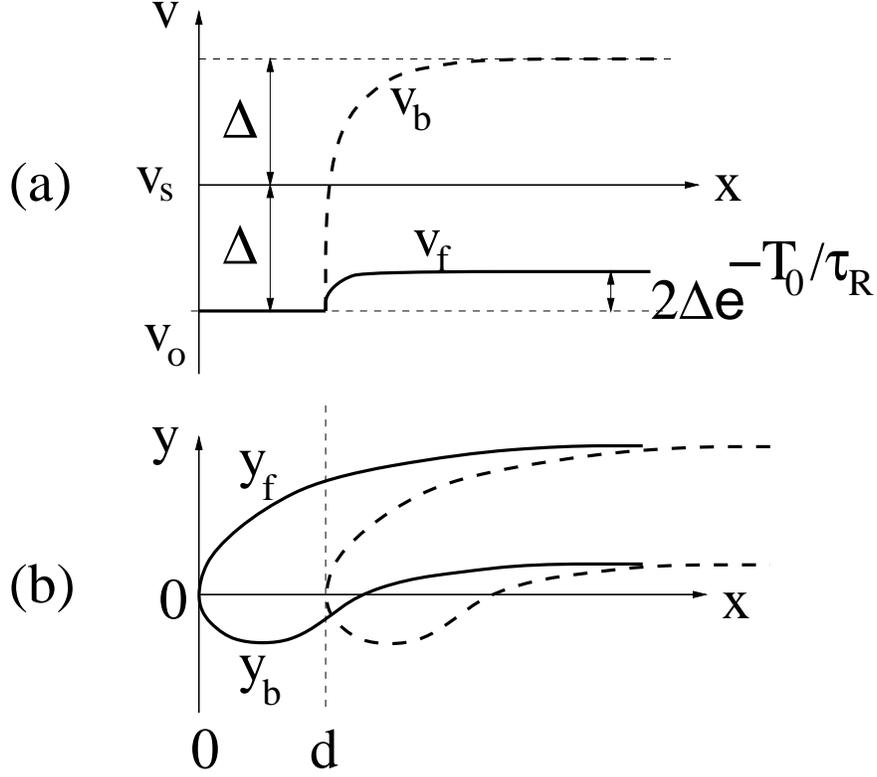,width=.7\textwidth}}
\bigskip
\caption{Schematic plot illustrating: 
(a) the variation, 
of the controller field $v$ on the instantaneous 
wavefront, $v_f$ (solid line), resulting from the previous passage
of the spiral wave at the same angular position with the
tip displaced radially outwards by $d$. The dashed line in
(a) indicates the variation of $v$ on the waveback, $v_b$, at the
time of the previous passage of the spiral. The solid and dashed line in (b)
represent the spiral boundary at the present time (solid line)
and at its previous passage (dashed line). Note that 
the excitability averaged along the instantaneous 
wavefront is higher than for steady-state rotation 
due to the radial displacement after one rotation. Our 
formalism provides a rigorous procedure for calculating
how the instantaneous tangential velocity of the
wave tip changes in response to this spatially
varying excitability.}
\label{vfield}
\end{figure}

We consider successive passages of the wave tip by the angular position
$\theta$. The successive radial displacements of the tip are $\cdots, \epsilon
q(\theta
-2\pi)/c_0\,, \epsilon q(\theta)/c_0\,, \epsilon q(\theta +2\pi)/c_0\,,\cdots$. Let us consider the passage at
the position $R_0+\epsilon q(\theta)/c_0$
in the Cartesian coordinate system $(x,y)$  
attached to the wave tip (see Fig.~\ref{vfield})
in which we choose to measure  
lengths in unit of $\epsilon/c_0$ .
The controller concentration
$v_f(x;\theta)$
in the medium just ahead of the front interface is related by the
controller recovery kinetics Eq.(\ref{inkin}) to the controller
concentration left 
just behind the back interface $v_b(x;\theta-2\pi)$ at the previous passage.
At dominant order in $\epsilon/(c_0 R_0)$, one can neglect the tip width
compared to the core perimeter and the time interval between two passages of
the spiral by the same angular position can be taken equal to the steady
spiral period $T_0$,
\begin{equation}
v_f(x;\theta)-v_0 = \exp(-\frac{T_0}{\tau_R})\ [v_b(x+
q(\theta)-q(\theta-2\pi);\theta-2\pi)-v_0]
\label{vfvbp}
\end{equation}
In (\ref{vfvbp}),  note that $v_f(x;\theta)$ is related to $v_b(x+
q(\theta)-q(\theta-2\pi);\theta-2\pi)$ since
the argument in $v_f$ refers to a frame attached to the
wave tip with origin at 
$R_0+\epsilon q(\theta)/c_0$ whereas
 the origin of the coordinate for $v_b$ is at
$R_0+\epsilon q(\theta-2\pi)/c_0$.

The controller concentration $v_f$ and $v_b$ at the same passage are also simply
related by the controller production equation in the excited region 
(\ref{exkin}),
\begin{equation}
v_b(x;\theta)=v_f(x;\theta) +\epsilon \frac{y_f(x)-y_b(x)}{ c_0 c_t\tau_e}
\label{vfvbs}
\end{equation}
Iterating back in time (\ref{vfvbp}) and (\ref{vfvbs}),
 we see that $v_f(x;\theta)$ depends in principle on the positions
of the tip, at all previous passages by the angular position $\theta$. However,
the memory of the position $q(\theta-n2\pi)$ is suppressed by the n-th power
of the small parameter $\exp(-T_0/\tau_R)$.
Therefore, to dominant order the
controller concentration only depends on the position of the tip at the previous
passage
\begin{equation}
v_f(x;\theta)= v_0+\epsilon\ 
e^{-T_0/\tau_R}\ \frac{y_f(x+d(\theta))
-y_b(x+d(\theta))}{c_t c_0\tau_e} \Theta(x+d(\theta))
\label{vfyp}
\end{equation}
where we have defined the relative displacement of the
tip  between its two passages $d(\theta)=q(\theta)-q(\theta-
2\pi)$. $\Theta$ is the usual Heavyside step function, $\Theta(x)=0$ for
$x<0$ and $\Theta(x)=1$ otherwise.
Eq.~(\ref{vfyp}) determines the controller concentration on the front interface
at $\theta$ as a function of $q(\theta)-q(\theta-2\pi)$. It is now an easy
task to
generalize the previous computations and obtain the tip tangential velocity
corresponding to this concentration.

As for steady interacting spirals, we obtain for the front interface in the
tip region,
\begin{equation}
\frac{d^2 y_f}{dx^2}=\frac{c_t}{c_0}
\left[1+ (\frac{dy_f}{dx})^2\right]-\left[1-\frac{\alpha}{c_0}  
(v_f(x;\theta)-v_0)\right]\left[1
+ (\frac{dy_f}{dx})^2\right]^{3/2}
\label{eq58}
\end{equation}
The only difference with Eq.~(\ref{eq41}) is that $\delta v_f=v_f(x;\theta)-v_0$
is now given by Eq.~(\ref{vfyp}). Expanding Eq.~(\ref{eq58}) around the
critical finger shape, $y_f=Y_f+\delta y_f$, one obtains as before,
\begin{eqnarray}
{\cal L}_f(\delta y_f)&=& \frac{\delta c_t}{c_0} \left[1+\left(\frac{dY_f}{dx}
\right)^2\right]
\nonumber\\
&+& B_c e^{-T_0/
\tau_R}\left [Y_f(x+d(\theta))-Y_b(x+d(\theta))\right ] \Theta(x+d(\theta))
 \left[1+\left(\frac{dY_f}{dx}\right)^2\right]^{3/2}
\end{eqnarray}
with the solution
\begin{equation}
\delta y_f = \frac{\delta c_t}{c_0}\, \eta_1 + B_c \exp(-T_0/\tau_R)\,
 \eta_{v,d(\theta)}
\end{equation} 
The linear operator ${\cal L}_f$ is defined by (\ref{oplinfront}),
$\eta_1$ is defined in (\ref{eta1}) and $\eta_{v,d}$ is the
solution of
\begin{equation}
{\cal L}_f (\eta_{v,d})= \left [Y_f(x+d)-Y_b(x+d)\right ]\Theta(x+d)
\left[1+\left(\frac{dY_f}{dx}\right)^2\right]^{3/2}
\label{etavd}
\end{equation}
which generalizes Eq.~(\ref{etav0}),
with the boundary conditions at $x=0$, $\eta_{v,d}(0)=0, \eta_{v,d}(x)
\sim  \sqrt{x/2}[Y_f(d)-Y_b(d)]\  \Theta(d)$ for $x\ll 1$.

Similarly, the back interface equation in the tip region is
\begin{equation}
\frac{d^2 y_b}{dx^2}=\frac{c_t}{c_0}
\left[1+ (\frac{dy_b}{dx})^2\right]+\left[1-\frac{\alpha}{c_0}
(v_f(x;\theta)-v_0)-\frac{\epsilon\alpha}{c_t c_0^2 \tau_e}
(y_f(x)-y_b(x)) \right]\left[1
+ (\frac{dy_b}{dx})^2\right]^{3/2}
\label{eq60}
\end{equation}
After linearization around the back interface of the critical finger,
$y_b(x)=Y_b(x)+\delta y_b(x)$, one obtains for the correction $\delta y_b$,
\begin{eqnarray}
&&{\cal L}_b(\delta y_b)= \frac{\delta c_t}{c_0} \left[1+\left(\frac{dY_b}{dx}
\right)^2\right]
\nonumber\\
&&+ 
\left[[B_c\,\frac{\delta c_t}{c_0}-\delta B]
[Y_f(x)-Y_b(x)]-B_c e^{-T_0\tau_R}[Y_f(x+d(\theta))-
Y_b(x+d(\theta))]\Theta(x+d(\theta))
\right.
\nonumber\\
& &
  - B_c\,
\delta y_f  
\left]
\left[1+\left(\frac{dY_b}{dx} 
\right)^2\right]\right.^{3/2}
\label{eq61}
\end{eqnarray}
Multiplying both sides of (\ref{eq61}) by the zero-mode $\xi(x)$ of the
adjoint of ${\cal L}_b$ and integrating from $x=0$ to $+\infty$, gives
\begin{equation}
\frac{\delta c_t}{c_0}[I_1+B_c(I_3-I_2)]=\delta B\  I_3+ B_c
e^{-T_0/\tau_R}[I_{3,d(\theta)} + B_c I_{v,d(\theta)}]
\end{equation}
where the definite integrals $I_1,I_2,I_3$ have been defined in (\ref{i123})
and $I_{3,d}, I_{v,d}$ are given by
\begin{eqnarray}
I_{3,d}&=&
\int_0^{+\infty}\!\!dx\,\xi(x)\, \left[Y_f(x+d)-Y_b(x+d)\right]\Theta(x+d)
\left[1+\left(\frac{dY_b}{dx}\right)^2\right]^{3/2}\nonumber\\
I_{v,d}&=&\int_0^{+\infty}\!\!dx\, \xi(x)\,\eta_{v,d}
 \left[1+\left(\frac{dY_b}{dx}
\right)^2\right]^{3/2}
\label{ivdi3d}
\end{eqnarray}
Finally, this gives the tangential tip velocity as a function of the tip 
displacement
\begin{equation}
\frac{\delta c_t}{c_0}= \frac{\delta B}{K} + \frac{B_c}{K}\  
 e^{-T_0/\tau_R}\ [J+F(q(\theta)-q(\theta-2\pi))]
\label{cfin}
\end{equation}
where the constants $K\simeq .630$ and $J\simeq
1.872$ are defined in Eq.~(\ref{eqK}) and Eq.~(\ref{eqJ}). 
The function $F(d)\equiv
[(I_{3,d}+ B_c I_{v,d})/I_3 -J]$
 vanishes at
$d=0$ and is plotted in Fig.~\ref{figF} \cite{noFm}. 

Comparing Eq.(\ref{cfin}) with Eq.(\ref{eq45})
for the steady case shows that the change in tangential velocity due to the tip
displacement is
\begin{equation}
\delta c_q= c_0 \frac{B_c}{K}\ 
 e^{- T_0/\tau_R}\ F(q(\theta)-q(\theta-2\pi))
\label{dcq}
\end{equation}

\subsection{Computation of the tangential velocity in other cases}

We conclude this section by emphasizing that, although
the present kinematic theory is quite general, the precise
expression for the tangential tip velocity that is to be used in
in conjunction with Eq. \ref{eqm1} depends on the application at hand. 
For example, Eq. \ref{dcq} above is valid for
self-interacting spirals without external forcing and is therefore
perfectly suited to analyze meander in the next section,
or interacting multi-armed spirals with a 
minor modification given in section \ref{marms}.
For the non-self-interacting spiral with an 
excitability that varies slowly in space or time (section VII.A), 
one can use directly the results for steady-state rotation
(Eq. \ref{cvsb}), whereas under the 
action of an external field (section VII.B) one needs to compute
a different expression for the tangential velocity. 
The general procedure, however, is clear. In each case, the
tangential velocity depends on the average controller variable
in the tip region and can be computed 
from a solvability condition.

\section{Meander}
\label{seclinstab}

In this section we analyze the classic
meandering instability in a linear and
nonlinear regime.
Substitution of (\ref{dcq}) in (\ref{eqm1})
expresses the r.h.s. of (\ref{eqm1})
 as a function of
$q(\theta)-q(\theta-2\pi)$ and provides the differential equation with delay
governing the tip motion
\begin{equation}
\frac{d^2q}{d\theta^2}+ q= m F(q(\theta)-q(\theta-2\pi)). 
\label{wavetip}
\end{equation}
The parameter $m$ is given by
\begin{equation}
m= \frac{3 B_c}{2 b K} \left(\frac{R_0 c_0}{\epsilon}\right)^{5/3}\ 
\exp(-T_0/\tau_R).
\label{mexp0}
\end{equation}
Values of $m$ of order unity are reached when $(R_0 c_0/\epsilon)^{5/3}
\exp(-T_0/\tau_R)\sim O(1)$. In this parameter 
regime, one can use the simple formula (\ref{asyRom}) to estimate the spiral
parameters since in (\ref{rint}) the correction term 
(the second term on the r.h.s)
is of order $(R_0 c_0/\epsilon)^{2/3}
\exp(-2\pi R_0/c_0\tau_R)\sim O(1)$ compared to the first term on the r.h.s. and
therefore smaller by $\epsilon/(R_0 c_0)$. This provides the explicit expression
of $m$ in terms of the parameter $B$ which characterizes the medium
\begin{equation}
m= \frac{3 B_c (b\,K)^{3/2}}{2(B_c-B)^{5/2}} \exp\left[-\frac{2\pi\epsilon}{
c_0^2 \tau_R}\left(\frac{b\,K}{B_c-B}\right)^{3/2}\right].
\label{mexp}
\end{equation}

\subsection{Linear stability analysis and instability criterion}
We begin by studying the 
linear stability of Eq.~(\ref{wavetip}) around q=0 that is, 
the linear stability of steady rotation.
For $q\ll 1$, one obtains
\begin{equation}
\frac{d^2q}{d\theta^2}+ q= \alpha\ (q(\theta)-q(\theta-2\pi)). 
\label{eqlin}
\end{equation}
where we have introduced $\alpha=m\,F'(0)$. Seeking $q$ under the form
$q=A \exp(\sigma\theta)$ gives the eigenvalue equation
\begin{equation}
\sigma^2+1=\alpha\  [1-\exp(-2\pi\sigma)]
\label{eigeneq}
\end{equation}

For any $\alpha$, $\sigma_\pm =\pm i $
 are isolated solutions of (\ref{eigeneq}).
They simply correspond to the two
translation modes of the spiral : for a steady spiral which is
slightly displaced from the origin and centered at $(x_0,y_0)$ with $x_0\ll R,
y_0\ll R$, the distance of the wavetip to the origin varies sinusoidally as
$q=|z_0 +R_0 \exp(i\theta)|-R_0=x_0 \cos\theta +y_0 \sin\theta$.

The other solutions of (\ref{eigeneq}) vary with $\alpha$. For small $\alpha>0$,
the r.h.s. of (\ref{eigeneq}) is comparable
to its l.h.s. only if the real part of 
$\sigma$ is large and negative, that is $Re(\sigma)\sim -1/2\pi \ \ln(\alpha)$.
Therefore, for small $\alpha$, all eigenvalues 
(different from the two translation modes) have a negative real part and
the steady rotation is stable. As $\alpha$ is increased, the eigenvalues
moves continuously in the complex plane. An instability occurs when the real
part of some of them traverses zero and becomes positive. This happens at the
critical value $\alpha=\alpha_c$ where Eq.~(\ref{eigeneq}) has a purely
imaginary root $\sigma=i\Omega$, namely for
\begin{eqnarray}
\alpha_c\, [1-\cos(2\pi\Omega)]&=&1-\Omega^2
\label{rpeig}\\
\alpha_c \sin(2\pi\Omega)&=&0
\label{ipeig}
\end{eqnarray}
Eq.~(\ref{ipeig}) requires that $\Omega$ be a half integer. Eq.~(\ref{rpeig})
can therefore be rewritten as $1-\Omega^2=\alpha_c [1-(-1)^{2\Omega}]$,
the only solution of which is, for $\alpha_c>0$, $\Omega=\pm 1/2, \alpha_c=
3/8$.

We therefore conclude that for $0<\alpha<\alpha_c$ all eigenvalues different
from $\sigma_\pm $ have a negative real part. As $\alpha$ increases past
$\alpha_c=3/8$, a couple of eigenvalues traverses the imaginary axis and
acquires a positive real part. The value 
$\alpha=\alpha_c$ is thus the threshold
of a Hopf bifurcation and corresponds to the meander 
onset with a frequency ratio
at threshold $\omega_2/\omega_1=1/2$. This ratio is consistent
with the extrapolation to infinite core radius 
of numerical simulation results as shown in 
Fig.~\ref{omvsr}.

It is interesting to note that as $\alpha$ is further increased, the frequency
of the two linearly unstable modes decreases and the two unstable eigenvalues
become purely real for $\alpha>\alpha_r$ ($\alpha_r$ is simply determined
as the value of $\alpha$ for which Eq.~(\ref{eigeneq}) has a doubly degenerate
root, $\alpha_R=\sigma_r/\pi\ \exp(2\pi\sigma_r)$ with $\sigma_r^2 +1 =\sigma
/\pi\ (\exp(2\pi\sigma_r-1)$ which gives $\sigma_r\simeq .375$ and $\alpha_r
\simeq 1.260$). This may explain why a previous analysis performed at small
$\epsilon$, but away from $\partial M$ \cite{klr},
yielded only real unstable modes instead of complex conjugate eigenvalues
as expected from a Hopf bifurcation.

Given the expression (\ref{mexp}) of the constant $m$, the criterion for
meander onset $\alpha_c=m\ F'(0) =1/2$ implies that, for small
$\epsilon$, the meander boundary $\Delta_m(\epsilon)$
lies close in the $(\epsilon, \Delta)$ plane
to the critical finger boundary $\Delta_c(\epsilon)$
(see Fig.~\ref{flowergarden})
with $\Delta_c-\Delta_m\sim \epsilon^{5/9}/\ln^{2/3}(\epsilon^{5/9}/F'(0))$.

In the pure sharp boundary description with no diffusion of the controller
$v$ field, the behavior of the function $F$ is nonanalytic at short distance
$F(d)\sim -.576 d\, \ln(|d|)$ for $d\ll 1$ as shown in Appendix \ref{appsolv}.
This implies that, in this context,
the onset of meander occurs right at the critical finger boundary. However,
when one starts from the full reaction-diffusion equation
(\ref{eqrd1},\ref{eqrd2}),
 the interface has a finite width of the order of $\epsilon$. This
eliminates any short distance nonanalyticity and cut-off the divergence of
$F'(d)$ at $d\sim\Delta\sim\epsilon^{1/3}$. 
This gives the estimate $F'(0)\sim -\ln(\epsilon)$ and $\Delta_c-\Delta_m\sim 
\epsilon^{5/9}/|\ln(\epsilon)|^{2/3}$.

The non-analyticity of $F$ also disappears if the slow field $v$ diffuses
that is, if instead of (\ref{veq}) one has
\begin{equation}
\partial_t v=\gamma\,\epsilon\,\nabla^2v+g(u^\pm (v),v)~~~~~{\rm in}
~~{\cal D}^\pm 
\end{equation}
For a sufficiently small diffusion constant $\gamma$, one can neglect entirely
the diffusion in the excited region $\cal{D}^+$ and consider only a radial 
diffusion of $v$ in $\cal{D}^-$. The controller concentration on the front
spiral interface is then a smoothened version of (\ref{vfyp})
\begin{eqnarray}
v_f-v_0&=&\frac{\epsilon \ e^{-T_0/\tau_R}}{c_0 c_t\,\tau_E}
\int_0^{+\infty}\frac{dx'}{\sqrt{\pi}\,\ell_D}\ 
e^{-(x+d(\theta)-x')^2/\ell_D^2}
\ \left(
Y_f(x')-Y_b(x')\right)
\label{vf}\\
v_b&=&v_f+ \epsilon
\left(Y_f(x)-Y_b(x)\right)/(c_0 c_t\,\tau_E) \label{vb}
\end{eqnarray}
The finite diffusion length $\ell_D=\sqrt{4\epsilon\gamma T_0}\ c_0/\epsilon$
removes the short distances analyticity
and gives a finite first derivative to $F$ at the origin which decreases
with increasing $\ell_D$ as plotted in Fig.~\ref{derfvsl}
 (see also Appendix \ref{appsolv}).
This decrease of stability  with a decrease of $\ell_D$ qualitatively agrees
with the numerical results of \cite{kesskupf}. Of course, diffusion controls
the stability only if $\ell_D$ is much larger than the interface width (or the
width of the tip boundary layer). When it is much smaller, stability is 
controlled
by finite interface width effects as discussed above.
When the two effects have comparable magnitude,
the numerical results of
\cite{kesskupf} suggest that more complex stability diagrams 
are possible (i.e. there is a region of reentrant stability). It
would be interesting to see if this could be explained
by a more complete computation
of $F$ taking into account both finite 
interface effect and diffusion of $v$.

We conclude this subsection with a simple
interpretation of the 
obtained results. 
The 
existence and magnitude of the instability threshold 
can be understood
by considering a displacement of the wave tip 
by a small distance $d$ 
towards the outside of its
steady circular trajectory. 
Since the outside of the
core is slightly less excitable than the inside, this outward
displacement 
will cause the spiral tip to propagate in a less excitable medium and
to rotate on a new larger radius $R_i=R_0+\delta R_i>R_0$. The fact that
$\delta R_i>0$
by itself is not sufficient to create an instability. It is
only if $\delta R_i$ is larger than $\sim d$ that the displacement
of the wave tip can be amplified and meander can appear. The excitability
change due to the displacement $d$ is 
$|\delta B|\sim d/R_{tip}\,  e^{-T_0/\tau_R}$ 
where the exponential factor simply reflects
the global attenuation of excitability variations between two passages of
the wave. This excitability change leads to a variation of
the rotation radius 
$\delta R_i\sim dR_0/dB \ \delta B$.
Thus, $\delta R_i/d\sim (dR_0/dB)/R_{tip}\,  e^{-T_0/\tau_R}  \sim m$ 
and the onset of meander 
occurs for  
$m$ of order unity in agreement with the
above stability analysis.
The period doubling like character 
of the unstable motion (i.e. $\omega_2/\omega_1=1/2$) can also be
attributed to the radial gradient of excitability at the edge of the
spiral core.
A wave tip
displaced outward from the center at a given 
passage, will propagate,
at
its next passage,
in a medium more excitable than the one produced by steady rotation. 
This will 
cause the tip to execute this second turn on a smaller radius and thus,
to propagate again in a less excitable medium and with a larger
radius at the next cycle,
leading to  the period doubling behavior.
As we shall discuss in section IX, this picture is
modified by finite core effects that roughly 
make trajectories of larger radius
take a longer
time to complete one rotation.
This effect causes the spiral tip to return sooner inside the core and,
in turn, leads
$\omega_2/\omega_1$ to increase away from 1/2 with
decreasing $R_0$.

\subsection{Nonlinear dynamics}

We now carry out a standard weakly nonlinear
analysis of the wavetip equation of motion (\ref{wavetip})
and show that the bifurcation
to meander is supercritical in agreement with existing  numerical
studies of reaction-diffusion models \cite{Hopf,hopfak}. This analysis also
allows us to characterize more precisely
the epi-cycle-like trajectories of 
the wave tip in the large
core limit. Next, we integrate Eq. \ref{wavetip}
numerically and explore the nonlinear regime
further away from the bifurcation point.

\subsubsection{Weakly nonlinear analysis}

To carry out the weakly 
nonlinear analysis, we first expand the 
function $F$ on the r.h.s. of Eq. \ref{wavetip} up to cubic
terms, which yields the equation 
\begin{equation}
\frac{d^2q}{d\theta^2}+q\,=\,\alpha \Delta q
+\Gamma (\Delta q)^2-\beta (\Delta q)^3 \label{wnl1}
\end{equation}
where we have defined
\begin{equation}
\Delta q\equiv q(\theta)-q(\theta-2\pi)
\end{equation}
and the constants 
\begin{eqnarray}
\alpha &=&mF'(0)\\
\Gamma &=& mF''(0)/2 \\
\beta &=& -mF'''(0)/6
\end{eqnarray}
In writing (\ref{wnl1}), we have supposed that the non-analyticity of
$F$ in the pure sharp boundary limit has been taken care of either by
taking
into account finite interface width effects or by a small diffusion of
$v$ (e.g. for $\ell_D=1$ one has $ F'(0)\simeq 1.12,\,F''(0)=2.8
10^{-2},\,
F'''(0)\simeq -1.1$). Note however that Eq.~(\ref{wavetip}) is well
defined
even for the non-analytic sharp boundary $F$. We shall comment in the next
subsection
on the small amplitude behavior in this case.
Eq. \ref{wnl1} is valid in a regime where the parameter
\begin{equation}
\mu \equiv \alpha-\alpha_c,
\end{equation}
which defines the distance above the onset ($\alpha_c=3/8$)
of the meandering instability is small.
Next, we seek perturbatively
for time periodic solutions of Eq. \ref{wnl1} of
the form
\begin{equation}
q(\theta)\,=\,q_0\,+
\,\sum_{n=1}^{\infty}\,A_n\,e^{i n\Omega \theta}\,+\,c.c.,
\label{expand}
\end{equation}
where as before $\Omega=\omega_2/\omega_1$ is the ratio
of the Hopf frequency at the meander bifurcation and the
primary angular rotation frequency.
Substituting Eq. \ref{expand} into Eq. \ref{wnl1}
and focusing on the first two modes ($n=1$ and $n=2$),
we obtain at once that
\begin{eqnarray}
(-\Omega^2+1)A_1&=&\alpha (1-\zeta)A_1 + 2\Gamma A_2 \bar A_1(1-\bar \zeta)(1-
\zeta^2)
\nonumber \\
& &~~~~~~~~~-3\beta (1-\zeta)^2(1-\bar \zeta)A_1|A_1|^2\label{sol1}\\
(-4 \Omega^2+1)A_2&=&\Gamma (1-\zeta)^2A_1^2\,+\,\alpha(1-\zeta^2)A_2
\nonumber \\
& &~~~~~~~~~-2\beta (1-\zeta)(1-\bar \zeta)(1-\zeta^2)A_2|A_1|^2\label{sol2}
\end{eqnarray}
where we have defined
\begin{equation}
\zeta=e^{-i2\pi \Omega}
\end{equation}
and $\bar \zeta$, $\bar A_n$, denote the complex conjugates of $\zeta$,
$A_n$, respectively. Eliminating $A_2$ 
between the above two relations, and
neglecting the terms proportional to $(1-\zeta)^2(1-\bar \zeta) A_2|A_1|^2$ 
on the
r.h.s. of Eq. \ref{sol1} 
(which can be checked to be of higher order at the end),
we obtain that
\begin{equation}
\Omega^2-1+\alpha(1-\zeta)+
\left[2\Gamma^2\frac{(1-\bar \zeta))1-\zeta)^2(1-\zeta^2)}{1-4\Omega^2-\alpha(1-\zeta^2)}
-3\beta(1-\zeta)^2(1-\bar \zeta)\right]|A_1|^2=0 \label{a1eq}
\end{equation}
The condition that the real and imaginary parts of the l.h.s.
of the above equation must vanish independently provide two independent
relations that determine $\Omega$ and $A_1$. 
Next, expanding Eq. \ref{a1eq} to first order in the frequency shift
$\Omega-1/2$, we obtain
\begin{equation}
-3/4+2\alpha+(1-i 2\pi \alpha_c)(\Omega-1/2)
-\left[24\beta+\frac{16 i \Gamma^2 \pi }{1+i \alpha_c \pi}\right]
\,|A_1|^2\,=\,0 
\end{equation}
The conditions that the real and imaginary parts of the above equation
must vanish lead after simple algebraic manipulations to the relations
\begin{eqnarray}
|A_1|&=&\sqrt{c_1 \mu } \label{a1exp} \\
\Omega-1/2 &=& -c_2 \label{fshi} \mu 
\end{eqnarray}
where $c_1$ and $c_2$ are constants defined by
\begin{eqnarray}
1/c_1&=&12\beta+\frac{\Gamma^2}{3}\,\frac{9\pi^2+32}{1+(3\pi/8)^2}\\
c_2&=&\frac{64 c_1 \Gamma^2}{3\left[1+(3\pi/8)^2\right]}
\end{eqnarray}
Eq. \ref{sol2} implies that at leading order in $\mu $,
\begin{equation}
A_2=\frac{\Gamma\,A_1^2}{(1/2-\Omega)(1+i\,3\pi/8)}\label{A1A2}
\end{equation}
or, using Eqs. \ref{a1exp} and \ref{fshi},
\begin{equation}
|A_2|\,=\,\frac{3}{64 \Gamma}\,\sqrt{1+(3\pi/8)^2}\label{a2amp}
\end{equation}
In addition, substituting Eq. \ref{expand}
into Eq. \ref{wnl1}, one obtains for $n=0$ that 
$q_0=8\Gamma |A_1|^2=8\Gamma c_1 \mu $.
It is simple to work out that 
higher order terms in the present expansion must scale as
$A_n \sim \mu ^{n/2}$ for $n$ odd and
$A_n \sim \mu ^{n-2}$ for $n$ even.
Note that the expansion of (\ref{wavetip}) leading to
(\ref{wnl1}) remains justified because
$\Delta q(\theta)$ vanishes
as $\mu \rightarrow 0$ even though $A_2$ remains
of order unity (i.e. $A_2[\exp(i2\Omega\theta)-
\exp(i2\Omega(\theta-2\pi)]\sim c_2\mu $ in this limit).

Let us now examine the meander 
trajectory of the wave tip. For this purpose it
is convenient to define the dimensionless coordinate 
$Z=X+iY=R e^{i\theta}/ R_{tip}$,
which is scaled by
the tip radius $R_{tip}=\epsilon/c_0$, and is given by
\begin{eqnarray}
Z=X+iY&=&(\rho_0+q)e^{i(\theta +\psi)} \label{zzz}\\
d\psi/d\theta&=&-(q-q_0)/\rho_0\label{psieq}
\end{eqnarray}
where we have defined the scaled steady-state
radius $\rho_0=R_0\/R_{tip}$. We have subtracted the
$\theta$-independent part of $q(\theta)$ which gives a  
shift of $\omega_1$ of $O(q_0/\rho_0)$ (Eq.~(\ref{eqct2}).
 Since $1/\rho_0\ll 1$, we can expand the above relations
to first order in $\psi$,
which yields
\begin{equation}
X+iY=\left(\rho_0+q-i\int (q-q_0)\, d\theta \right)\,e^{i(1-q_0/\rho_0)\theta+i\psi_0}
\label{genexp}
\end{equation}
Since the phase factor $\psi_0$
corresponds to a translation of the center of rotation, we
can set $\psi_0=0$,
which yields the relation
\begin{equation}
X+iY=\rho_0e^{i\theta}+
\,\sum_{n=1}^{\infty}\left[\bar A_n\left(1+\frac{1}{n\Omega}\right)
\,e^{i (1-n\Omega) \theta}\,+\,
A_n\left(1-\frac{1}{n\Omega}\right)\,e^{i (1+n\Omega)\theta}\right]
\label{generalxy}
\end{equation}
where the amplitudes $A_n$ 
dictate the meandering motion of the tip.

Note that in deriving 
Eq. \ref{generalxy} we have only assumed
that $q/\rho_0$ is small, such that this equation is not
restricted to the asymptotic large core limit
where $\Omega=1/2$ at the bifurcation.
In fact, in the weakly excitable limit that
is typically accessible in simulation, 
$\Omega$ is larger than $1/2$ at the bifurcation point
due to finite core radius corrections $\sim 1/\rho_0$ that modify
Eq. \ref{wavetip} as discussed in section \ref{smallcore} (see e.g.
Eq.~(\ref{eqlinmod})).
In this case, the bifurcation is not resonant (i.e. $2\omega_2\ne
\omega_1$), and $A_2\sim \mu $ near onset.
Eq. \ref{generalxy} implies that in this generic case,
relevant for usual simulations and experiments,
the motion of the tip can be described by keeping only the terms proportional
to $A_1$ and $\bar A_1$ in Eq.~(\ref{generalxy}) that is  a
three-radius epi-cycle (or epi-epi-cycle)
\begin{equation}
X+iY= \rho_0e^{i\theta}+\rho_1\,e^{i (1-\Omega) \theta-i\theta_1}\,-\,
\rho_2\,e^{i (1+\Omega)\theta+i\theta_1}~~~~~~~~~~(\Omega>1/2)
\label{epi3}
\end{equation}
where $\theta_1$ is an arbitrary phase, $\rho_1\sim \sqrt\mu $, and
\begin{equation}
\rho_2/\rho_1=(1-\Omega)/(1+\Omega) \label{rhoratio}
\end{equation}
The fact that $\rho_2/\rho_1$ vanishes as
$\Omega\rightarrow 1$ may provide an 
explanation for why the meander trajectories in simulations
of reaction-diffusion models of excitable media
have been traditionally well fitted by a simple
epi-cycle (Eq. \ref{epi3} with $\rho_2=0$).
In Ref \cite{hopfak}, it was argued 
that meander trajectories should generally
be epi-epi-cycles close to the onset of instability.
It was left unexplained, however,
why the ratio $\rho_2/\rho_1$ turns out to be very small.
For the simulation of the FN of Ref. \cite{hopfak},
$\Omega\approx 0.782$, in which case Eq. \ref{rhoratio}
predicts that $\rho_2/\rho_1\approx 0.12$.
This ratio is roughly consistent with the ratio of the amplitudes
of the peaks of $1+\Omega$ and $1-\Omega$ in
the power spectrum of $X(t)$ in Fig. 4 of \cite{hopfak}.
Here, Fig. \ref{tipcomp} illustrates that two-radius and three-radius
epicycle trajectories are very close even when $\Omega$ departs
significantly from unity. Such a small difference is probably hard
to resolve experimentally.

Let us now examine the meander trajectory predicted by
Eq. \ref{generalxy} in the asymptotic limit where $\Omega=1/2$, which
is more difficult to reach in simulation and experiment.
The main difference in this case is
that $A_2$ is $O(1)$ because the bifurcation is resonant, i.e.
$A_1^2 e^{i 2\Omega \theta}$ act as a 
periodic drive of the wave tip at 
the primary frequency $\Omega=1$.
Inserting the results of the weakly nonlinear
analysis, Eqs. \ref{a1exp}-\ref{a2amp},
into Eq. \ref{generalxy}, we obtain that 
\begin{equation}
X+iY= \rho_0e^{i\theta}+
\rho_1\,e^{i \theta/2-i\theta_1}\,-\,
\rho_2\,e^{i 3\theta/2+i\theta_1}\,+\,\rho_3
e^{i\Omega^*\theta+i(-2\theta_1+\tan^{-1}3\pi/8)}~~~~~~(\Omega=1/2)
\label{epi4}
\end{equation}
where we have defined
$\Omega^*\equiv 1-2\Omega=2c_2\mu $, $\theta_1$ is 
an arbitrary phase, and
\begin{equation}
\rho_1=3\sqrt{c_1\mu },~~~~
\rho_2=\rho_1/3,~~~~~\rho_3=3 \sqrt{1+(3\pi/8)^2}/(32\,
\Gamma)
\end{equation}
Consequently, the effect of the 
resonance when $\Omega=1/2$ is to add a slow component of motion 
with frequency $\Omega^*\sim \mu $ around a circle of radius
$\rho_3$ of $O(1)$.  Steady-state rotation is approached smoothly
when $\mu \rightarrow 0$, even though
$\rho_3$ remains finite, because $\Omega^*$ 
vanishes in this limit.
Finally, we note that $\rho_3$ diverges as
$1/\Gamma$ in the limit $\Gamma\rightarrow 0$.
The tangential velocity of the tip 
around the circle of radius $\rho_3$, however,
scales as $\Omega^*\rho_3 \sim \Gamma$ 
and vanishes in this limit, 
which is therefore well-behaved. 

\subsubsection{Numerical integration of the wavetip equation}

Eq. \ref{wavetip} was integrated numerically using the algorithm
described in Appendix A. We used both the function
$F$ plotted in Fig.\ref{figF}, and the simple analytical form
\begin{equation}
F(x)={\rm tanh}(x-a)+{\rm tanh(a)}, \label{asymth}
\end{equation}
This form has qualitatively the same shape as the
calculated function $F$, which is plotted 
for different $\ell_D$ in Fig. \ref{figF}, and
yields a qualitatively similar nonlinear behavior.
For this reason, all the results presented here are
for this simplified form of $F$ defined by Eq. \ref{asymth}
for the choice of parameter $a=0.2$.
As noted earlier, calculated function $F$ is non analytic
at the origin in the singly diffusive sharp boundary model
and behaves as $-.576 q \ln(|q|)$. When this is used in Eq.~(\ref{wavetip}) for
the tip motion, as noted previously, a steady rotation is unstable for all
$m$ (Eq.~(\ref{mexp})) however small since the slope of $F$ at the origin
diverges. The growth of the modulation as one moves away from threshold
is however much slower than in
the analytic case, the amplitude of the modulation being of order
$~\exp(-cst/m)$. It is interesting to note that requiring this amplitude
to be larger than the interface width $\epsilon$, as a criterion for meander
threshold in a real small-$\epsilon$ model, gives $m \sim cst/ |\ln(\epsilon)|$
quite similarly to what was obtained previously by cutting off the slope
of $F$ at the scale of the interface width. Away from onset, however, this
non-analyticity does not modify much the nonlinear behavior.
For this reason, we shall not treat this case separately.

The results of the numerical integration of Eq.
\ref{wavetip} are illustrated in Figs. \ref{qplotab} and
\ref{qplot}. We have found it 
convenient to plot $q(\theta)-q(\theta-2\pi)$,
instead of $q(\theta)$ because the latter quantity 
contains a component $\sim e^{i\theta}$ that only yields
a translation of the center of rotation.
We have checked that
the amplitude of oscillation and the frequency shift of
$\Omega$ from $1/2$ increase quantitatively 
for small $\mu $ as predicted
by the weakly nonlinear analysis. 
Fig. \ref{qplot} shows that the oscillations
become more nonlinear with increasing distance
from the bifurcation point, but remain periodic
with a frequency close to $1/2$. The fact that
the frequency is rather insensitive to $m$ can be 
understood by remarking that $F$ (calculated or
approximated by Eq. \ref{asymth} with $a$ small)
is close to being an odd function of its argument.
For $F$ exactly odd ($\Gamma=mF''(0)=0$), the weakly nonlinear
analysis of the last section predicts 
that $A_n=0$ for all $n$ even and that there is no nonlinear
frequency shift, i.e. $\Omega=1/2$ 
for any value of $\mu >0$.
One would therefore naturally expect to
to find that $\Omega$ remains close to $1/2$, even far from onset, 
when $F$ deviates slightly from an odd function.

Finally, it is worth noting
that hyper-meander (i.e. chaotic meander)
is not contained in the large core limit. This is
consistent with the fact hypermeander has 
been observed numerically in the 
opposite parameter range of high excitability \cite{Win2}.
In this range, the shape of the spiral boundary
is not constant in time on the scale of $R_{tip}$.
It therefore seems likely that the dynamics on this scale
plays an important role in hyper-meander.

\section{Spiral motion under external action}
Motion of spiral waves can be induced by modulating the medium
excitability in space or time or by adding an external field. It is
not difficult to extend the approach of section \ref{secdyn} to
describe these effects simply and quantitatively in the large core
limit.
\subsection{Variation of the medium excitability}

We consider first the effect of 
spatial and temporal modulations of the
excitability (obtained by changing 
$\Delta$ and/or $\epsilon$ into space and/or time).
Such a modulation will generally produce a variation of both 
the planar front velocity $c_0$ and a variation $\delta B(z,t)$
of the parameter $B$ characterizing the medium.
We assume that this variation is
small enough to be treated as a perturbation, that
$\delta B(z,t)$ varies slowly in time (i.e. on the scale of the
spiral rotation period) and in space (i.e. on the scale of the spiral
core) and that $B$ is close enough to $B_c$ (i.e. $\partial R$) 
so that the spiral self-interaction can be neglected.
The radius of curvature of the tip trajectory will then depart
from its unperturbed value $R_0$, $R_i=R_0+\delta R_i$ with
\begin{equation}
\frac{\delta R_i}{R_0}=\frac{3}{2}\frac{\delta B}{B_c-B}
\end{equation}
and the variation of $c_0$ gives a subdominant contribution
for $B$ close to $B_c$. Substituting the above expression
into Eq.~(\ref{eqm0}), we obtain at once
\begin{equation}
\ddot{q}+\omega_1^2 q= \frac{3}{2} \frac{c_0 R_0}{\epsilon}
\frac{\omega_1^2 \delta B(z,t)}{B_c-B} 
\label{eqmvb} 
\end{equation}

Integration of Eq.~(\ref{eqmvb}) gives the spiral tip motion resulting
from a given space time variation of excitability. As a simple illustration,
we show that a global  periodic variation of excitability at the spiral
frequency induces a spiral drift \cite{agl}. When $\delta B= A \cos (\omega_1
t +\phi)$, the r.h.s. of Eq.~(\ref{eqmvb}) is resonant with the natural
oscillation modes of the l.h.s., the translation modes, and induces their 
growth
\begin{equation}
q(t)=\frac{3}{4} \frac{c_0 R_0}{\epsilon} \frac{\omega_1 t}{B_c-B} A
\sin(\omega_1 t+\phi)
\label{qoscb}
\end{equation}

A simple way to understand the motion described by (\ref{qoscb}) is
to remember that
for a steady spiral centered close to the origin (compared to the radius of its
 core), at $z_0=x_0+i y_0$, the distance of the wave tip to the origin varies
periodically as 
\begin{equation}
|R_0 \exp(i\omega_1 t) + z_0|\simeq R_0 + x_0 \cos(\omega_1 t) +
 y_0 \sin(\omega_1 t)
\label{trans}
\end{equation}
Comparing the two expressions shows that Eq.~(\ref{qoscb})
describes a linear drift of the spiral
\begin{equation}
z_0= \frac{3}{4} \frac{c_0 R_0}{\epsilon} \frac{A}{B_c-B}\, \omega_1 t\,
(-i\exp(-i\phi))
\label{pedrift}
\end{equation}
The drift direction depends on the relative phase between the spiral rotation
and the periodic modulation of excitability : the spiral drifts 
perpendicularly
to the direction ($\exp(-i\phi)$)
of the spiral tip at the maximum excitability viewed from
the spiral center. One can note that
our derivation of (\ref{pedrift}) 
is simple, but, of course, it breaks down when the spiral center is no
longer close to the origin and the 
linearization giving (\ref{eqm0}) and thus (\ref{eqmvb}) becomes
illegitimate. The remedy is standard: a nicer looking derivation is obtained
by introducing a the very start of the derivation
of (\ref{eqm0}) the spiral center $z_0$ and parameterizing the
wave tip as $z=z_0+ (R_0+\epsilon q(t)/c_0) e^{i\omega_1 t + \psi(t)}$.
The slow variation of $z_0$ with time is obtained by requiring that it 
cancels the secular term on the r.h.s of Eq.~(\ref{eqmvb}).

A time independent excitability which varies slowly in space is another
simple case. The parameter $B(z)$ in Eq.~(\ref{eqmvb}) should be evaluated
at the spiral tip position. As the spiral tip turns around the spiral core,
$B$ varies harmonically in time at the spiral rotation period and
the spiral drifts. Since the
direction of maximum excitability viewed from the spiral center
is along the gradient of $B$, one concludes that the spiral drifts
perpendicularly to the gradient of $B$, along an iso-excitability line.

\subsection{Drift in an external field and filament tension}
\label{efield}
It has been reported in previous experimental \cite{ag,stein,bel}
and theoretical studies \cite{mit,krins} that a spiral drifts when
it is submitted to a constant
external field. Interestingly, the spiral was found
to drift at a non zero angle with the applied external field. 
In presence of an external field $\mathbf{E}$ which couples to the activator
$u$, the activator reaction-diffusion (\ref{eqrd1}) becomes
\begin{equation}
\partial_t u=\epsilon\,\nabla^2u+f(u,v)/\epsilon
-\mathbf{E.\nabla} u
\label{fb1e}
\end{equation}
A simple way to determine the effect of $\mathbf{E}$ is to view the wave 
dynamics in a frame M which moves at velocity $\mathbf{E}$. In
such a frame, the supplementary gradient term in Eq.~(\ref{fb1e}) disappears
and $u$ simply obeys the field-less Eq.~(\ref{eqrd1}). However,
the controller
equation is modified. It reads, in the excited region, 
\begin{equation}
\partial_t v=1/\tau_e 
+\mathbf{E.\nabla} v
\label{eqve}
\end{equation}
The gradient term in Eq.~(\ref{eqve}) modifies the relation between the
tangential tip velocity and the medium parameters.
 As shown below, one obtains instead of (\ref{eqK})
\begin{equation}
c_t= c_0 +c_0 \frac{B-B_c}{K} +\gamma_{\parallel}\, E_{\parallel}+
\gamma_{\perp}\,  E_{\perp}
\label{eqce}
\end{equation}
where $ E_{\parallel}$ and $ E_{\perp}$ are the external
field component respectively parallel 
and orthogonal  to the tangential tip
velocity (measured in the frame M). Our sign convention
is that $ E_{\perp}>0$ when it points toward the excited region
of the spiral tip.  The numerical coefficients 
$\gamma_{\parallel}$ and $\gamma_{\perp}$ are determined below from a
solvability condition, as we have now done several times. Before detailing this
computation, we show that the spiral drift is a simple consequence of 
Eq.~(\ref{eqce}).
 As above, 
the wave tip motion is determined by Eqs.~(\ref{eqct},\ref{eqri})
where now $z=x+iy$ denotes the position of the wave tip in the frame M and
$c_t$ is given by (\ref{eqce}) and depends on the angle between 
the instantaneous velocity (in the frame M) and the external field
$\mathbf{E}$. The form of the function $R_i[c_t]$ is a consequence
of the front interface dynamics determined by $(\ref{eqrd1})$ which applies
in the frame M. Therefore, it still has the large core asymptotic form 
(\ref{rivsc}).
Writing $c_t=c_t^0+\delta c_E$ in (\ref{eqce}) as a constant part
$c_t^0$ independent of the external field and a small external field
dependent part $\delta c_E=\gamma_{\parallel} E_{\parallel}+
\gamma_{\perp}  E_{\perp}$, we can again copy the  analysis of
subsection (\ref{kin}) and simply replace $\delta c_q$ by $\delta c_E$.
For a perturbed wave tip circle motion 
$z=(R_0+\epsilon q(t)/c_0) e^{i\omega_1 t + \psi(t)}$, 
this gives instead of (\ref{eqm0})
\begin{equation}
\ddot{q}+\omega_1^2 q= \omega_1^2 \delta c_E
\frac{c_0 R'_i[c_t^0]}{\epsilon}=\omega_1^2\frac{3}{2} \frac{c_0 R_0}{\epsilon}
\frac{\delta c_E}{c_0-c_t^0}
\label{eqme1}
\end{equation}
For definiteness, we suppose that the field $\mathbf{E}$ is parallel to the
$x$-axis which gives,
to lowest order in the perturbation, 
$ E_{\parallel}=-E \sin(\omega_1 t)$ and
$ E_{\perp}= E \cos(\omega_1 t)$. So, Eq.~(\ref{eqme1}) is again found
to be the equation
of an harmonic oscillator forced at its natural frequency and the amplitude
$q$ of the oscillation diverges in time
\begin{equation}
 \frac{\epsilon}{c_0 R_0} q(t)= \frac{3}{4}\frac{E}{c_0-c_t}
\, \omega_1 t \,\,[ \gamma_{\parallel} \cos(\omega_1 t) +\gamma_{\perp} \sin(
\omega_1 t)]
\label{eqd0}
\end{equation}
Comparing Eq.~(\ref{eqd0}) with the expression of $q$ for a translated
spiral (\ref{trans}), one concludes that
(\ref{eqd0}) describes a spiral drifting away from the origin at constant
velocity with
\begin{eqnarray}
x_0 &=& \frac{3}{4}\frac{E}{c_0-c_t}\, \gamma_{\parallel}\,
R_0\, \omega_1 t
\nonumber\\
y_0 &=& \frac{3}{4}\frac{E}{c_0-c_t} \,\gamma_{\perp}\,
R_0\, \omega_1 t
\label{edrift}
\end{eqnarray}
The spiral drift angle $\theta_D$ with the external field is therefore
\begin{equation}
\tan(\theta_D)=\gamma_{\perp}/\gamma_{\parallel}
\end{equation}

Several remarks can  be made :\\
i) formally, $\theta_D$ is the angle between the drift velocity
and the external field in the M frame. However, the drift
velocity in the large core limit is dominantly produced by the time
dependent variation of the spiral radius and is much larger than the
velocity difference between the lab. frame and the M frame. Terms
of the same order as the velocity difference between the two frames
have been neglected in obtaining (\ref{eqd0}). It therefore
makes no sense to correct $\theta_D$ for this velocity difference.\\
ii) A constant field produces a spiral drift because the r.h.s. of
the components of the external field in the tip frame,
$ E_{\parallel}=-E \sin(\omega_1 t)$ and
$ E_{\perp}= E \cos(\omega_1 t)$, oscillate at the resonant frequency
$\omega_1$. A sinusoidal external field oscillating at $\omega_e$ has 
components in the tip frame at $\omega_e+\omega_1$ and $\omega_e-
\omega_1$. A spiral drift is therefore induced by an external field
when it
oscillates at {\em twice} the spiral frequency  
($\omega_e=2\omega_1$), as noted in previous studies \cite{apmun}.\\
iii) As said previously, the 
derivation of (\ref{edrift}) 
breaks down when the spiral center is no
longer close to the origin and the linearization giving (\ref{eqd0}) becomes
illegitimate. This can be cured as said above, by introducing from the start 
the spiral center the motion of which is determined through the requirement
that no secular terms appear on the r.h.s of Eq.~(\ref{eqme1}).\\

It remains to obtain (\ref{eqce}) and
compute the parameters $\gamma_{\parallel}$ and
$\gamma_{\perp}$. We consider the spiral  (in the M frame) in a Cartesian
coordinate system attached to the wave tip as in Fig~.\ref{figcoord}. 
As before,
the front interface $y_f(x)$ simply
obeys  Eq.~(\ref{fingfront}) in the tip region. However, 
the controller concentration on the back interface is changed by the
external field (Eq.~(\ref{eqve})) and this modifies the back equation
(\ref{fingback}).
 
We begin by computing the controller concentration on the back interface.
The time dependence of the field components can be neglected
since it is on the scale of the rotation
period, $R_0/c_0$  which is much longer in the large core
limit than the time scale of interest, the spiral width traversal time
$\epsilon/c_0^2$.
Eq.~(\ref{eqve}) thus  shows that in the excited region $v$ obeys
\begin{equation}
v(t,x- E_{\perp} t,y-E_{\parallel} t)=v(0,x,y) +t/\tau_e
\end{equation}
The concentration $v_b(x)$ on the back interface
at the point $(x,y_b(x))$ is related
to the controller concentration $v_0$ on the front interface at the point
$(x_f,y_f(x_f)+c_t t(x))$  at a previous time $t(x)$
with
\begin{eqnarray}
x_f=x-E_{\perp} t(x)
\nonumber\\
y_f(x_f)+ c_t\,\, t(x)=y_b(x)- E_{\parallel}
\end{eqnarray}
$x_f$ and $t(x)$ are functions of $x$, the considered point of the 
back interface which can be determined perturbatively for small external field.
Writing $x_f=x+\delta x, t(x)=t_0(x)+\delta t(x)$ , one obtains
$t_0(x)=[y_b(x)-y_f(x)]/c_t, \delta x=- E_{\perp} t_0(x) $ and
$\delta t(x)=t_0(x) (- E_{\parallel}+ E_{\perp} dy_f/dx|_x)/c_t$.
Therefore the controller concentration at abscissa $x$ on 
the back interface is equal to 
\begin{equation}
v_b(x)=v_0-t(x)/\tau_e=v_0 +\frac{y_f(x)-y_b(x)}{c_t \tau_e}[1+
(- E_{\parallel}/c_t + E_{\perp}/c_t\,\, \frac{dy_f}{dx}|_x)]
\label{vbme}
\end{equation}
The last term is the modification of $v$ on the back interface coming from the
external field. 

When (\ref{vbme}) is taken into account, the back equation in the tip region
reads (using as before space variables scaled by $\epsilon/c_0$),
\begin{equation}
\frac{d^2 y_b}{dx^2}= [\cdots]_{old} 
- B\frac{c_0}{c_t}\left[y_f(x)-y_b(x)\right]\left[
- E_{\parallel}/c_t + E_{\perp}/c_t\,\,  
\frac{dy_f}{dx}|_x\right]
\left[1
+ (\frac{dy_b}{dx})^2\right]^{3/2}
\label{rfingbacke}
\end{equation}
where $[\cdots]_{old}$ denote the terms on the r.h.s. of Eq.~(\ref{rfingback}).
When 
the front and back equations are 
linearized around the critical finger as $y_f(x)= Y_f(x)+\delta y_f(x),
y_b(x)= Y_b(x)+\delta y_b(x)$
one obtains as before $\delta y_f(x)= \eta_1 \delta c_t/c_0$ (Eq.~(\ref{linrfront}) and
 (\ref{eta1})) and a modified equation for $\delta y_b(x)$,
\begin{equation}
{\cal L}_b(\delta y_b)= [\cdots]_{old} 
- B_c\left[Y_f(x)-Y_b(x)\right]\left[
- E_{\parallel}/c_0 + E_{\perp}/c_0\,\, \frac{dY_f}{dx}|_x
\right]
\left[1
+ (\frac{dY_b}{dx})^2\right]^{3/2}
\label{linrbacke}
\end{equation}

Integrating both sides of Eq.~(\ref{linrbacke}), one obtains
the solvability condition which replaces Eq.~(\ref{eq23})
\begin{equation}
\frac{\delta c_t}{c_0} [I_1 + B_c (-I_2+I_3)]-\delta B\, I_3=
B_c (- E_{\parallel}/c_0\, I_3 + E_{\perp}/c_0\, I_{\perp})
\label{cte}
\end{equation}
where the constants $I_1,I_2,I_3$ have previously been defined
(Eq.~(\ref{i123})) and the
new constant $I_{\perp}$ is given by the following integral
\begin{equation}
I_{\perp}=\int_0^{+\infty}\!\!dx\,\xi(x)\, \left[Y_f(x)-Y_b(x)\right]
\frac{dY_f}{dx}\ 
\left[1+\left(\frac{dY_b}{dx}\right)^2\right]^{3/2}\simeq 8.431
\label{iperp}
\end{equation}

Eq.~(\ref{cte}) shows that (\ref{eqce}) holds 
with the following expressions for
$\gamma_{\parallel}$ and $\gamma_{\perp}$,
\begin{eqnarray}
\gamma_{\parallel}&=&-\frac{B_c}{K}\simeq -1.177
\nonumber\\
\gamma_{\perp}&=& \frac{B_c I_{\perp}}{K I_3}\simeq 1.287
\end{eqnarray}
Changes of spiral core radius are the dominant effect in the
large core limit and lead to a drift opposite to the field ($\gamma_{\parallel}
<0$) as qualitatively argued in (\cite{krins}). We quantitatively
find here that a counterclockwise rotating spiral drifts
at an angle
of about $132.5^{\circ}$ with the field direction
in good agreement with previous 
simulations \cite{krins}
as well as our own, as shown in Fig.~\ref{edrift.fig}
(the sign of $\gamma_{\perp}$ and of the drift
angle would be opposite for a clockwise rotating spiral).

Finally, we note that the curvature induced motion of a weakly curved 3D
scroll wave \cite{keen3d,Biktension} 
filament is directly related to spiral drift in an electric
field. For a 3D filament $(x_0(s),y_0(s),z_0(s))$, we can choose
a coordinate system with its third axis aligned with the filament tangent
at $s$. Locally, the activator field can be written  $u(x-x_0(s),y-y_0(s);t)$
with u(x,y;t) a two dimensional spiral wave. The two
dimensional laplacian in Eq.~(\ref{eqrd1}) acting
on such a solution gives $\nabla_{2D}^2u -(x''\partial_x +y''\partial_y)u=
\nabla_{2D}^2u-\kappa N.\nabla u$ where $\kappa$ is the filament curvature
and $N$ the filament normal with $\kappa N$ directed toward the filament center
of curvature. Therefore, $\epsilon \kappa N$ acts as  an external field
$E$ in the normal $(x,y)$ plane. Since $\gamma_{\parallel}<0$ and a
spiral drifts opposite to the field direction, one concludes that curvature
is destabilizing in the large core limit (negative line tension) and
that a scroll ring grows. Moreover, it propagates normally to the
plane of the ring at a velocity proportional to its expansion
velocity since $\gamma_{\perp}>0$.
The other laws governing filament motion
can similarly
be deduced by reducing the 3D dynamics to an effective 2D process. We defer,
however, a detailed study of 3D dynamics in the large core limit to a future
publication.

\section{Multiarmed spirals}
\label{marms}

In this section we extend our analysis to the situation
where several thin excited regions or `spiral arms'' rotate 
around a common core. Our main finding is that
such multiarmed spiral waves 
are always linearly unstable in the large core limit. 
We confirm this finding by numerical simulation 
of the FitzHugh-Nagumo model for two-arm and three-arm spirals.
A different conclusion has been reached in ref.\cite{vas} where
multiarmed spiral waves were found by numerical simulation
of the FN model, with a well-prepared initial condition,
to be stable over windows of
parameters in the large core limit.
We shall comment 
at the
end of this section,
on the possible origin of this disagreement. 

Let us denote by $q_j(\theta)$ the
coordinate of the tip of the
$j$ th spiral arm. We make the arbitrary choice that
rotation is counter-clockwise and take
the index $j\in [0,N-1]$
to increase clockwise. 
The equation for the phases, $\psi_j=\theta_j-\omega_1t$,
are given by
\begin{equation}
d\psi_j/dt=-(\epsilon/c_0R_0)\omega_1 q_j~~~~~~(j=0,...,N-1)
\end{equation}
For simplicity, we consider an initial condition where the angular positions
of the $N$ spiral arms are uniformly distributed.
To lowest order in $\epsilon/(c_0R_0)$,
one can assume that
the spiral arms rotate at constant angular velocity
and that the phase difference between two successive arms remains constant:
$\psi_j-\psi_{j-1}=2\pi/N$.
The equation that governs
the motion of a given arm, say arm $j$,
is essentially the same as the one governing
the motion of a one-arm spiral, except that 
this arm interacts with the exponential
recovery tail of the controller field $v$ of arm $j-1$,
instead of its own recovery tail. Consequently, the
equation of motion for arm $j$ is simply obtained
by replacing the interaction term 
$mF(q(\theta)-q(\theta-2\pi))$ on the r.h.s. of Eq. \ref{wavetip}
by $m_NF(q_j(\theta)-q_{j-1}(\theta-2\pi/N))$, with $m_N$
defined in terms of the reduced period $2\pi R_0/N$.
For a spiral with $N$ arms, the wave tips 
are governed by the $N$ coupled equations
\begin{equation}
\frac{d^2q_j}{d\theta^2}+q_j=m_NF\left(q_j(\theta)-q_{j-1}(\theta-2\pi/N)\right)
~~~~~~~~~(j=0,...,N-1) \label{multieq}
\end{equation}
where
\begin{equation}
m_N=\frac{3B_c(bK)^{3/2}}{2(B_c-B)^{5/2}}
\exp\left[-\frac{2\pi\epsilon}{c_0^2N\tau_R}
\left(\frac{bK}{B_c-B}\right)^{3/2}\right]\label{mdef}
\end{equation}
and $F$ is the same function as for a one-arm spiral. 

\subsection{Linear stability}

Let us first analyze the linear stability 
of an $N$-arm spiral. Linearizing Eqs. \ref{multieq}, we obtain
\begin{equation}
\frac{d^2q_j}{d\theta^2}+q_j=\alpha
\left(q_j(\theta)-q_{j-1}(\theta-2\pi/N)\right)
~~~~~~~~~(j=0,...,N-1) \label{multilin}
\end{equation}
where we have defined $\alpha\equiv m_NF'(0)$. 
The symmetry of the above system 
of linear equations implies that its
solutions must be of the discrete Floquet-Bloch form 
\begin{equation}
q_j=\hat q\exp(ik_nj+\Omega_n\theta)~~~~~~
\end{equation}
where $k_n$ is the discrete Bloch wavector that takes on the values
\begin{equation}
k_n=\frac{2\pi n}{N}~~~~~~~~(n=0,...,N-1)
\end{equation}
Substituting the above form into Eq. \ref{multilin}, we obtain
the eigenvalue equation 
\begin{equation}
\Omega_n^2\,+\,1=\alpha
\left[1-\exp\left(-\frac{2\pi}{N}(\Omega_n+in)\right)\right]
~~~~~~~~(n=0,...,N-1)\label{eigenmulti}
\end{equation}
that determines the allowed values of $\Omega_n$ 
for each mode $n$ and hence its stability.
The two global translational modes, which are exact solutions
of Eq. \ref{eigenmulti} for arbitrary $\alpha$, 
correspond to $\Omega_1=-i$
and $\Omega_{N-1}=i$. We restrict ourselves to considering the
$2N-2$ other
modes which correspond to the coupled translations of the individual spiral
arms. The
eigenvalues corresponding to these modes can be
calculated perturbatively by expanding
$\Omega_n$ in a power series in $\alpha$
about $\pm i$. For brevity of notation,
let us denote by $\Omega^+_n$ the $N-1$ eigenvalues obtained
by expanding about $\Omega_1=+i$ for $n=0,2,...,N-1$, and
by $\Omega_n^-$ the ones obtained by expanding about 
$\Omega_{N-1}=-i$ for $n=0,1,...,N-2$.
Substituting the power series expansions
\begin{equation}
\Omega_n~=~\pm i\,
+\,\alpha \Omega_{n(1)}^\pm \,+\,\alpha^2 \Omega_{n(2)}^\pm \,+\,...
\label{power}
\end{equation}
into Eq. \ref{eigenmulti} we obtain after simple
algebraic steps 
\begin{equation}
\Omega_{n(1)}^\pm \,=
\,\pm \frac{1}{2}\,\sin\left(\frac{2\pi(n\pm 1)}{N}\right)\mp i\,\sin^2\left(
\frac{\pi(n\pm 1)}{N}\right)\label{result}
\end{equation}
Since the leading term in the expansion (\ref{power}) is
purely imaginary, the stability is determined by the sign
of the real part of $\Omega_{n(1)}$. Eq. \ref{result} implies
that ${\rm Re}(\Omega_{n(1)}^-)>0$ for $n=0$ or $n>N/2+1$,
and ${\rm Re}(\Omega_{n(1)}^+)>0$ for
$n<N/2-1$, and therefore that N-arm spirals
are always unstable for $N>2$.
For the special case 
$N=2$ and $n=0$, Eq. \ref{result} implies
that ${\rm Re}(\Omega_{n(1)}^+)=0$, in which case the stability
is determined by the sign of the real part of the next
order term in the expansion, ${\rm Re}(\Omega_{n(2)}^\pm )$.
The calculation at order $\alpha^2$ yields that
$\Omega_{0(2)}^\pm =\pi/2\mp i/2$
and therefore that 
${\rm Re}(\Omega_{0(2)}^\pm )=\pi/2>0$. Thus 
the symmetric ($n=0$) mode is always linearly 
unstable for a 2-arm spiral.
In contrast, for the antisymmetric ($n=1$) mode,
$\Omega_1=\pm i$ remains solution for 
arbitrary $\alpha$.
We conclude that 
$N$-arm spiral waves are always linearly 
unstable for $N> 1$ in the large core limit. 

The nature of the linearly
unstable tip trajectories are simple
to deduce from the above results. To be concrete, let us
consider 2-arm and 3-arm spirals that we shall study in simulations
below. For $N=2$, aside from the two translational modes,
there are two unstable modes corresponding
to the complex conjugate pair
\begin{equation}
\Omega_0^\pm =\pi\alpha^2/2\,
\pm \,i\left(1-\alpha- \alpha^2/2\right)~~~~~~~(N=2) \label{Neq2}
\end{equation}
Since this pair corresponds to $n=0$,
the two tips will move symmetrically (with 
equal radial displacements)
about a fixed center of rotation. Furthermore, since the
imaginary part of $\Omega_0^\pm $ is slightly less than unity,
the two tips will oscillate in and out of the unperturbed
steady-state circle of rotation with a period
slightly larger than the basic period $T_0$, and with an amplitude
of oscillation that grows exponentially in time.
For $N=3$, there are four modes aside from the two global
translational modes: a 
complex conjugate pair with a negative real part, which is
stable, and the unstable complex conjugate pair 
\begin{equation}
\Omega_0^\pm =\sqrt{3}\alpha/4
\pm \,i\,\left(1-3\alpha/4\right)~~~~~~~(N=3) \label{Neq3}
\end{equation}
obtained by evaluating Eq. \ref{result} for $N=3$,
where the tips move with equal radial displacements.
As for $N=2$, the finite imaginary part slightly smaller
than unity implies that the tips will exhibit
exponentially growing oscillations 
with a period slightly larger than $T_0$.

In addition, $\alpha$ is
typically much smaller than unity in the large
core limit since the spiral period is large compared to the
recovery time, $\tau_R\ll T_0/N$, and the spiral 
arms are only weakly coupled via the controller field $v$. 
Therefore, the instability of a multiarmed spiral
should generically develop on a time
scale much longer than $T_0$,
especially for $N=2$ since the real part of $\Omega_0^\pm $ 
scales as $\alpha^2$, instead as $\alpha$ for $N>2$.

\subsection{Numerical simulations}

In order to test the above predictions, we 
investigate numerically the stability of spiral
waves with two and three arms in the FN model.
We restrict ourselves to a range of parameters
where a one-arm spiral is linearly stable and rotates rigidly.
We construct an initial condition for an
$N$-arm spiral, denoted by ($u_N,v_N$), by simply rotating
$N-1$ times by $2\pi/N$ a one-arm 
spiral wave, which yields the
expression
\begin{eqnarray}
u_N(r,\theta)&=&
\sum_{j=0}^{N-1} u\left(r,\theta-2\pi j/N\right)\,-\,(N-1)u_0 \label{uN}\\
v_N(r,\theta)&=&
\sum_{j=0}^{N-1} v\left(r,\theta-2\pi j/N\right)\,-\,(N-1)v_0 \label{vN}
\end{eqnarray}
where $(u_0,v_0)$
are as before the resting values of $u$ and $v$.
Since the simulations are performed in Cartesian coordinates,
and the edges have a negligible effect,
each rotation of $2\pi/N$ is simply carried out by running the
simulation of a one arm spiral for a time equal to $T_0/N$.
The initial condition defined by (\ref{uN}) and (\ref{vN}) deviates 
from the true steady-state solution of an $N$-arm spiral 
by an amount proportional to $v-v_0$ 
on the wave fronts, which is exponentially
small in the large core limit. Therefore, this initial
condition can be considered as a slightly perturbed
$N$-arm spiral solution and is ideal
for the present purposes.

Results of the simulations are shown
in Figs. \ref{2a3arms}
where  we plot the normalized 
radial displacement of the wave tips, $(r_j(t)-R_0)/R_0$, 
which corresponds to $\epsilon q_j/c_0$ in our analysis. 
We calculated the position of the $N$ wave tips by looking for
the points of zero normal velocity along the spiral
boundary defined by $u=0$. This is equivalent to
looking for the $N$ intersections of the curves
$u=0$ and $\partial_t u=0$.
We measured $r_j(t)$ from the instantaneous center defined by
$\bar x(t)=\sum_{j=1}^N x_j(t)/N$ and $\bar y(t)=\sum_{j=1}^N y_j(t)/N)$.
All the main qualitative features predicted by our analysis
are observed in the simulations. (We have not attempted 
a detailed quantitative comparison because 
our predictions are strictly valid 
outside the range of our simulations.) Firstly, 
during the initial instability, the center of 
rotation ($\bar x(t), \bar y(t)$) remains fixed in time and the
radial displacements are equal for all tips. This implies
that the symmetric $n=0$ is the most unstable one.
Secondly, the radial displacements exhibit exponentially
amplified oscillations, with the amplification rate
depending sensitively on the steady-state period $T_0$ and the number of
arms, which both determine the parameter, $\alpha=F'(0)m_N$,
entering in the predicted amplification rates
(i.e. the real parts of $\Omega_0^\pm $ in Eqs. \ref{Neq2} and \ref{Neq3}). 
In particular, Fig. \ref{2a3arms} shows
that the amplification is much slower
in (b) than (a), which agrees with the fact
that $T_0$ is about $1.46$ times larger in (b)
than (a). In addition, for the same parameters, the three-arm spiral in (c)
is destabilized much faster than the two arm spiral in
(b), in agreement with the fact that $m_N$ defined by Eq. \ref{mdef} is 
larger for $N=3$ than $N=2$. Lastly, the period of the radial
oscillation is slightly larger than $T_0$ as predicted by
our analysis. This can be seen for example in Fig. \ref{2a3arms}(b)
where the radial displacement of the tips exhibits 
48 peaks over a time lapse of $50 T_0$.

One interesting question is whether the instability
of the symmetric mode saturates in a nonlinear regime.
To explore this question, we have 
integrated Eq. \ref{multieq}
numerically for the symmetric mode
by letting $q_1(t)=q_2(t)=...q_N(t)\equiv q(t)$,
in which case Eq. \ref{multieq} reduces to a single
equation for $q(t)$.
We investigated different values of $N$ and $m$ for
the function $F$ defined by Eq. \ref{asymth} with $a=0.2$. The
results are shown in Fig. \ref{multiarmq} for $N=2$
and $N=3$, the plots for higher $N$ being qualitatively
identical to the plot for $N=3$. These plots show
that the bifurcation is subcritical. For all $N\ge 2$, the 
amplitude of oscillation increases linearly in time in
the nonlinear regime. This comes about because in the
forced harmonic oscillator equation for $q$
the amplitude of the resonant forcing term $F$ saturates
when $q$ becomes of order one. For small $m_N$, averaging
the forcing term over one period of the harmonic motion, gives the mean energy
increase of the oscillator and accounts for the phenomenon.
 Interestingly, the cross-over from
the linear to the nonlinear regime is qualitatively
different for $N=2$ and $N>2$. For $N>2$, the slope of the envelope
of the oscillations increases monotonously in time until it
reaches a constant value in the nonlinear regime.
Whereas for $N=2$, the slope of the envelope 
increases non-monotonously with time.
The FN simulation for $N=2$ 
shows qualitatively the same non-monotonous
increase of the envelope of radial oscillations with time as
obtained by integrating the wave 
tip equation, as can be seen by
comparing Fig. \ref{2a3arms}(b) and Fig. \ref{multiarmq}(a)). 
This shows that even relatively fine details of the nonlinear 
instability of multiarmed spiral waves are captured by our analysis.
In Fig. \ref{2a3arms}(a), the oscillations grow too rapidly 
to their saturated values to observe this cross-over.

One important consequence of 
the absence of a weakly nonlinear saturation of
the unstable symmetric mode is that the 
distance of closest approach between
the wavetips (which occurs at the minimum of
each oscillation) decreases with time. The resulting
highly nonlinear regime is obviously not described by
the wavetip equation (\ref{multieq}), which is only valid for small
radial displacements of the wavetip compared to $R_0$.
Results of the FN simulations show the complexity of
the dynamics in this regime, as illustrated by Figs. 
\ref{multiarmt} and \ref{2armframes}.

To conclude, let us contrast 
our results to those 
of Ref. \cite{vas}
where
the stability of
multiarm spiral waves 
was studied in a slightly different 
version of FN kinetics, but in a similar regime
of weak excitability. When starting from
sufficiently well-prepared initial conditions,
multiarmed spiral waves were found to be stable when
the period $T_0$ was large enough
to accommodate a finite number of arms around a single core.
Moreover, it was observed that a spiral with $N$ arms became
unstable and decayed into a spiral with $N-1$ arms when a
transition line was crossed by decreasing $T_0$ in the plane of $T_0$
and the refractory period (defined as the minimum interval between
waves in response to the lowest stimulus exciting the medium), 
with a separate line for each $N$. 
 The main difference in our predictions is
that steadily rotating
multiarmed spiral waves are always linearly unstable for $N\ge 2$
for any parameters in this plane.
Note however, that steadily rotating
multiarmed
spirals were not observed in \cite{vas} when starting from
randomly broken arms. 

We have actually checked that the instability predicted by
our analysis, and observed in our FN simulations, also
occurs in the FN kinetics studied 
in \cite{vas}. This is illustrated
in Fig. \ref{2armvasiev} for a two-arm spiral
and $k_g=5.2$, other parameters being chosen
the same as in Ref. \cite{vas}.
The main difficulty in observing
this instability is that it develops extremely slowly
when the spiral period is much larger than the refractory period,
in which case $m_N$ defined by Eq. \ref{mdef} becomes
exponentially small, and the time to observe the 
instability exponentially large, as a function of the
ratio of the two periods. For example, for the parameter
of Fig. \ref{2armvasiev}, the destabilization of the two-arm 
spiral already occurs over a timescale 
of about 10 rotations. For the value $k_g=5$ reported
in Fig. 3 of \cite{vas}, $T_0$ is about twice
larger than for $k_g=5.2$. Hence, the instability cannot be
seen on a time scale of a few rotations.

\section{Toward smaller core radii: a discussion}
\label{smallcore}
We have seen in section \ref{seclinstab} 
that the large core equation of motion
(\ref{wavetip}) leads to a meander onset frequency $\omega_2$ which is
equal to half the basic spiral frequency, quite independently of
the detailed form of the function $F$. It is interesting to identify
the main subdominant effects which leads $\omega_2/\omega_1$ to depart
from $1/2$ for smaller core radius (as shown in Fig.~\ref{omvsr}). 
The following two assertions underly the large core result:\\
i) the tangential velocity and spiral tip rotation rate only depend
on the instantaneous characteristics of the medium in which the spiral
tip propagate (i.e. the relaxation of the tip velocity and rotation rate
can be taken to be instantaneous),\\
ii) the angular tip position is slaved to time ($\theta=\omega_1 t$) i.e. 
the time interval
between two successive passages of the spiral tip by the same
angular position $\theta$ can be taken to be $2\pi/\omega_1$ and one can
neglect the dependence of this time interval on the spiral path.\\ 
A systematic discussion of corrections to
the large core limit is beyond the scope of this article. We content ourselves
here, in showing that corrections to i) and ii) both affect the value of
$\omega_2/\omega_1$ at onset. As discussed below,  taking into
account the non-instantaneous relaxation
(i.e. corrections to i)) formally appear to
give the dominant correction to the large core limit results. However, 
corrections to ii), although subdominant, seem the most 
important
for the parameter range of Fig.~\ref{omvsr} and account semi-quantitatively
for the numerical results.

We begin by discussing i). The motion of the spiral tip is determined
from the two relations (\ref{eqct}),(\ref{eqri}). The tangential tip velocity
is determined by the dynamics of the close tip region $\sim \epsilon/R_0$,
which is fast and independent of the spiral core size. The determination
of the radius of curvature of the spiral tip trajectory involves however
the dynamics of a whole intermediate region $\sim (R_0 R^2_{tip})^{1/3}$
and, as discussed in Appendix \ref{appdyn}, this happens on a time scale $t_d$
with 
$\omega_1 t_d/\sim (R_{tip}/R_0)^{1/3}$. So one expects that this
instantaneous radius of curvature, which we denote here by $\tilde R_i$ to
distinguish it from the steady state value $R_i$,
 adapts on a time scale $t_d$ to changes in
medium conditions. 
Short of solving Eq.~(\ref{riccatidyn}),
a crude model of this
effect is obtained by replacing 
the instantaneous Eq.~(\ref{eqri}) by
\begin{equation}
t_d \frac{d\tilde R_i}{dt}+\tilde R_i=R_i[c_t(\{v\})]
\label{eqrimod}
\end{equation}
This gives instead of Eq.~(\ref{eqm1}), the couple of equations
\begin{eqnarray}
\frac{d^2q}{d\theta^2}+q &=&\delta \tilde R_i c_0/\epsilon
\label{relaxq}
\\
\omega_1 t_d\  \frac{d (\delta \tilde R_i)}{d\theta} +
\delta \tilde R_i &=&\delta R_i 
\label{relaxr}
\end{eqnarray}
where $\delta R_i$ is given by Eq.~(\ref{dri},\ref{dcq}), as previously.
 At the linear level,Eq.~(\ref{relaxr}) simply becomes 
\begin{equation}
\omega_1 t_d\  \frac{d (\delta \tilde R_i)}{d\theta} +
\delta \tilde R_i=\alpha\ \frac{\epsilon}{c_0} [q(\theta)-q(\theta-2\pi)]
\label{relaxlin}
\end{equation}
Searching for the eigenmodes of (\ref{relaxq},\ref{relaxlin}) under the form
$q=A \exp(\sigma \theta)$, gives the modified eigenvalue equation
\begin{equation}
(\sigma^2+1)(1+\sigma \omega_1\tau_d)=\alpha\  [1-\exp(-2\pi\sigma]
\label{relaxdisp}
\end{equation}
The meander threshold is determined by requiring that
Eq.~(\ref{relaxdisp}) has purely imaginary roots
$\sigma=i \Omega$ besides the two translation modes $\sigma=\pm i$. 
Perturbation around the large core ($t_d=0$) result give the modification 
to the meander frequency at onset
\begin{equation}
\Omega=\frac{1}{2}-\frac{\omega_1\tau_d}{2\pi}
\label{correlax}
\end{equation}
So relaxation effects lower the frequency ratio $\omega_2/\omega_1$ below
$1/2$ and cannot account for the numerical observations reported in 
Fig.\ref{omvsr}.

In contrast, we show that improving on ii) lead to corrections in agreement
with the numerical data.
We parameterize the spiral tip position as in
section (\ref{kin}) as $z=(R_0+\epsilon q/c_0) \exp[i(\omega_1 t +\psi(t))]$.
The angular tip position is 
\begin{equation}
\theta=\omega_1 t +\psi(t)
\label{tipangpos}
\end{equation}
Beyond
leading order, $\psi(t)$ is not negligible in (\ref{tipangpos})
and the spiral period of rotation $T$
depends on the spiral tip path. Eq.~(\ref{eqct2}) gives $\dot{\psi}=-
\omega_1 q/R_0$ to dominant order (near the meander onset the other term in
(\ref{eqct2}) is of higher order,
$\delta c_q/c_0\sim (\epsilon/c_0 R_0)^{5/3}$ using Eq.~(\ref{eqm1},\ref{dri})
 and
it can be neglected). This implies that it actually takes a time $T=T_0+
\Delta T$ longer (shorter) than the period $T_0$ of the steady spiral to return
to the same $\theta$ for outward (inward) displacements
\begin{equation}
\Delta T= T_0 \ \frac{\epsilon}{c_0 R_0}\ \int_{\theta-2\pi}^{\theta}
\frac{d\phi}{2\pi} q(\phi)
\label{dtvsq}
\end{equation}
This reduces (increases)
the interaction with the previous tip by $\sim \Delta T/\tau_R
\exp(-T_0/\tau_R)$ and causes the spiral tip to return sooner inside (outside)
the core. This leads $\omega_2/\omega_1$ to move away from $1/2$ toward
unity. 
In order to explicitly show this, we compute
the variation of the tip trajectory radius of rotation
$\delta R_i$ due to the spiral displacement taking (\ref{dtvsq}) into account.
Comparing Eq.~(\ref{cfin}) and
(\ref{dtvsq}) gives 
\begin{equation}
\frac{\delta R}{R_0}=
\frac{3 B_c}{2 b K} \left(\frac{R_0 c_0}{\epsilon}\right)^{2/3}\
\exp(-T_0/\tau_R))\left [ F(q(\theta)-q(\theta-2\pi))-\frac{\Delta T}{\tau_R}
\frac{2 K +J B_c}{B_c}\right ]
\label{drimod}
\end{equation}
A modified version of (\ref{wavetip}) is obtained by substituting
(\ref{drimod}) in (\ref{eqm0})
\begin{equation}
\ddot{q}+\omega_1^2 q= m \,\omega_1^2  \left\{ F[q(\theta)-q(\theta-2\pi)]-
\frac{\Delta T}{\tau_R}\frac{2K+J B_c}{B_c}\right\}
\label{eqmmod}
\end{equation}
where the constant $m$ is given by (\ref{mexp}) and $\Delta T$ depends on the
tip trajectory (Eq.~(\ref{dtvsq}). The linear version of (\ref{eqmmod}) is
(where we can replace time by angular position),
\begin{equation}
\frac{d^2q}{d\theta^2}+ q= \alpha\ (q(\theta)-q(\theta-2\pi)) -\beta
\int^{\theta}_{\theta-2\pi}\!\! \frac{d\phi}{2\pi}\  q(\phi)
\label{eqlinmod}
\end{equation}
where $\alpha=m F'(0)$ as before and $\beta=m (T_0/\tau_R)
(\epsilon/c_0 R_0)
(2K+2J B_c)/B_c $. The eigenmodes of (\ref{eqlinmod}) are of the form
$q(\theta)=A \exp(\sigma\theta)$ where $\sigma$ is a solution of
\begin{equation}
\sigma^2+1=(\alpha-\frac{\beta}{2\pi\sigma})(1-\exp(-2\pi\sigma))
\label{dispmod}
\end{equation}
The meander onset corresponds to the critical value $\alpha_c$ where
(\ref{dispmod}) has  purely imaginary roots $\sigma=i \Omega$
(besides the two translation modes $\sigma=\pm i$). For small
$\beta$, first-order perturbation around the $\beta=0$ values gives
\begin{eqnarray}
\alpha_c &=& \frac{3}{8}-\frac{4\beta}{3\pi^2}
\\
\Omega &=& \frac{1}{2}+\frac{8 \beta}{3 \pi^2}
\label{corpermod}
\end{eqnarray}
This shows that the correction term in (\ref{eqmmod}) lowers the threshold
for the meander instability (i.e. plays a destabilizing role) . More
importantly, it increases the frequency ratio $\Omega=\omega_2/\omega_1$ at
meander onset, as announced. 

The frequency shift predicted by (\ref{corpermod})
can be compared to the numerical results of Fig.~\ref{omvsr}. Using the
lowest order threshold estimate $m F'(0)=3/8$, one obtains
$\Omega-1/2\simeq\epsilon/(c_0 c_t \tau_R F'(0))$. With the estimate
$F'(0)\sim -.58 \ln(\Delta)$,  the frequency shift is found to
be of the same order
of magnitude as the one measured. This semi-quantitative agreement
leads us to think that, for the parameters of
Fig.~\ref{omvsr}, the correction (\ref{corpermod})
is the main effect and that 
the correction (\ref{correlax}) is still
numerically smaller than
(\ref{corpermod}).
Of course, for spirals of sufficiently large core this should cease to be
true, 
the correction (\ref{correlax}) should become dominant 
and $\omega_2/\omega_1$ is expected to drop below $1/2$ before
ultimately reaching its asymptotic value. Unfortunately,  a numerical
check of this
non-monotonic behavior would require  simulating spirals of
very large core radius. This appears a difficult task with present-day 
computers.

\section{Conclusion}
We have developed an analytical approach to spiral waves close to the line
$\partial R$ where the spiral rotates around a large core  and in the 
free boundary limit where the  medium exhibits an abrupt response to a stimulus
($\epsilon\ll 1$). The main ingredient of our analysis has been to note
that in this limit the entire wave tip can be treated as an essentially rigid
body, the slow motion of which is controlled by the local spatial gradient of
excitability in the medium in a way that can be precisely deduced from the
starting reaction-diffusion equations. This has provided a simple understanding
of the spiral tip motion and a precise reduction of its dynamics to that of
a single point. This has allowed us to describe the Hopf bifurcation nature
of the meander instability and to derive simply, but with precise asymptotic
estimates, spiral drift due to spatial or temporal variation of excitability,
or due an imposed external field. This last computation determines in particular
the drift angle of the spiral with the external field and also the parameters
governing the motion of an average scroll wave filament (curvature has been
found to be destabilizing in  
the large core regime). In addition, our analysis has allowed us
to elucidate a generic instability of multiarmed spiral
waves that was previously missed in 
numerical simulations in the large core limit because it 
develops very slowly.

The present analysis can be compared with several previous analytical
approaches which have
provided insights into spiral wave dynamics. As already noted,
a phenomenological
kinematical model 
of spiral wave dynamics\cite{mikzyk}
has been proposed several years ago
and has succeeded in capturing many aspects of spiral wave motion. 
It
differs from the present approach not only because its parameters need to
be adjusted and cannot be obtained from the underlying
reaction-diffusion equation but also more fundamentally 
because, here the 
dynamics of the spiral tip is reduced to an ordinary differential equation
(ODE)
and drives the motion of
the rest of the spiral arm whereas in the kinematical model of \cite{mikzyk}
the tip motion follows from that of the whole curve.
Moreover, the tip
motion is described here in a different way,
by the tip rotation rate and not by a growing
or retracting velocity as in \cite{mikzyk}. 
Another notable approach is based on normal
forms \cite{Cod2}.
 As in our case, the tip motion is described by ODE. The normal from
approach postulates the existence of a Hopf bifurcation and it describes its
coupling to the spiral translation modes and the resulting tip motion
based on general symmetry arguments, 
close to the resonant case where the meander frequency $\omega_2$
is equal to the basic spiral rotation frequency $\omega_1$. The present
approach is  restricted to a particular limit but makes more specific
predictions. Besides
providing determined parameters in the reduced equation which gives, 
for instance, the drift angle with an external field,
it has the advantage, in our view, to provide an understanding of the physical
mechanisms responsible for the very existence of spiral waves and of their
dynamics, be it meander or drift due to external action.

Extensions of the present work can be considered in several directions:\\
- It
would be interesting to extend the analysis to slightly more excitable media
to capture hypermeandering or at least the change from inward to outward petals
(i.e. the line $\omega_1=\omega_2$). This would require going beyond our 
adiabatic approximation and considering the dynamics of the 
intermediate region.\\
- The large core nature of the spiral rotation (i.e. the
proximity of the line $\partial R$)  is an essential element of
our approach but several of our arguments do not really require sharp 
front and back interfaces (i.e. $\epsilon \ll 1$).
 This is certainly true for the $-3/2$ divergence
of the spiral radius divergence near the line $\partial R$ which only
requires that the spiral normal velocity and curvature be related by an eikonal
equation of the type of Eq.~(\ref{eikonal}) on a sufficiently large scale.
This is also  the case for the validity of the
adiabatic approximation. Thus, it appears
that a computation of critical fingers and of the allied
solvability conditions at finite $\epsilon$ would provide an extension
of our reduced description to the neighborhood of the full line $\partial R$.
This would not accurately describe 
meander(since $\partial R$ and $\partial M$ are
close only for $\epsilon \ll 1$) but would allow a simple quantitative 
description of spiral drift and other phenomena along this line.\\
- Finally, it appears possible to extend some of our calculations to scroll
waves in 3D as succinctly described for filament motion in 
section (\ref{efield}).
Hopefully, this will not only provide definite coefficients in the
average filament equations of motion,
 but it will also provide a better understanding
of the dynamics and instabilities of 3d scroll filaments \cite{win3d,fenkar}.

{\bf Acknowledgments} We are grateful to B. Pier for performing some 
computations at an early stage of this work. 
The work of A.K. is supported by the American Heart Association.

\appendix
\section{The tip boundary layer}
\label{apptipbl}
It has been noted in section (\ref{seccritfing}) that the solutions of the free
boundary problem (\ref{eikonal},\ref{exkin})
 are continuous as well as their first two derivatives
but have a discontinuous third derivative at their tip. We show in this appendix
that this weak non-analyticity can be taken care of by introducing a boundary
layer of size $\epsilon/\sqrt{c_0}\sim \epsilon^{5/6}$ 
near the wave tip (i.e. smaller than the
tip radius of curvature of size $\epsilon/c_0\sim \epsilon^{2/3}$ and larger
than the interface width $\sim \epsilon$).

We restrict ourselves to  analyzing the case of a critical finger. We take
the interface width $\epsilon$ as length unit. We first consider the sharp
interface case.
We find it convenient to
parameterize the interface as $x=h(y)$ instead of $y=y_{f/b}(x)$ as in the main part
of this article. 
 In the vicinity of the wave tip ($h'(y)\ll 1$), Eq.~(\ref{eikonal})
reduces to
\begin{equation}
c_0\  h'(y)=c(v)-h''(y)
\label{eikontip}
\end{equation}
The non analyticity of the interface is a direct consequence of the 
non-analyticity of $c(v)$,
\begin{eqnarray}
c(v)=c_0=-\alpha(v_0-v_s),\ y>0
\nonumber\\
c(v)=c_0-2y \frac{\alpha\epsilon}{c_0\tau_e}+\cdots, \ y<0
\label{csharp}
\end{eqnarray}
Eq.(\ref{eikontip},\ref{csharp}) gives
\begin{eqnarray}
h(y)&=&\frac{1}{c_0}[\frac{1}{2} (y c_0)^2 -\frac{1}{6} (y c_0)^3 +\cdots ]
,\ y>0
\nonumber\\
h(y)&=&
\frac{1}{c_0}[\frac{1}{2} (y c_0)^2 -\frac{1}{6} (1-2 B_c) (y c_0)^3 +\cdots
], \ y<0
\label{disc3d}
\end{eqnarray}

When one takes into account the finite width of the interface
$c(v)$ becomes a rapidly but smoothly varying function in the wave tip
neighborhood. For a shape moving at $c_0$ along the $y$-direction
the two reaction-diffusion equations (\ref{eqrd1},\ref{eqrd2}) become
\begin{eqnarray}
-c_0 \partial_y u&=&\nabla^2 u +f(u,v)
\label{equ1}\\
-c_0 \partial_y v &=& \epsilon g(u,v)
\label{eqv1}
\end{eqnarray}
For notational simplicity, we consider functions $f$ and $g$ of the form
$f(u,v)=F(u)-v,\ g(u,v)=u-\eta$ with the stall concentration  $v_s=0$ and
the corresponding rest state at $u=0,v=0$. We choose
$F(u)=-A u(u-1)(u-2)$ for illustrative purposes which gives $\Delta=2 A \eta$.
To study the tip neighborhood, it is convenient to use
instead of $x$, the displaced coordinate $z=x-h(y)$ where $x=h(y)$ is
a line in the interface transition region (for definiteness, one can take
an iso-$u$ line, for instance the line $u=1$ with the above choice of $F$).
Eq.~(\ref{equ1}) then reads
\begin{equation}
-c_0 [\partial_y-h'(y)\partial_z]u=\partial_z^2 u +
[\partial_y-h'(y)\partial_z]^2
u +F(u)-v
\label{equ2}
\end{equation}
The controller field $v$ is assumed to remain close to the stall concentration
and 
in the tip neighborhood $h'$ is small. Thus, at dominant order, 
Eq.~(\ref{equ2}) reduces to 
\begin{equation}
\partial_z^2 u+F(u)=0
\label{equ22}
\end{equation}
which has a standing front solution $u^{(0)}(z)$ which goes from
$u^{(0)}=0$ at $z=-\infty$ to $u^{(0)}=u_{\infty}$ at
$z=+\infty$  ( $u^{(0)}(x)=1 +\tanh(x\sqrt{A/2})$ for
the above choice of $F$). At next order, one obtains,
\begin{equation}
\partial_z^2 u^{(1)}+F'(u^{(0)})u^{(1)}=(c_0 h'(y)+h''(y))\partial_z u^{(0)}
-h'^2(y)\partial_z^2 u^{(0)} +v
\label{equ3}
\end{equation}
Integrating both members of Eq.(\ref{equ3}) with the zero-mode $\partial_x
u^{(0)}$ gives the solvability condition,
\begin{equation}
c_0 h'(y)=-h''(y)-c(y)
\label{eiksmooth}
\end{equation}
with
\begin{equation}
c(y)=\frac{\int dz\ v\partial_z u^{(0)}}{\int dz\
 (\partial_z u^{(0)})^2}
\label{solv1}
\end{equation}
When $v$ has negligible variations in the interface width, Eq.~(\ref{solv1})
gives back the sharp interface result with $c(v)=-\alpha v$ and
\begin{equation}
\alpha=\frac{u_{\infty}}{(\int dz\ (\partial_z u^{(0))})^2}
\end{equation}
On the contrary, in the tip region, $v$ varies in the interface transition
region and the integral term in (\ref{solv1}) needs to be more carefully
evaluated.
To lowest order, the field $v$ on the interface is obtained by integrating
Eq~.(\ref{eqv1})
\begin{equation}
v(z,y)= v_0 +\frac{\epsilon}{c_0} \int_y^{+\infty}\!\! dy_1\  u_0(z+h(y)-h(y_1))
\label{vexp1}
\end{equation}
When (\ref{vexp1}) is substituted in (\ref{solv1}), one obtains   a smooth
function  $c(y)$ 
\begin{equation}
c(y)=-\alpha v_0 -\frac{\epsilon\alpha u_{\infty}}{c_0}
\int^{+\infty}_{y}\!\! dy_1
T(h(y)-h(y_1))
\label{csmooth}
\end{equation}
with
\begin{equation}
T(w)=\int_{-\infty}^{+\infty} \! dz \frac{\partial_z u_0 \ u_0(z+w)}{
u_{\infty}^2}
\label{texp}
\end{equation}
Eq.~(\ref{texp}) gives a smoothly varying
function (for instance 
 with the above choice of $F$,
$T(w)=1/2\ [\exp(w \sqrt{A/2})/\sinh(w \sqrt{A/2})-
w \sqrt{A/2}/\sinh^2(w \sqrt{A/2})]$)
instead of the
Heavyside function
of the sharp interface limit.
To make further progress, we assume (and check 
afterwards) that $h(y)=c_0 y^2/2 +\eta(y)$ with $\eta$ 
a small correction in a neighborhood
of the spiral tip that can be neglected in evaluating the integral term 
in Eq.~(\ref{csmooth}), 
\begin{equation}
\int^{+\infty}_{y}\!\! dy_1
T(h(y)-h(y_1))\simeq \int^{+\infty}_{y}\!\! dy_1 T(c_0 y^2/2 -c_0 y_1^2/2)
\equiv \frac{1}{\sqrt{c_0}}S(y\sqrt{c_0})
\label{estint}
\end{equation}
Eq.~(\ref{eiksmooth}) then gives for the tip profile correction
\begin{equation}
c_0^2 y +c_0 \eta'(y)= -\,\frac{\epsilon\alpha u_{\infty}}{c_0^{3/2}}
\ S(y\sqrt{c_0})-\eta''(y)
\label{shacor1}
\end{equation}
Comparing the different terms, one obtains that a consistent 
scaling
 is
$y\sim 1/\sqrt{c_0}$ and $\eta\sim \sqrt{c_0}$ which give the size of the
boundary layer (note that, here,
 our unit of length is the interface width $\epsilon$)
and the magnitude of the shape correction in the boundary layer.
This legitimates the neglect of $\eta$ in ($\ref{estint}$).
In the scaled
variables, $y=Y/\sqrt{c_0}$ and $\eta(y)=\sqrt{c_0}H(y\sqrt{c_0})$, the equation for the tip profile correction is
\begin{equation}
\frac{d^2H}{dY^2}+B_c \ S(Y)+Y=0
\end{equation}
where the function $S(Y)$ is defined by (\ref{estint}) and (\ref{texp})
(the second term on the l.h.s of (\ref{shacor1}) is of higher order).
The behaviors of $S$ at infinity, $S(Y)\rightarrow 0$ at $Y=+\infty$
and $S(Y)\rightarrow -2Y$ at $Y=-\infty$,
 give the corresponding asymptotic behaviors
of $H$, $H(Y)\rightarrow -Y^3/6$ at $Y=+\infty$ and $H(Y)\rightarrow 
-Y^3 (1-2 B_c)/6$ at $Y=-\infty$. These precisely match the 
different small $y$ behaviors (Eq.~(\ref{disc3d}))
 of the sharp interface  description. It shows that $H(Y)$ interpolates
smoothly between these different behaviors.

\section{Dynamics of the intermediate region}
\label{appdyn}
The analysis of spiral dynamics that we have developed in the main part of
this article makes a crucial use of an 'adiabatic' assumption. Namely, that
for changes of medium parameter on the time scale of the spiral rotation
period $T\sim R_0/c_0$, the instantaneous motion of the spiral tip can be taken to be that
of a spiral tip moving in an steady medium  with characteristics invariant in
time and identical to those of the changing medium at the considered time.
In the large core limit, the spiral is described by matching three regions :\\
- a close tip region on the scale of the tip radius $R_{tip}=\epsilon/c_0$ which determines
the spiral tip tangential velocity,\\
- an intermediate region  of size $(R_0 R_{tip}^2)^{1/3}$ which determines
the instantaneous radius of curvature of the tip trajectory\\
- and finally, an outer scale the dynamics of which is driven by the previous
two regions.\\
The close tip region relaxes on a time scale which is independent of the spiral
radius and which therefore clearly becomes short compared to the spiral period
$T$
for a spiral of sufficiently large radius. In this appendix, we show that the
intermediate region relaxes on a time scale $T\,(R_{tip}/R_0)^{1/3}$ which
is also much shorter than the rotation period $T$ in the large radius limit.
This justifies our adiabatic assumption.

We first write the dynamic equivalent of the static BCF equation (\ref{bcf}),
 that is the motion
of a curve governed by (\ref{eikonal}) using polar coordinates
\begin{equation}
r\,\frac{\partial\theta_f}{\partial t}=
 c_0 \left(1+\left(r\frac{d\theta_f}{dr}\right)^2\right)^{1/2}
+ \epsilon\,\left( \frac{d\theta_f}{dr} +\frac{\frac{d}{dr} (r 
\frac{d\theta_f}{dr})
}{1+\left(r\frac {d\theta_f}{dr}\right)^2}\right)
\label{bcfdyn}
\end{equation}
As for the static case, it is convenient in the intermediate region to introduce
the rescaled variables $y$ and $\xi$ with $\theta_f=\omega_1 t+ y \epsilon/(c_0 R_0)$ and
$r=R_0 +[2 R_0(\epsilon/c_0)^2]^{1/3} \xi$. Expanding the square root in 
(\ref{bcfdyn}) and keeping terms of the dominant order gives
\begin{equation}
\frac{R_0}{c_0} \left(\frac{\epsilon}{2 c_0 R_0}\right)^{1/3} 
\frac{\partial y}{\partial t} + \xi + a=
\frac{1}{2} \frac{d^2y}{d\xi^2}+\frac{1}{4}\left(\frac{dy}{d\xi}\right)^2
\label{riccatidyn}
\end{equation}
with
\begin{equation}
a= 2^{-1/3} \left(\frac{c_0 R_0}{\epsilon}\right)^{2/3} \, \frac{c_t-c_0}{c_0}
\label{adyn}
\end{equation}
Eq.~(\ref{riccatidyn}) is the dynamic equivalent of the static equation
(\ref{riccati}) determining the shape of the intermediate region. It shows
that the characteristic time to adapt to changes of $a$ (e.g. of $c_t$) for
$\xi$ of order unity (e.g. for the intermediate region) is $R_0/c_0 \ 
[\epsilon/(c_0 R_0)]^{1/3}$. 
It is shorter by the factor $(R_{tip}/R_0)^{1/3}$ than the 
rotation period as announced above. It is however worth pointing out than
this is larger than the time one may have guessed,
 namely the length of the intermediate
region divided by the velocity $c_0$. The reason is that in the 
intermediate region the interface is almost radial. As a consequence,
the advection velocity is much smaller than $c_0$ and advective
effects become comparable to diffusion-like effects due to surface tension
(i.e. the last two terms in (\ref{riccatidyn}) are of the same magnitude). 
\section{Solvability integrals and functions : some relations}
\label{appsolv}
In this appendix,
we recapitulate  the definitions and give additional information on
the several functions and integrals 
which have been introduced in the evaluation of solvability conditions.

The linear operators considered are ${\cal L}_f$ and ${\cal L}_b$ which
comes from the linearization of the front and back equations around the
critical finger in the tip region
\begin{eqnarray}
{\cal L}_f &=& \frac{d^2}{dx^2}+\left[-2 +3\left[1+\left(\frac{dY_f}{
dx}\right)^2\right]^{1/2}\right] \frac{dY_f}{dx} \frac{d}{dx}
\label{oplinfrontap}\\
{\cal L}_b &=& \frac{d^2}{dx^2} -a(x)\frac{d}{dx}-b(x)
\label{oplinbackap}\\
\end{eqnarray}
with,
\begin{eqnarray}
a(x)&=&
\left[2+3\left[1-B_c\left(Y_f\left(x\right)-Y_b(x)\right)
\right]\left[1+\left(\frac{dY_b}{dx}\right)^2\right]
^{1/2}\right] \frac{dY_b}{dx}\nonumber\\
b(x)&=& B_c \left(1+(\frac{dY_b}{dx})^2\right)^{3/2}
\label{eqabap}
\end{eqnarray}
$Y_f(x)$ and $Y_b(x)$ are the critical finger front and back interfaces
which satisfy (\ref{fingfront}) and (\ref{fingback}) with 
$B=B_c=0.5353\cdots$. The small $x$ behaviors of these different functions
are  
$Y_f(x)=\sqrt{2x}+x/3+\cdots$,  
$Y_b(x)=-\sqrt{2x}+x (1-2 B_c)/3+ \cdots$, $a(x)=-3/(2x) +\sqrt{2} B_c/\sqrt{x}
+ \cdots$ and $b(x)= B_c/(2x)^{3/2}+\cdots$.

The zero-mode $\xi(x),\ x\ge 0$, of the adjoint of ${\cal L}_b$
is the solution
of
\begin{equation}
{\cal L}_b^{\dag}(\xi)=\frac{d^2\xi}{dx^2}+\frac{d}{dx}(a(x)\xi)-b(x)\xi=0
\label{adjap}
\end{equation}
which tends to $0$ when 
$ x\rightarrow +\infty$. 
It is here normalized
by imposing the supplementary
condition $\sup_{x\ge 0}[\xi(x)]=1$. A local analysis determines the behavior
of $\xi(x)$ for small $x$, $\xi(x)=\xi'(0)\,[x-B_c/\sqrt{2} x^{3/2} \ln(x)+
\cdots]$. Eq.(\ref{adjap}) has been solved numerically by a finite-
difference scheme on a non-uniform grid (with a step size decreasing to zero
at small $x$). A graph of the obtained solution is shown in 
Fig.~\ref{figadj}. The computed value of the derivative of $\xi$
at the origin is $\xi'(0)\simeq 4.441$. An exact relation between $\xi'(0)$
and a weighted integral of $\xi$ is obtained by integrating (\ref{adjap})
between $x=0$ and $x=+\infty$,
\begin{equation}
\xi'(0)/2= \int_0^{+\infty}\!\! dx\ \xi(x) b(x)= B_c I_4
\label{relder}
\end{equation}
The verification of (\ref{relder}) serves as a check of our numerical
computation.

To evaluate the solvability conditions, besides $Y_f,Y_b$ and
$\xi$,
the solutions of the following inhomogeneous equations with the  linear 
operator ${\cal L}_f$ (Eq.~(\ref{oplinfrontap}))are needed, 
\begin{eqnarray}
{\cal L}_f (\eta_1)&=& 1+\left(\frac{dY_f}{dx}\right)^2, 
\ \eta_1(0)=0,\ \eta_1'(0)=1/3
\label{eta1ap}\\
{\cal L}_f (\eta_2)&=& \left[1+\left(\frac{dY_f}{dx}\right)^2\right]^{3/2}, 
\ \eta_2(0)=0,\ \eta_2(x)\sim \sqrt{x/2}\  {\mathrm{for}}\  x\ll 1
\label{eta2ap}\\
{\cal L}_f (\eta_{v,0})&=& (Y_f(x)-Y_b(x))
\left[1+\left(\frac{dY_f}{dx}\right)^2\right]^{3/2},\ \eta_{v,0}(0)=0,
\frac{d\eta_{v,0}}{dx} (0)=2/3  
\label{etav0ap}\\
{\cal L}_f (\eta_e)&=& \frac{dY_f}{dx} 
\left[1+\left(\frac{dY_f}{dx}\right)^2\right], 
\ \eta_e(0)=0,\ \eta_e(x)\sim \sqrt{x/2}\ {\mathrm{for}}\  x\ll 1
\label{etaeap}
\end{eqnarray}
They are plotted in Fig.~\ref{figetap}. 

For the evaluation of the different solvability conditions, it is useful
to compute the following integrals,
\begin{eqnarray}
I_1&=&\int_0^{+\infty}\!\!dx\,\xi(x)\left[1+\left(\frac{dY_b}{dx}\right)^2\right]
\simeq 2.771  
\nonumber
\\
I_2&=&\int_0^{+\infty}\!\!dx\,\xi(x)\eta_1(x)
\left[1+\left(\frac{dY_b}{dx}\right)^2\right]^{3/2}\simeq
3.814
\nonumber\\
I_3&=&\int_0^{+\infty}\!\!dx\,\xi(x)\, \left[Y_f(x)-Y_b(x)\right]
\left[1+\left(\frac{dY_b}{dx}\right)^2\right]^{3/2}\simeq 7.708
\nonumber\\
I_4&=&\int_0^{+\infty}\!\!dx\, \xi(x)
 \left[1+\left(\frac{dY_b}{dx}
\right)^2\right]^{3/2}\simeq 4.1476
\nonumber\\
I_5&=&\int_0^{+\infty}\!\!dx\, \xi(x) \eta_2(x)
 \left[1+\left(\frac{dY_b}{dx}
\right)^2\right]^{3/2}\simeq 6.306
\nonumber\\
I_6&=&\int_0^{+\infty}\!\!dx\,
\xi(x)\frac{dY_b}{dx}\left[1+\left(\frac{dY_b}{dx}\right)^2\right]
\simeq -2.118 
\nonumber\\
I_{v,0}&=&\int_0^{+\infty}\!\!dx\, \xi(x)\,\eta_{v,0}
 \left[1+\left(\frac{dY_b}{dx}
\right)^2\right]^{3/2}\simeq 12.553
\nonumber\\
I_{\perp}&=&\int_0^{+\infty}\!\!dx\,\xi(x)\, \left[Y_f(x)-Y_b(x)\right]
\frac{dY_f}{dx}\
\left[1+\left(\frac{dY_b}{dx}\right)^2\right]^{3/2}\simeq 8.431
\nonumber\\
I_e&=&\int_0^{+\infty}\!\!dx\,\xi(x)\eta_e(x)
\left[1+\left(\frac{dY_b}{dx}\right)^2\right]^{3/2}\simeq
4.476
\label{i123ap}
\end{eqnarray}

Exact relations between some of these integrals can be obtained by
using symmetry transformations of known action on the interfaces.
For instance, under dilation the critical finger front and back
become $Y_{f,\alpha}=Y_f(\alpha x)/\alpha$,$
\ Y_{b,\alpha}=Y_b(\alpha x)/\alpha$ and obey scaled versions
of Eq.(\ref{fingfront}) and (\ref{fingback}),
\begin{eqnarray}
\frac{d^2 Y_{f,\alpha}}{dx^2}&=&
\alpha\ \left\{\left[1+ (\frac{dY_{f,\alpha}}{dx})^2\right]-\left[1  
+ (\frac{dY_{f,\alpha}}{dx})^2\right]^{3/2}\right\}
\label{fingfrontdil}\\
\frac{d^2 Y_{b,\alpha}}{dx^2}&=&\alpha\ \left\{
\left[1+ (\frac{dY_{b,\alpha}}{dx})^2\right]
+\left[1- B_c\,\alpha\left[
Y_{f,\alpha}(x)-Y_{b,\alpha}(x)\right]\right]\left[1
+ (\frac{dY_{b,\alpha}}{dx})^2\right]^{3/2}\right\}
\label{fingbackdil}
\end{eqnarray}
An expansion around $\alpha=1$ gives $Y_{f,\alpha}=Y_f+(\alpha-1)
 \delta Y_f
+\cdots$ where $\delta Y_f(x)=x Y'_f(x)-Y_f(x)$. Similarly,
one has $Y_{b,\alpha}=Y_b+(\alpha1) \delta Y_b+\cdots$ with
$\delta Y_b(x)=x Y'_b(x)-Y_b(x)$. Expanding 
(\ref{fingfrontdil},\ref{fingbackdil}) in the same limit shows
that $\delta Y_f$ and $\delta Y_b$ obey the following
linear equations ,
\begin{eqnarray}
{\cal L}_f (\delta Y_f)&=&\left[1+ (\frac{dY_f}{dx})^2\right]-
\left[1
+ (\frac{dY_f}{dx})^2\right]^{3/2}
\label{eqdyf}
\\
{\cal L}_b (\delta Y_b)&=&
\left[1+ (\frac{dY_b}{dx})^2\right]
+\left[1- 2\, B_c\,\left[
Y_f(x)-Y_b(x)\right]-B_c\, \delta Y_f\right]\left[1
+ (\frac{dY_b}{dx})^2\right]^{3/2}
\label{eqdyb}
\end{eqnarray}
Eq.(\ref{eqdyf}) shows that $\delta Y_f=\eta_1-\eta_2$ (\ref{eta1ap},
\ref{eta2ap}) since $\delta Y_f =0$ 
(as can be checked from its explicit expression). Then, multiplying
both sides of (\ref{eqdyb}) by $\xi(x)$ and integrating from $x=0$ to
$+\infty$ gives the searched relations between the above integrals
\begin{equation}
I_1+I_4 -\ B_c\ (I_2+2 I_3-I_5)=0
\end{equation}

Using rotational symmetry in a similar manner, one obtains
that $\eta_e(x)=x+Y_f dY_f/dx-1$ and the other relation
\begin{equation}
I_6+ B_c (I_{\perp}-I_e)=0
\end{equation}

In the analysis of meandering, there appear
several functions of the tip displacement:
\begin{equation}
{\cal L}_f (\eta_{v,d})= \left [Y_f(x+d)-Y_b(x+d)\right ]\Theta(x+d)
\left[1+\left(\frac{dY_f}{dx}\right)^2\right]^{3/2},\ \eta_{v,d}(0)=0
\label{etavdap}
\end{equation}
\begin{eqnarray}
I_{3,d}&=&
\int_0^{+\infty}\!\!dx\,\xi(x)\, \left[Y_f(x+d)-Y_b(x+d)\right]\Theta(x+d)
\left[1+\left(\frac{dY_b}{dx}\right)^2\right]^{3/2}\nonumber\\
I_{v,d}&=&\int_0^{+\infty}\!\!dx\, \xi(x)\,\eta_{v,d}
 \left[1+\left(\frac{dY_b}{dx}
\right)^2\right]^{3/2}
\label{ivdi3dap}
\end{eqnarray}
For large $|d|$, these functions tend toward constant values, toward
$0$ when $d\rightarrow -\infty$ and $I_{3,d}
\rightarrow 2 I_4/B_c$ and $I_{v,d}\rightarrow 2 I_5/B_c$ when $d\rightarrow
+\infty$
Their behavior for small displacements ($|d|\ll 1$) is
nonanalytic. From the small $x$ behaviors of $Y_f, Y_b$ and $ \xi$, one
obtains, for $0<|d|\ll 1 $,
\begin{eqnarray}
I_{3,d}=I_3 -\frac{\xi'(0)}{2} d \ln(|d|) + O(d)
\\
\eta_{v,d}(x)=\eta_{v,0}(x) - d\ln(|d|) + O(d)
\label{expet}
\end{eqnarray}
The expansion
(\ref{expet}) of
$\eta_{v,d}$ for small $d$ gives for $I_{v,d}$
\begin{equation}
I_{v,d}=I_{v,0}-I_4\ d\ln(|d|)+ O(d)
\end{equation} 
The function $F$ which measures the spiral self-interaction, 
$F(d)=(I_{3,d}+B_c I_{v,d})/I_3 -J$, has therefore a singular expansion
for $0<d\ll 1 $,
\begin{equation}
F(d)\sim -\ \frac{2 B_c}{I_3}\ d\ln(|d|)\simeq -.576\  d\ln(|d|)
\end{equation} 
(where we have used (\ref{relder}) 
which shows that the singular contributions
of $I_{3,d}$ and $ B_c I_{v,d}$ are equal). For $d\rightarrow -\infty$
$F(d)$ tends toward $-J\simeq-1.872$ and for $d\rightarrow +\infty$
$F(d)$ approaches  $2(I_5+I_4/B_c)/I_3 -J\simeq 1.774$.

The singular behavior of $F(d)$ at small $d$ disappears when the controller
field diffuses. For small diffusion $(Y_f(x+d)-Y_b(x+d))\Theta(x+d)$ is 
simply replaced in (\ref{etavdap} \ref{ivdi3dap}) by the smoother
function $YS(x+d)$
\begin{equation}
YS(x;\ell_D)=\int_0^{+\infty}\frac{dx'}{\sqrt{\pi}\,\ell_D}
\exp(-\frac{(x-x')^2}{\ell_D^2})(Y_f(x')-Y_b(x'))
\end{equation}
For $\ell_D\ll 1$, one can check that
the singular behavior of the self-interaction
function is cut-off
at $d\sim\ell_D$ and the small distance behavior of $F(d;\ell_D)$ is
\begin{equation}
F(d;\ell_D)\sim -.576\  d \ln(\ell_D)
\label{ldsmall}
\end{equation}
 
In the other limit $\ell_D\gg 1$, for $x$ of order one,
$YS(x;\ell_D)\simeq 1/B_c +  2 x /(\sqrt{\pi} B_c \ell_D)$. The corresponding
behavior of 
$I_{3,d;\ell_D}$ and $I_{v,d;\ell_D}$ at small $d$ is
$I_{3,d;\ell_D}=I_{3,0;\ell_D}+ 2 I_4 d /(B_c\sqrt{\pi}\ell_D)+\cdots$,
$I_{v,d;\ell_D}=I_{v,0;\ell_D}+ 2 I_5 d /(B_c\sqrt{\pi}\ell_D)+\cdots$.
So the small distance behavior of the spiral self-interaction function is
, for $\ell\gg 1$ (but still smaller than the scale of the matching region),
\begin{equation}
F(d;\ell_D)\sim 2\frac{I_4+B_c I_5}{B_c\sqrt{\pi} I_3}\frac{d}{\ell_D}\simeq 
2.06 d/\ell_D
\label{ldlarge}
\end{equation}
The derivative at $d=0$ of the spiral self interaction function has been
computed numerically for intermediate values of $\ell_D$. It is plotted
in Fig.~\ref{derfvsl}.

\section{Numerical integration of the 
wave tip equation}

In this appendix, we describe a simple scheme to
integrate numerically the equation of motion for
the wave tip (\ref{wavetip}), which is convenient to
rewrite as a system of first order ordinary
differential equations
\begin{eqnarray}
\frac{dq}{d\theta}&=&p\label{ode1}\\
\frac{dp}{d\theta}&=&-q+mF(q(\theta)-q(\theta-2\pi)) \label{ode2}
\end{eqnarray}
The difficulty of integrating
this equation comes from the fact
that the translational invariance of the
underlying reaction-diffusion equations
remains present in Eq.~(\ref{wavetip}), and
hence in Eqs.~(\ref{ode1})-(\ref{ode2}),
which are invariant under the
transformation  
\begin{equation}
q(\theta)=q(\theta)+A\,e^{i\theta} + c.c. \label{symm}
\end{equation}
where $A$ is an arbitrary complex amplitude. It is therefore desirable to
develop a numerical scheme that {\it discretely} preserves this symmetry
in order to avoid spurious discrete effects resulting from the
coupling of this translational mode to other modes. To see how to
construct such a scheme, let us first consider the case
the second term on the r.h.s. of Eq.~(\ref{ode2}) is absent. In this
case these equations describe simple harmonic motion
with a constant energy $\sim |A|^2$. It is
well-known (and simple to show) that the
simple Euler explicit scheme
\begin{eqnarray}
q_{n+1}&=&q_n+h\,p_n\\
p_{n+1}&=&p_n-\,h q_n
\end{eqnarray}
where $h$ is the time (angle) step,
does not conserve energy, but rather pumps energy
into the motion. As a result, it leads to an unbounded
increase in $A$ at long time. Therefore, this scheme
violates in an obvious way
the symmetry that we would like to preserve here.
In contrast, the modified (Euler-Cromer) scheme
\begin{eqnarray}
p_{n+1}&=&p_n-\,h q_n\,\label{sch1} \\
q_{n+1}&=&q_n+h\,p_{n+1} \label{sch2}
\end{eqnarray}
exactly conserves the energy of harmonic motion. This can be
seen by substituting the ansatz
$p_n=p_0 r^n$ and $q_n=q_0 r^n$ into
Eqs.~(\ref{sch1}) and (\ref{sch2}). Nontrivial solutions then
exist only if
\begin{equation}
r = 1-h^2/2 \pm i h\sqrt{1-h^2/4}=e^{\pm i\Delta \theta}
\end{equation}
where
\begin{equation}
\Delta \theta = \tan^{-1}\left(
\frac{h\sqrt{1-h^2/4}}{1-h^2/2}\right) \label{interm}
\end{equation}
Hence, $q_n^*=Ae^{i n \Delta \theta}+c.c.$
is a solution of Eqs.~(\ref{sch1}-\ref{sch2}) with constant $A$.
Let us now extend this scheme to the case where the second term
on the r.h.s. of Eq.~(\ref{ode2}) is included, by simply letting
\begin{eqnarray}
p_{n+1}&=&p_n-\,h q_n\,+\,h\,m\,F(q_n-q_{n-N})\label{esch1}\\
q_{n+1}&=&q_n+h\,p_{n+1}\label{esch2}
\end{eqnarray}
Now the key point is that
$q_n^*=Ae^{i n \Delta \theta}+c.c.$
remains an exact solution of these equations
only if $q_n^*-q_{n-N}^*=0$, and thus
\begin{equation}
\Delta \theta= 2\pi/N
\end{equation}
This condition together with Eq.~(\ref{interm}) then
uniquely fixes the step $h$ for a given number of
time steps, $N$, per basic period of $2\pi$. After simple algebraic
manipulations, we find that $h$ should be equal to
\begin{equation}
h=2 \sin(\pi/N)
\label{esch3}
\end{equation}
In summary, our integration scheme is uniquely defined by Eqs.~(\ref{esch1})
and (\ref{esch2}) with $h$ given by Eqs.~(\ref{esch3}).
For an arbitrary value of $N$, this scheme is invariant
under the transformation
\begin{equation}
q_n\,=\,q_n \,+\,A e^{i n\Delta \theta}\,+\, c.c.,
\end{equation}
which is the direct discrete analog of Eq.~(\ref{symm}).
A
solution of a desired numerical accuracy can then be obtained
by choosing $N$ sufficiently large.

\newsavebox{\fag}

\newsavebox{\fbg}

\newsavebox{\fcg}

\newsavebox{\fdg}
\begin{figure}
\centerline{
\psfig{file=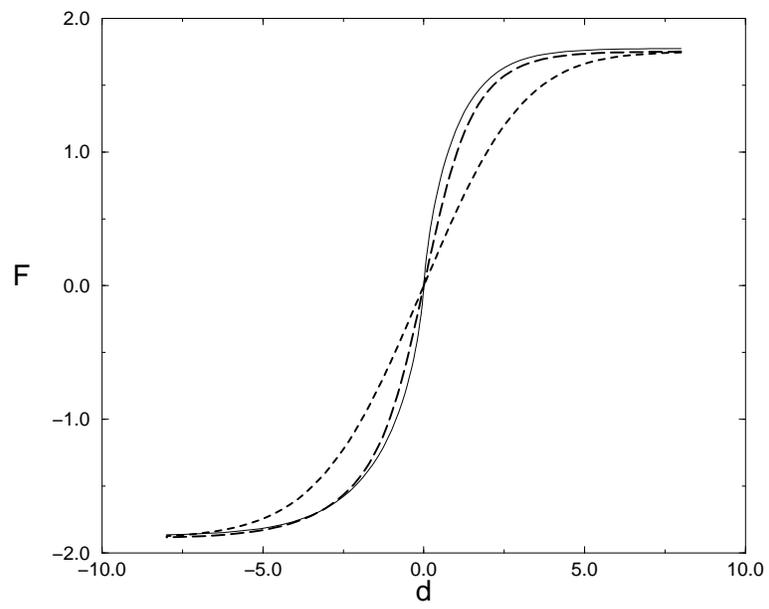,angle=-90,width=.7\textwidth}}
\bigskip
\caption{Graph of the of the spiral self-interaction function
$F(d)$ {\em{vs.}} tip displacement $d$ for different diffusion lengths :
$\ell_D=0$ (solid line), $\ell_D=1$ (long-dashed line) and $\ell_D=3$
(short-dashed line).
}
\label{figF}
\end{figure}

\newsavebox{\feg}
\begin{figure}
\centerline{
\psfig{file=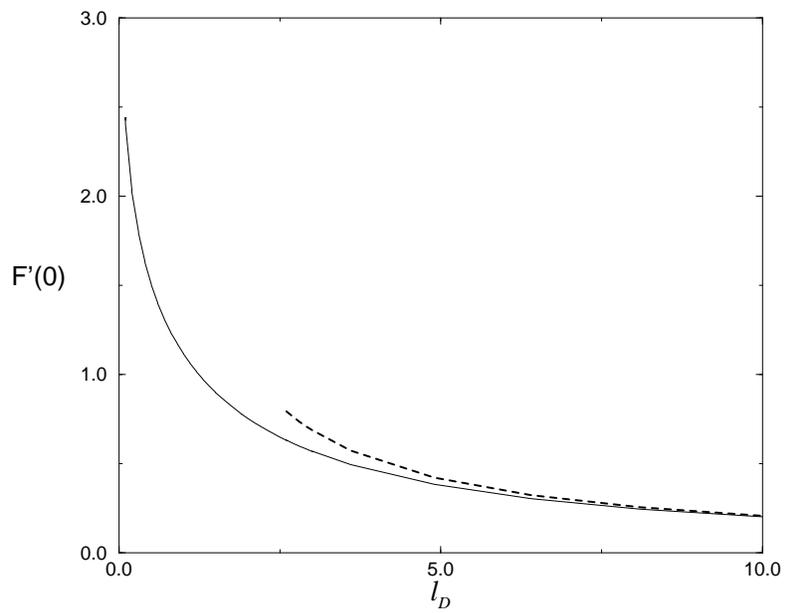,angle=-90,width=.7\textwidth}}
\bigskip
\caption{Derivative at $d=0$ of the spiral self-interaction function
$F(d;\ell_D)$ {\em vs.} the diffusion length $\ell_D$. The dashed line
shows 
the large $\ell_D$ 
(Eq.~\ref{ldlarge}) asymptotic behavior.
}
\label{derfvsl}
\end{figure}

\newsavebox{\ffg}
\begin{figure}
\centerline{
\psfig{file=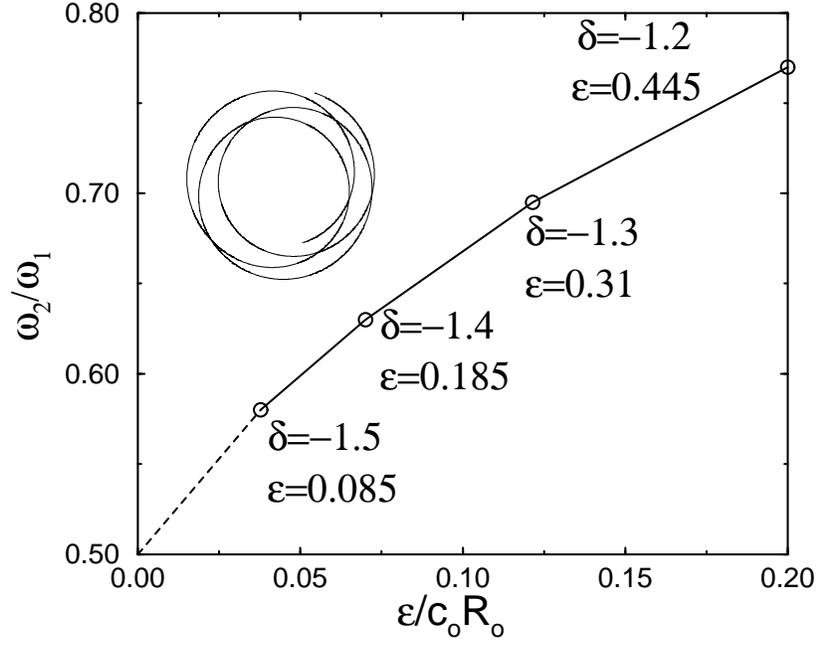,angle=-90,width=.7\textwidth}}
\bigskip
\caption{Plot of $\omega_2/\omega_1$ vs
$c_0/\epsilon R_0$
obtained by simulations
of FitzHugh-Nagumo kinetics with $f(u,v)=3u-u^3-v$
and $g(u,v)=u-\delta$ (solid line and circles).
$c_0=(\delta^3-3\delta)/2^{1/2}$.
The dash line represents the
extrapolation to the asymptotic limit $\omega_2/\omega_1=1/2$
predicted by our analysis.
These simulations were carried out using a second order accurate
direction implicit scheme
with $dx/\epsilon=0.33$ and $dt/\epsilon=0.1$. The insert
shows an example of a large core meander pattern
for $\epsilon=0.180$ and $\delta=-1.4$,
where $\omega_2/\omega_1\simeq 0.67$.}
\label{omvsr}
\end{figure}

\begin{figure}
\centerline{
\psfig{file=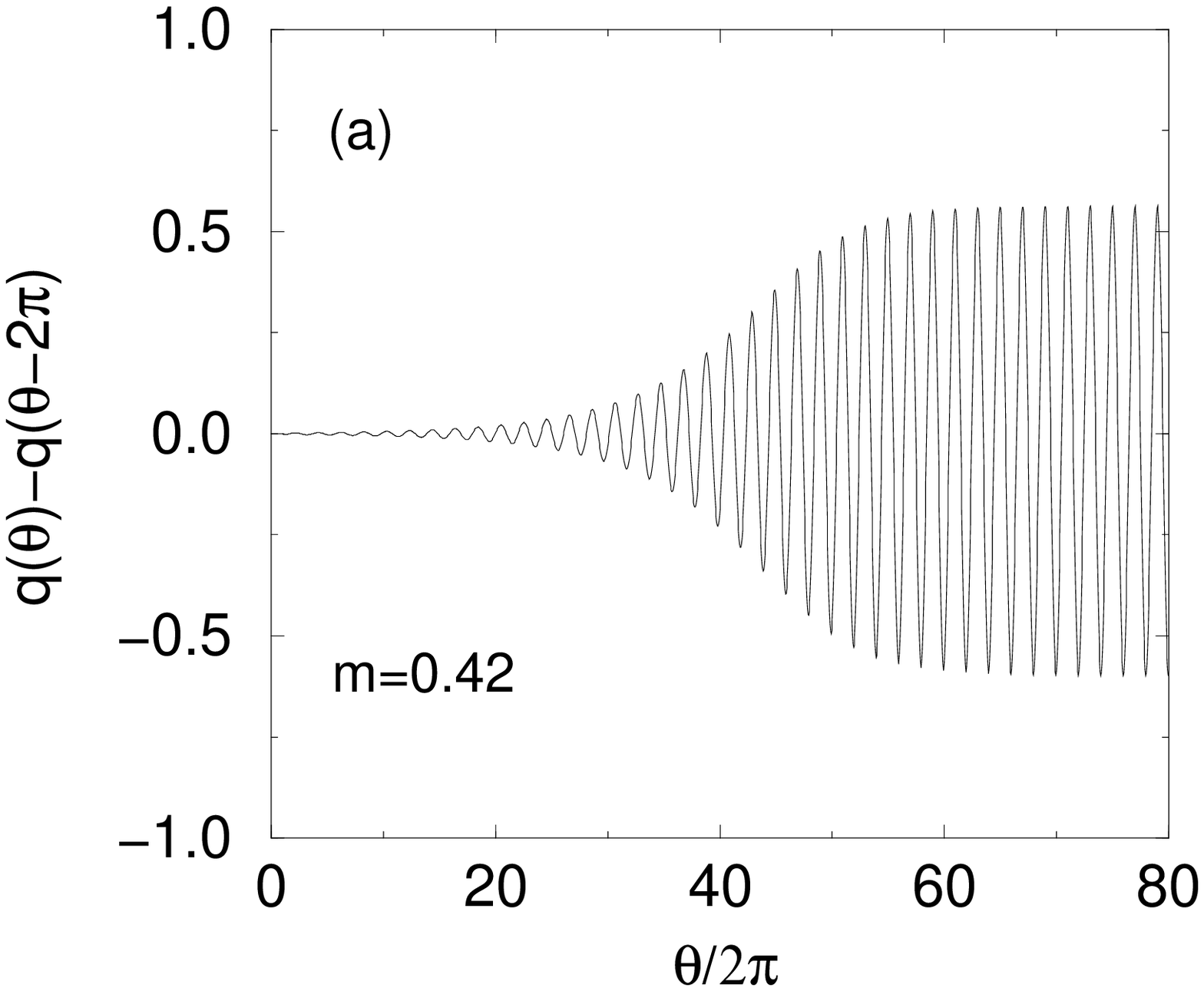,width=.45\textwidth}
\hfill
\psfig{file=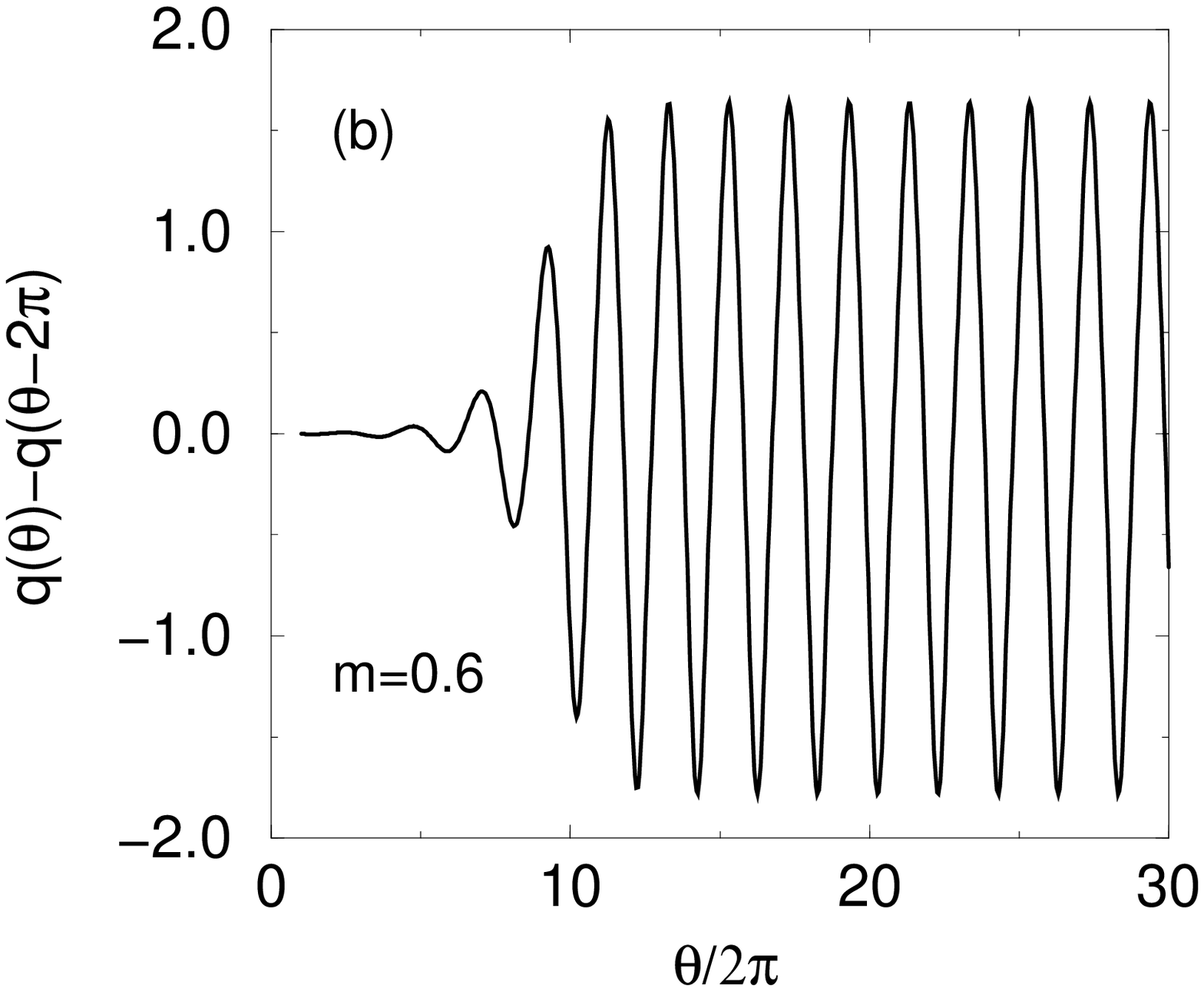,width=.45\textwidth}}
\caption{
Plot of $q(\theta)-q(\theta-2\pi)$ vs $\theta/2\pi$ obtained by numerical
integration of the wave tip equation with $F$ defined by
\protect Eq. \ref{asymth} and $a=0.2$;
(a) $m=0.42$, and (b) $m=0.6$. The onset of meander for this 
function $F$ corresponds to $m_c=0.3902$.}
\label{qplotab}
\end{figure}

\begin{figure}
\centerline{
\psfig{file=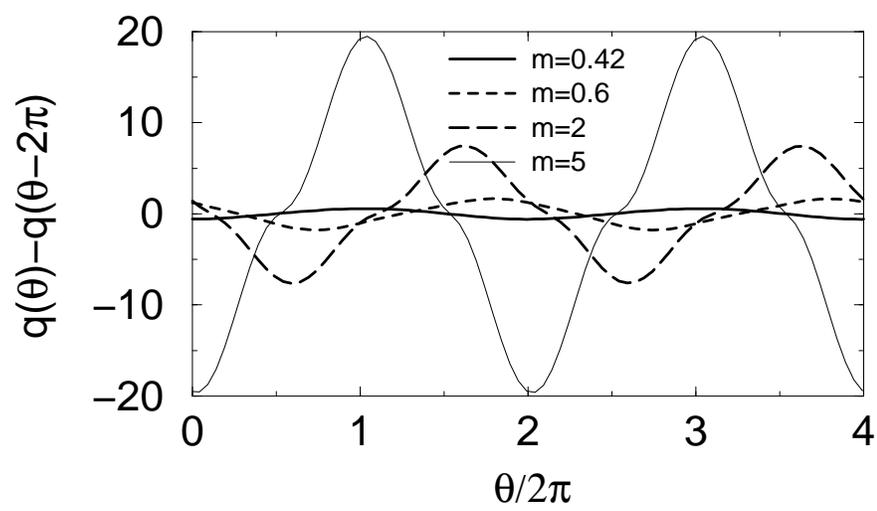,width=.85\textwidth}
}
\caption{
Plot showing saturated oscillations
of $q(\theta)-q(\theta-2\pi)$ vs $\theta/2\pi$ for different
$m$.}
\label{qplot}
\end{figure}

\begin{figure}
\centerline{
\psfig{file=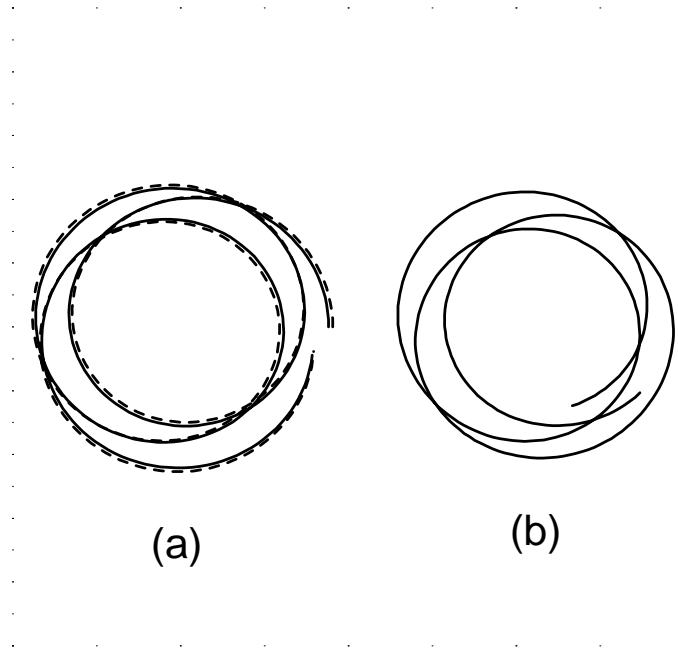,width=.85\textwidth}
}
\caption{
Comparison of large 
core meander trajectories obtained:
(a) by plotting a two-radius epi-cycle (solid line)
with $\Omega=0.735$ and $\rho_1/\rho_0=1/5$
and the predicted three-radius epicycle (dashed line) with $\Omega=0.735$,
$\rho_1/\rho_0=1/5$, and $\rho_2/\rho_1=(1-\Omega)/(1+\Omega)$,
and (b) by simulation of the FN model 
with $\epsilon=0.18$ and $\delta=-1.4$.
The value of $\Omega$ and the ratio $\rho_1/\rho_0$ used as input in (a)
were extracted from the simulation in (b). The total time in (a) and
(b) is about $3T_0$.}
\label{tipcomp}
\end{figure}

\newsavebox{\fed}
\begin{figure}
\centerline{
\psfig{file=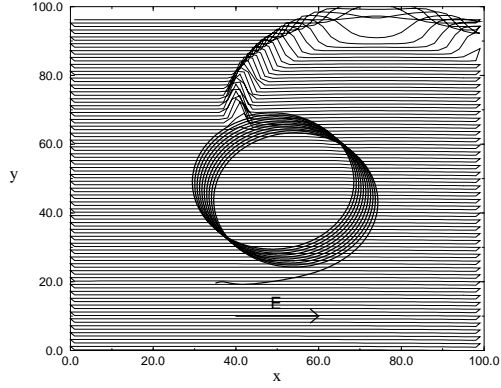,angle=-90,width=.45\textwidth}}
\bigskip
\caption{Simulation of the FN model of Fig.~\ref{flowergarden}
($\epsilon=0.185, \delta=-1.41$) with an external field added as in
Eq.~(\ref{fb1e}) with $E=1.0\ 10^{-3}$. The wave tip trajectory is shown
(bold line) as well as surface plots of $u$ showing the spiral position
at the end of the simulation. The spiral is found to drift at about
$135^{\circ}$ with the field in good agreement with the asymptotic prediction
of $132,5^{\circ}$
}
\label{edrift.fig}
\end{figure}

\newsavebox{\fddg}
\begin{figure}
\centerline{
\psfig{file=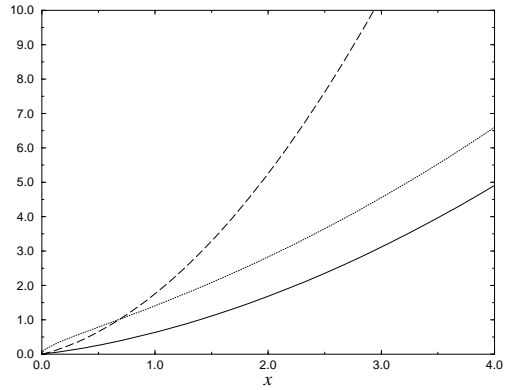,angle=-90,width=.45\textwidth}}
\bigskip
\caption{Graph of the functions $\eta_1(x)$ (solid line), $\eta_2(x)$ (dotted
line) 
 and $\eta_{v,0}(x)$ (dashed line)  defined in
Eq.~(\ref{eta1ap},\ref{etav0ap}). 
}
\label{figetap}
\end{figure}

\begin{figure}
\centerline{
\psfig{file=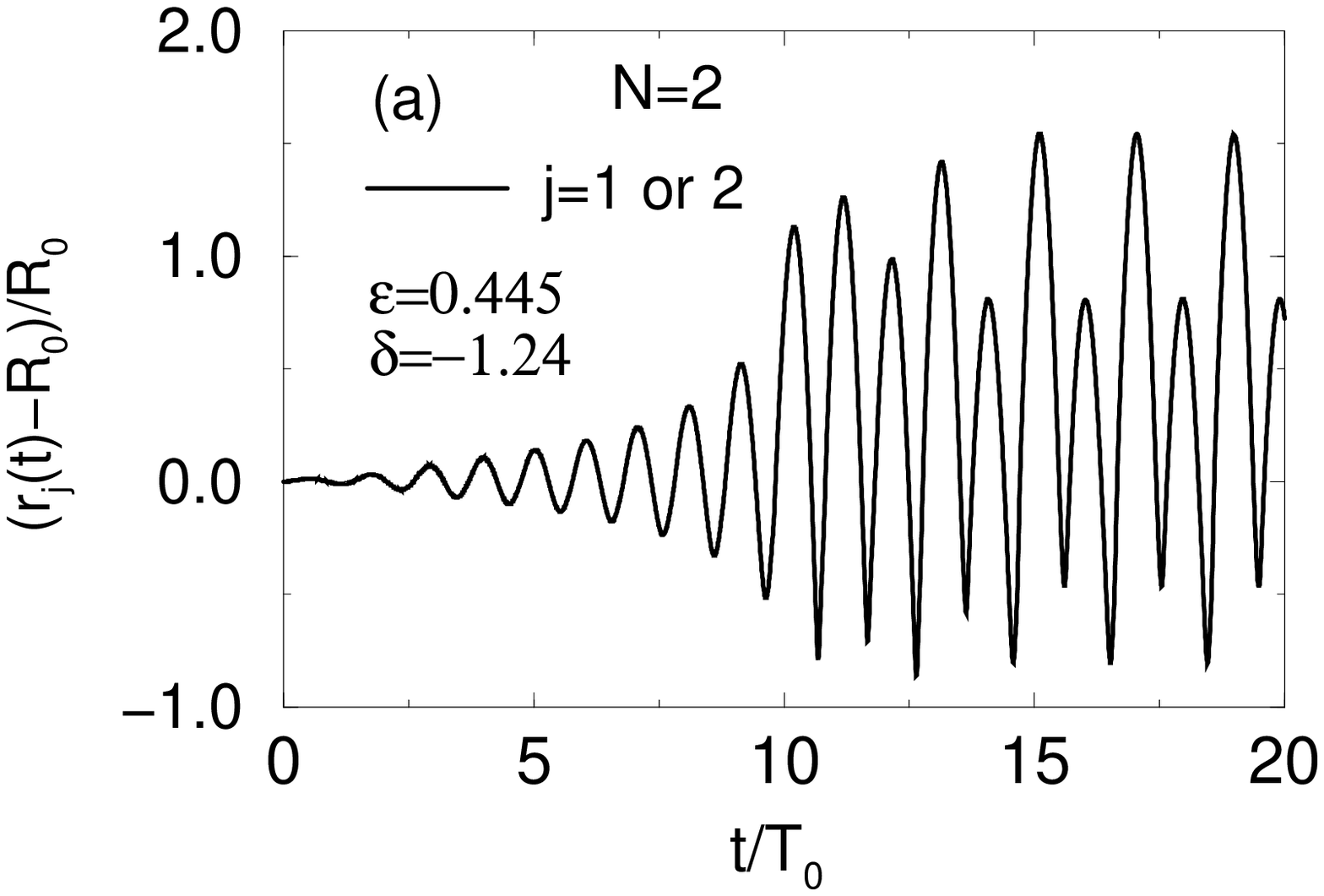,width=.4\textwidth}}
\centerline{
\psfig{file=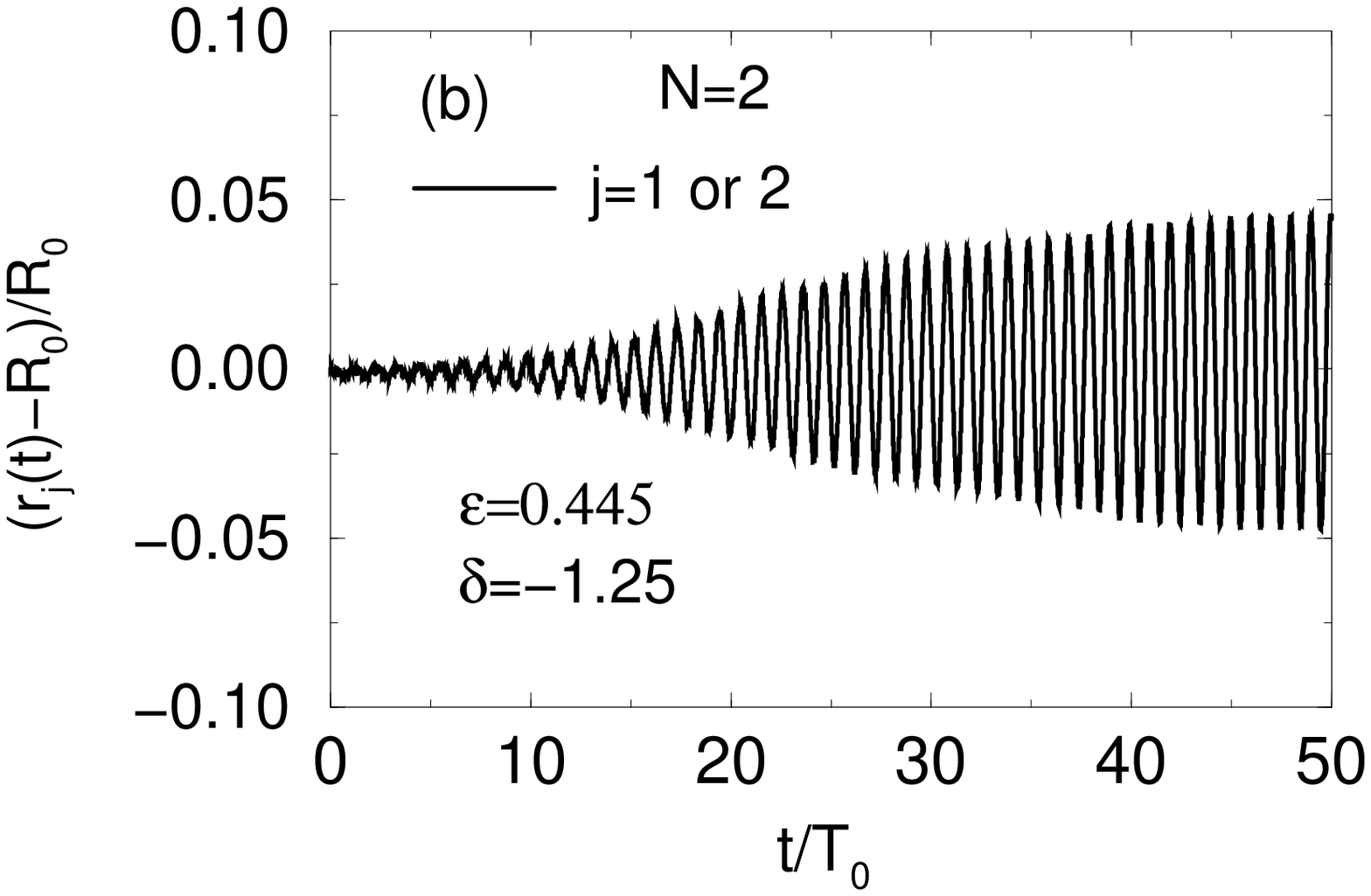,width=.4\textwidth}
\hfill
\psfig{file=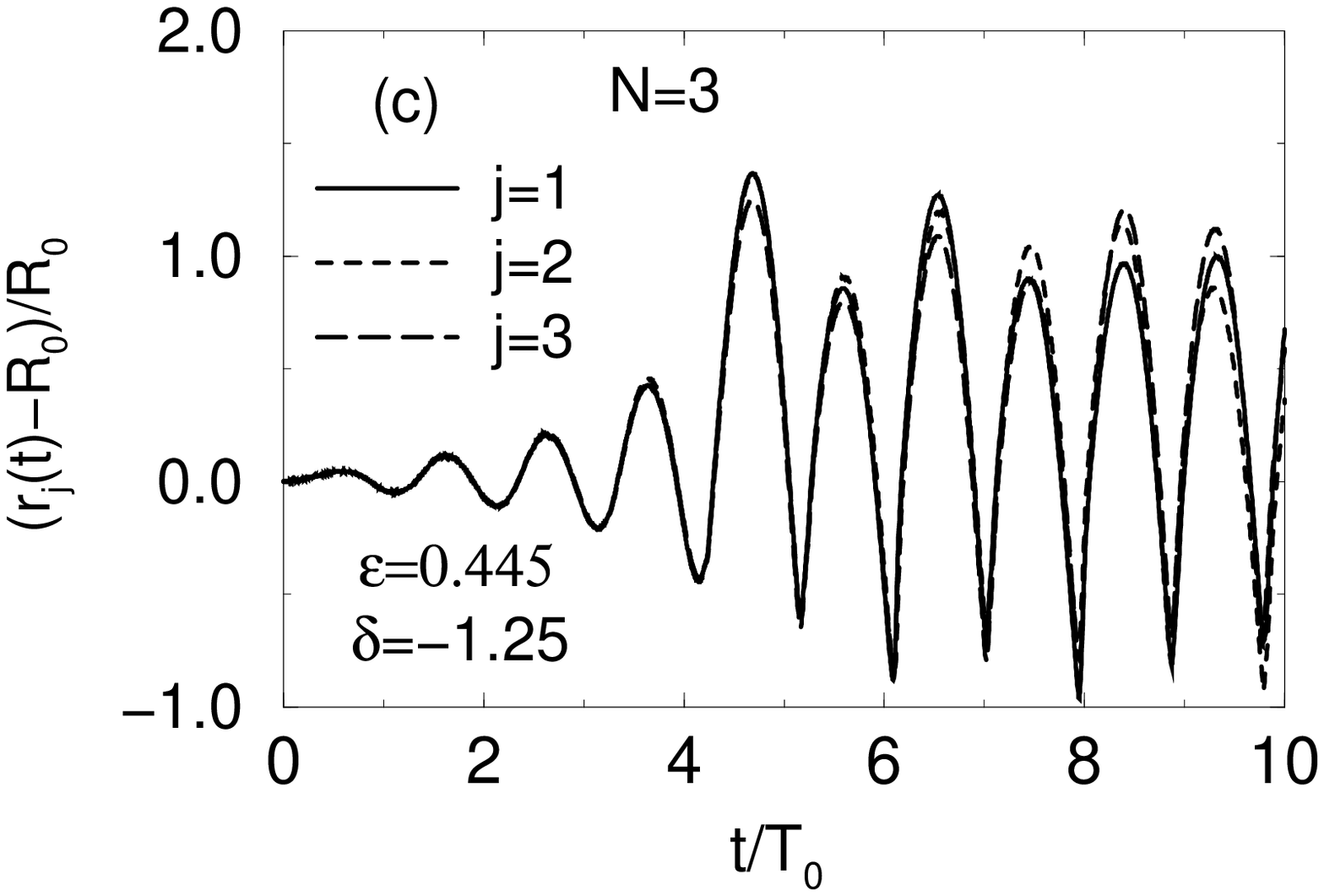,width=.4\textwidth}
}
\caption{Plots of the radial displacement of the
wave tips vs time for two-arm (a and b) 
and three-arm (c) spirals. The highly nonlinear symmetric meander
dynamics of three-arm spirals (Fig. \protect\ref{multiarmt}(b)) is
destabilized at large enough time leading to the elimination
of one arm at boundaries. In contrast, the symmetric meander dynamics
of two-arms spirals is stable on the time scale of
our simulations despite the 
collisions
illustrated in Fig. \protect\ref{2armframes}.
}
\label{2a3arms}
\end{figure}

\begin{figure}
\centerline{
\psfig{file=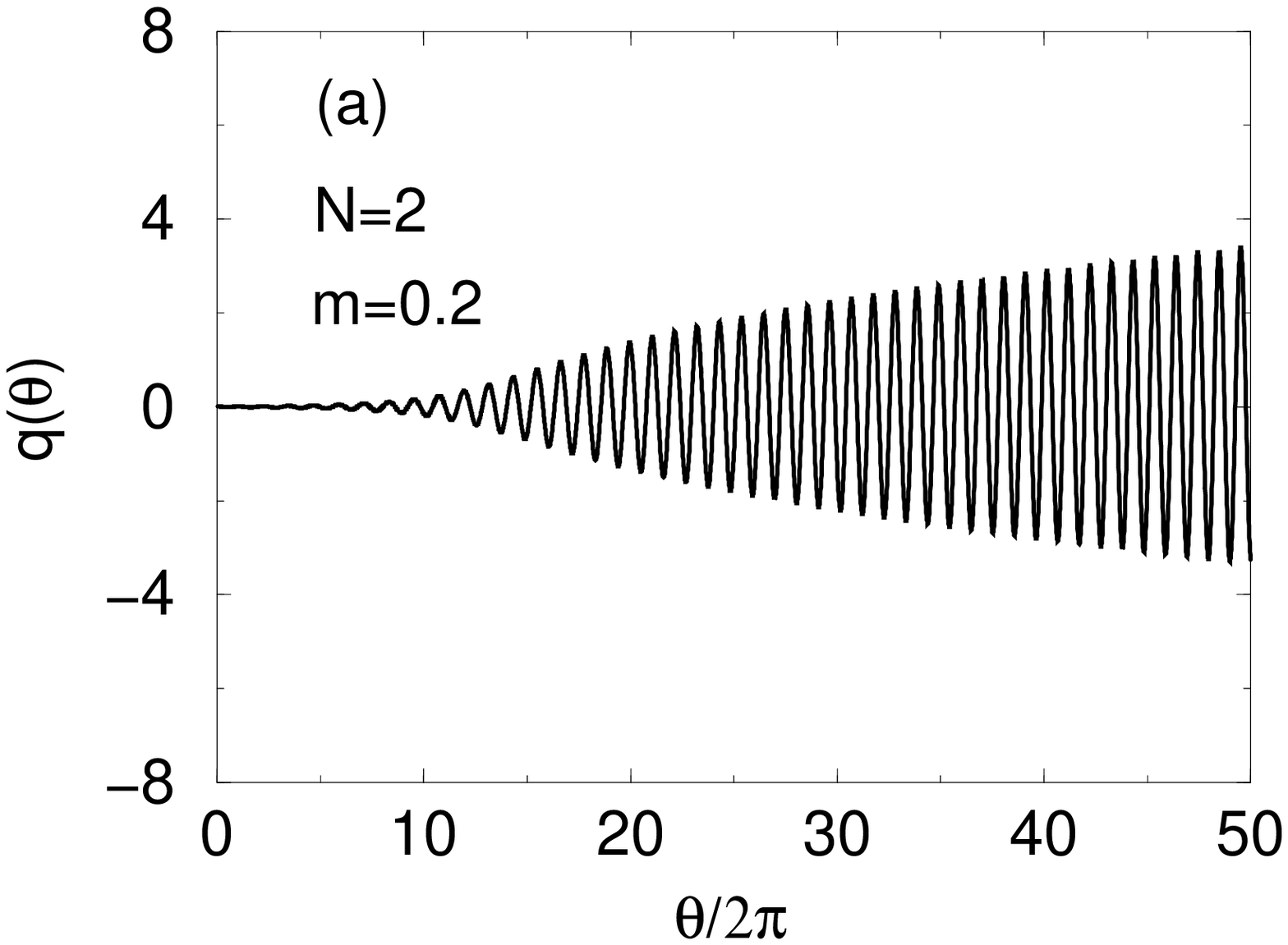,width=.45\textwidth}
\hfill
\psfig{file=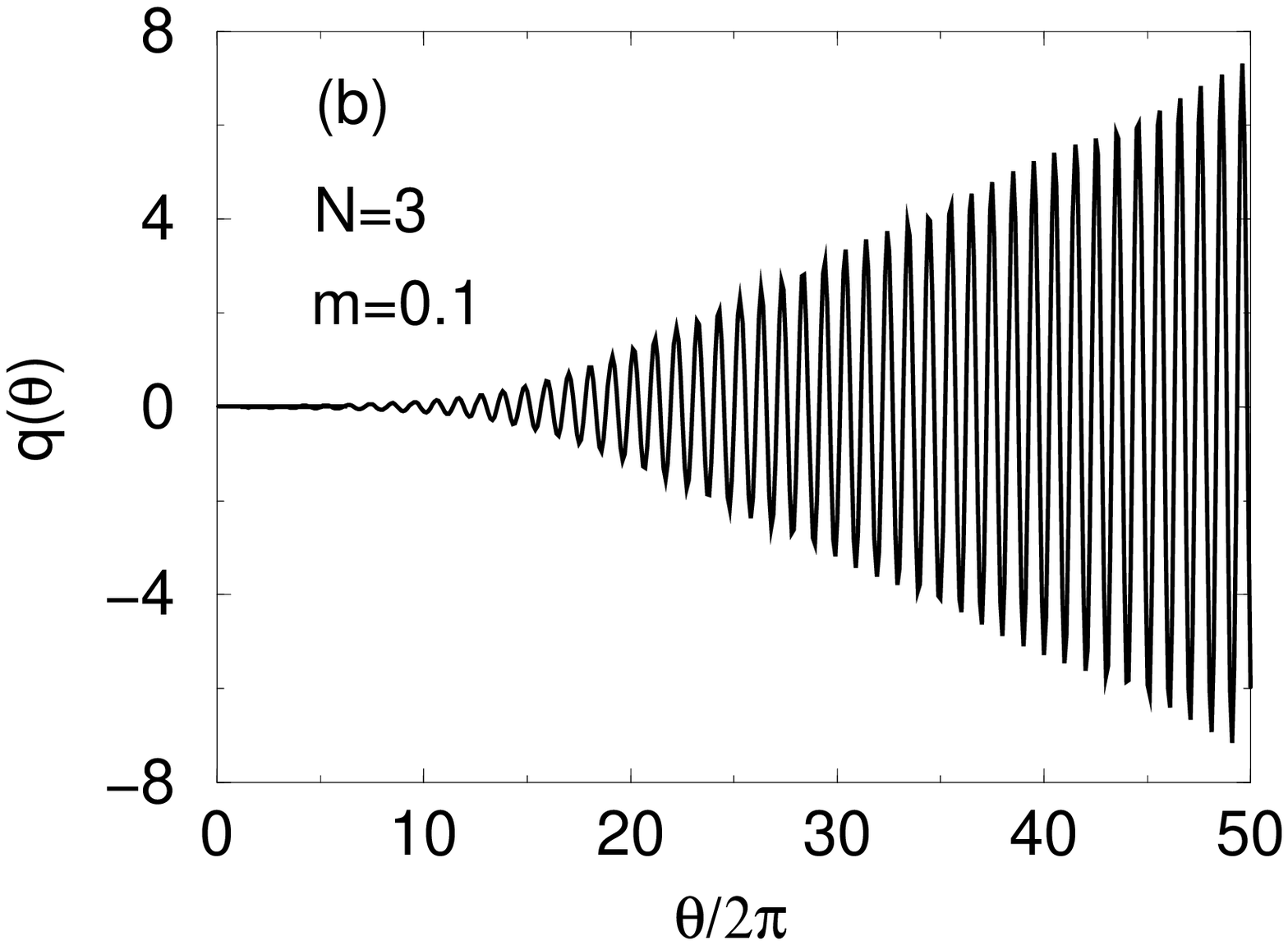,width=.45\textwidth}
}
\caption{Plot of $q(\theta)$ vs $\theta/2\pi$ obtained by
numerical integration of the wavetip equation with
$F$ defined by Eq. \protect\ref{asymth} and $a=0.2$; (a) two-arm
spiral for $m=0.2$, and (b) three-arm spiral for $m=0.1$.}
\label{multiarmq}
\end{figure}

\begin{figure}
\centerline{
\psfig{file=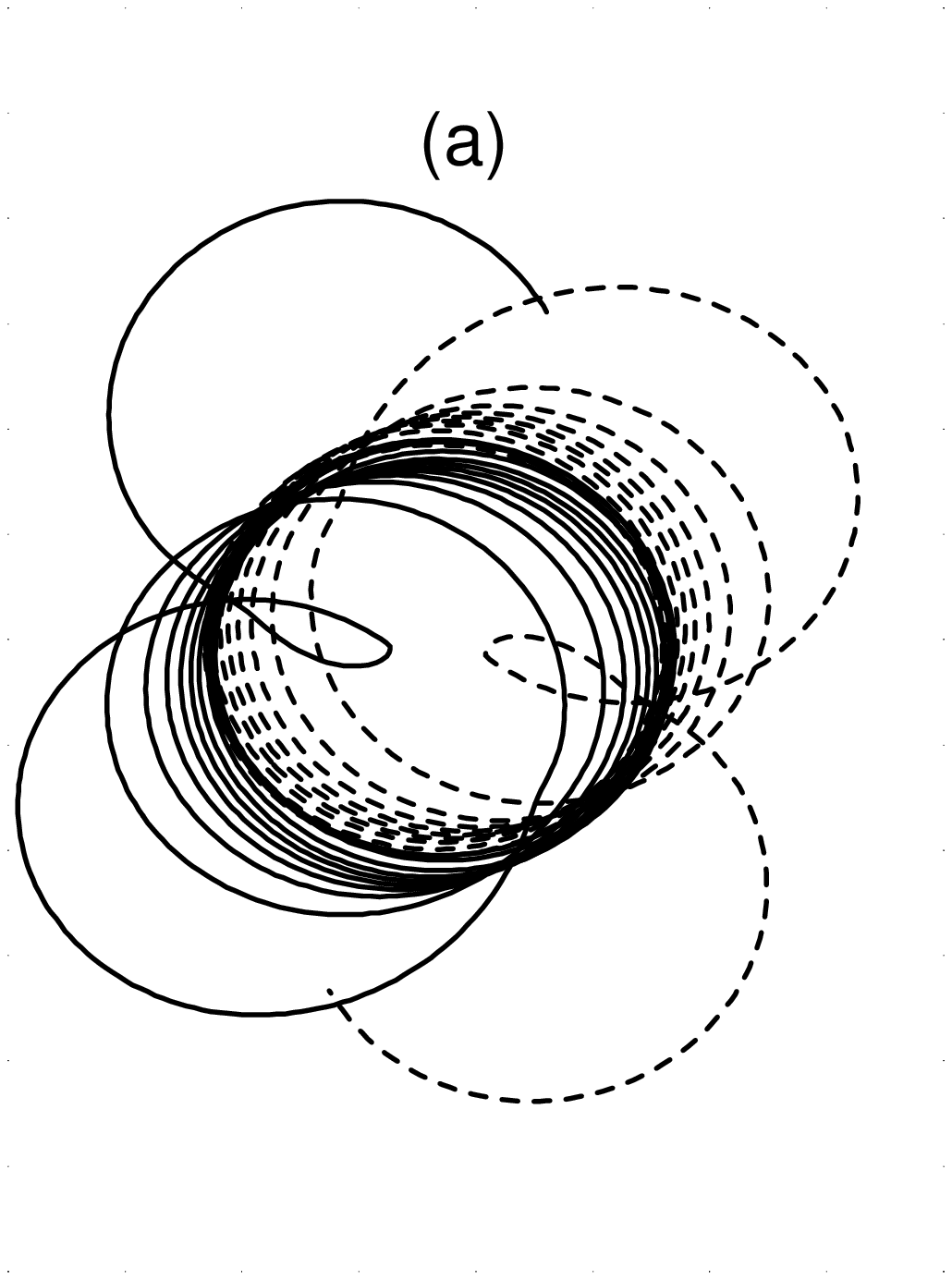,width=.55\textwidth}
\hfill
\psfig{file=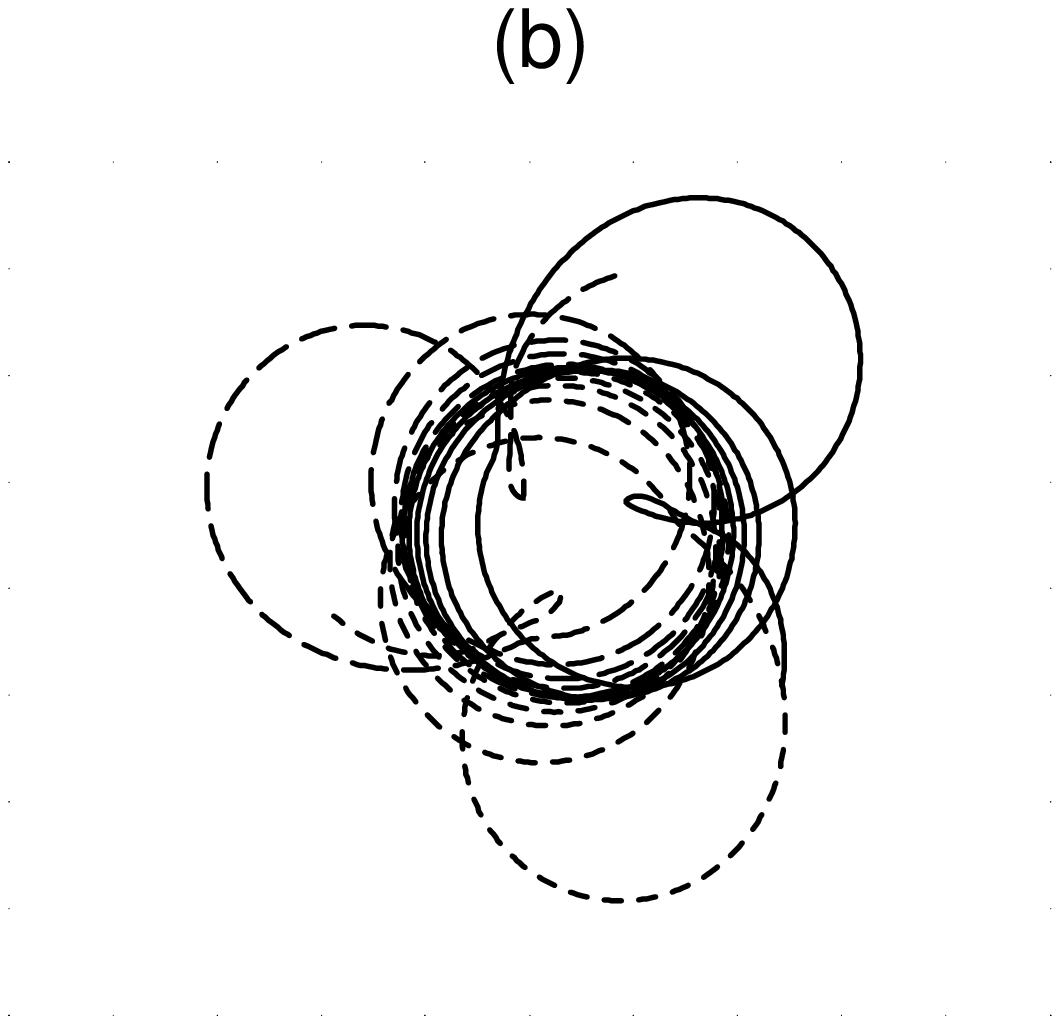,width=.55\textwidth}}
\caption{Simulations of the FN model
showing the wave tip trajectories during the initial
development of the instability of multi-arm spirals 
for: (a) a two-arm spiral with $\epsilon=0.445$ 
and $\delta=-1.24$, and (b) a
three-arm spiral with $\epsilon=0.445$ and $\delta=-1.25$.
Each line type (solid, dashed, or long-dashed) corresponds
to a different wave tip trajectory.}
\label{multiarmt}
\end{figure}

\newpage

\begin{figure}
\centerline{
  \psfig{file=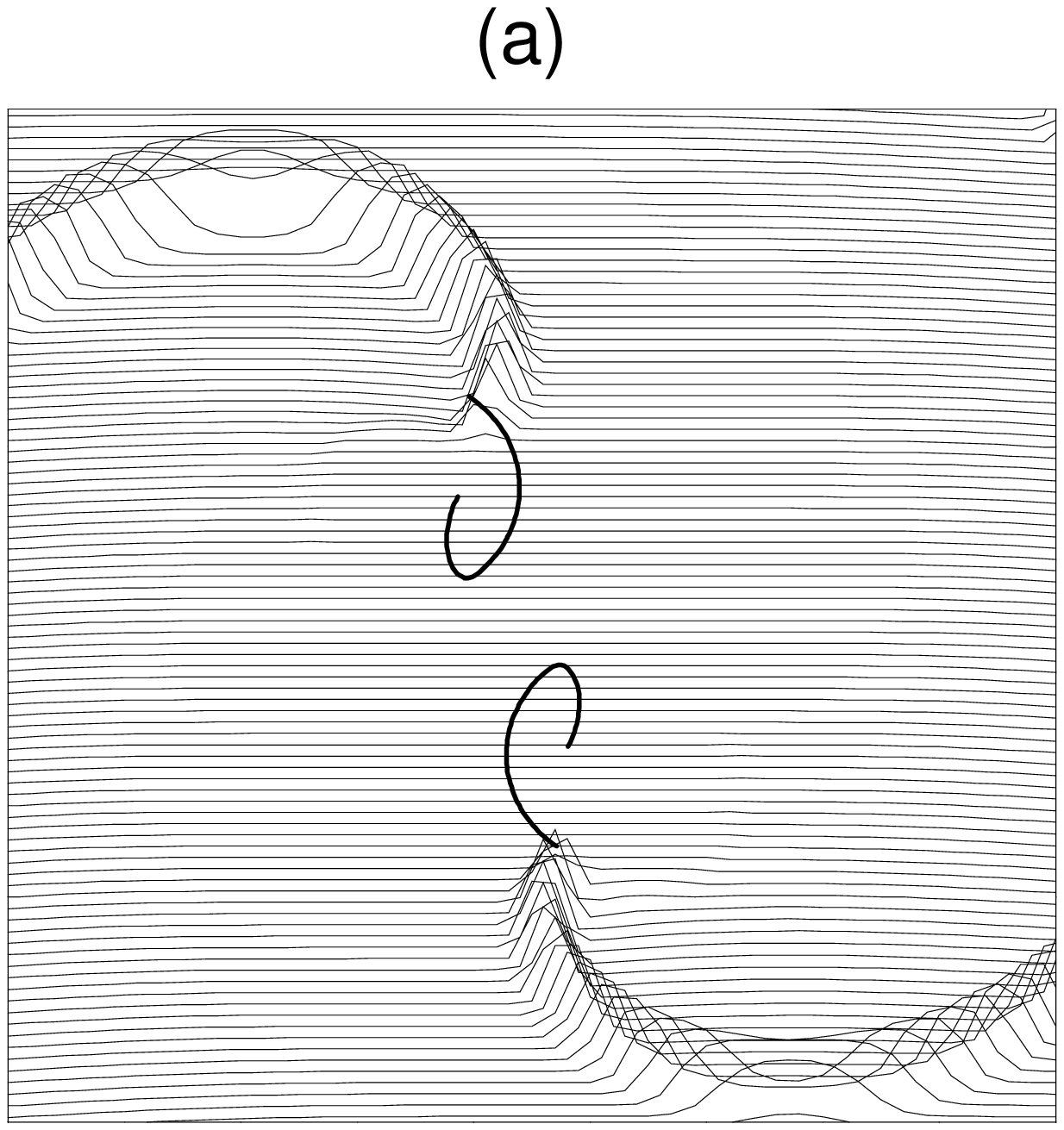,width=.4\textwidth}
\hfill
\psfig{file=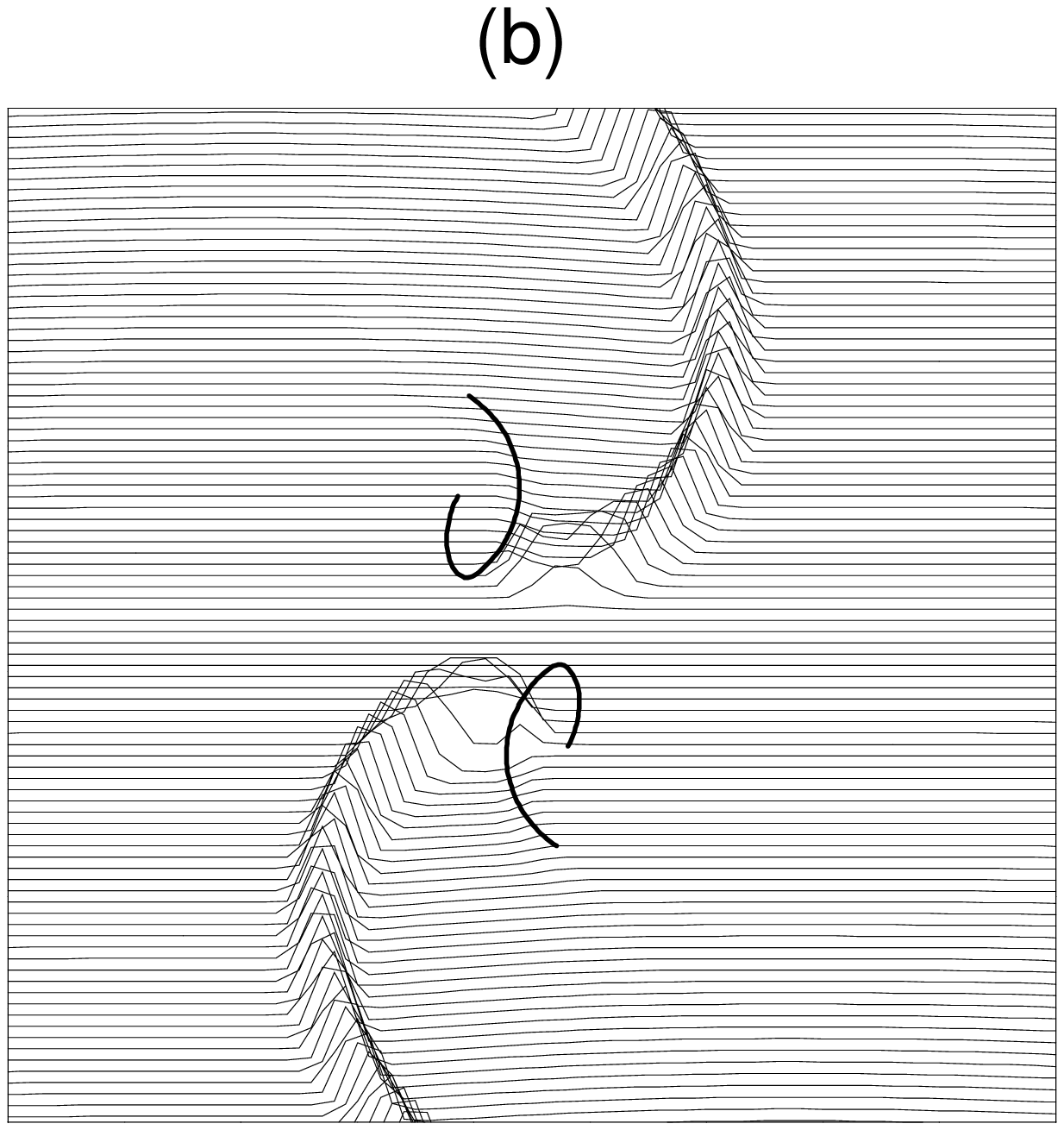,width=.4\textwidth}}
\centerline{
\psfig{file=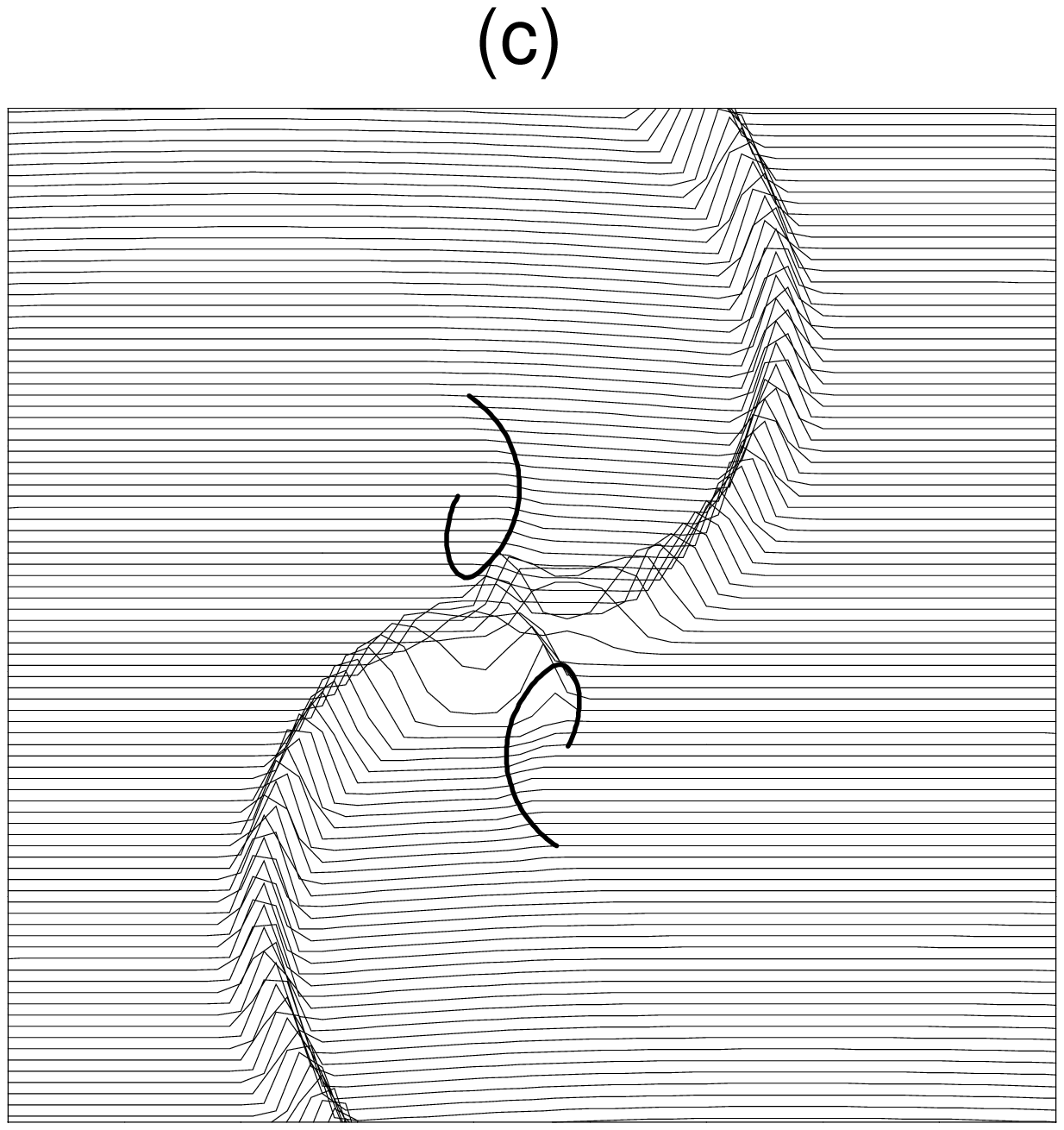,width=.4\textwidth}
\hfill
\psfig{file=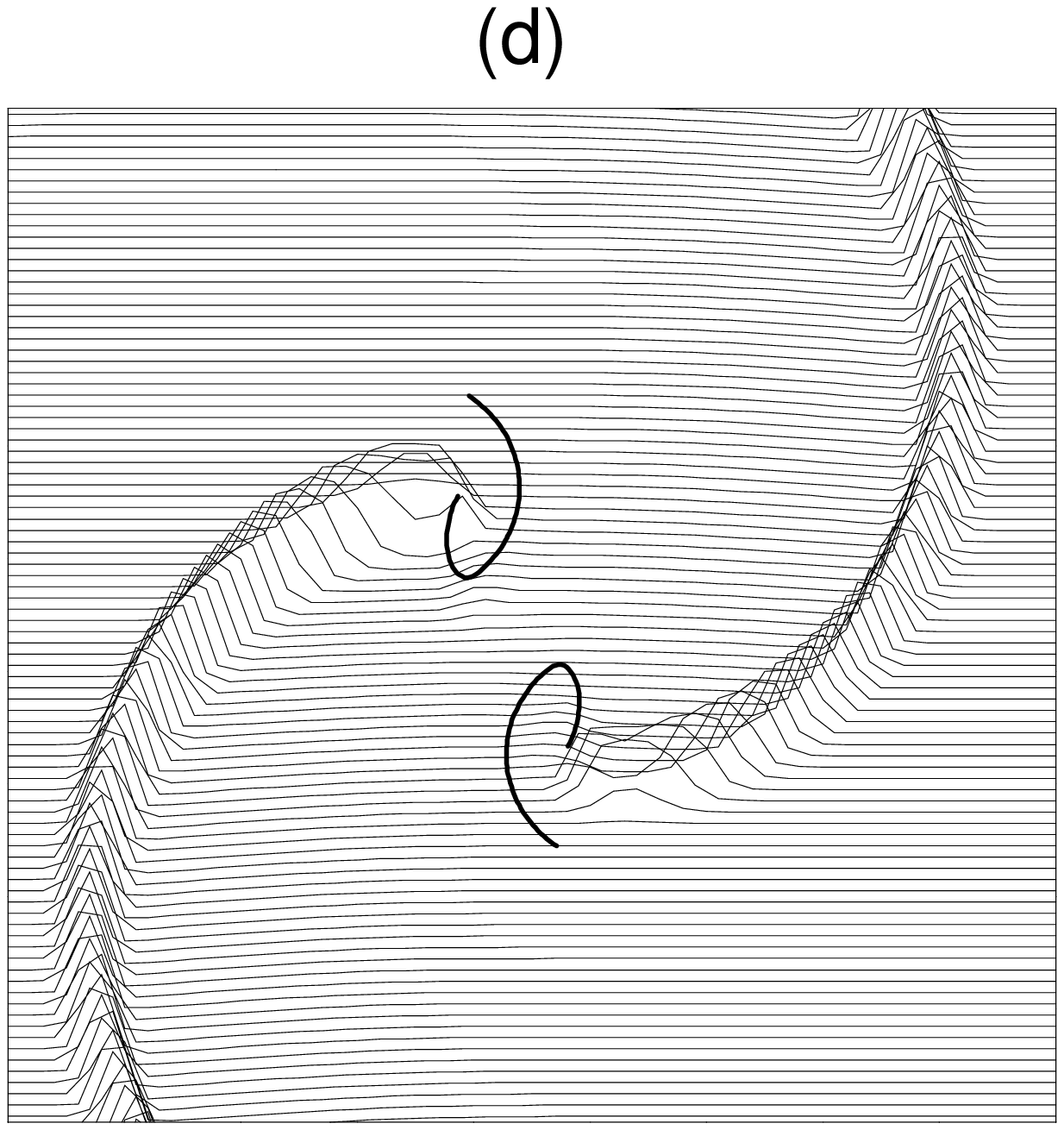,width=.4\textwidth}}
\caption{
Sequence of surface plots of $u$
and superimposed wave tip trajectories (thick solid lines)
illustrating the highly nonlinear collision 
of wave fronts occurring during a
symmetric meander pattern of a two-arm spiral.
Simulation parameters are $\delta=-1.24$
and $\epsilon=0.445$. Frames (b), (c), and (d)
are at $t/\epsilon=20$, 25, and 40, respectively, with
$t$ measured from frame $a$.
Note that an exchange of wave 
tips and spiral arms occurs during
the collision in (c), such that
the wave tip of the downward moving
arm in (b) is at the end of the upward 
moving arm in (d) and vice versa.
This exchange produces the
sharp pivot turns around the small 
inward meander petals.}
\label{2armframes}
\end{figure}

\begin{figure}
\centerline{
  \psfig{file=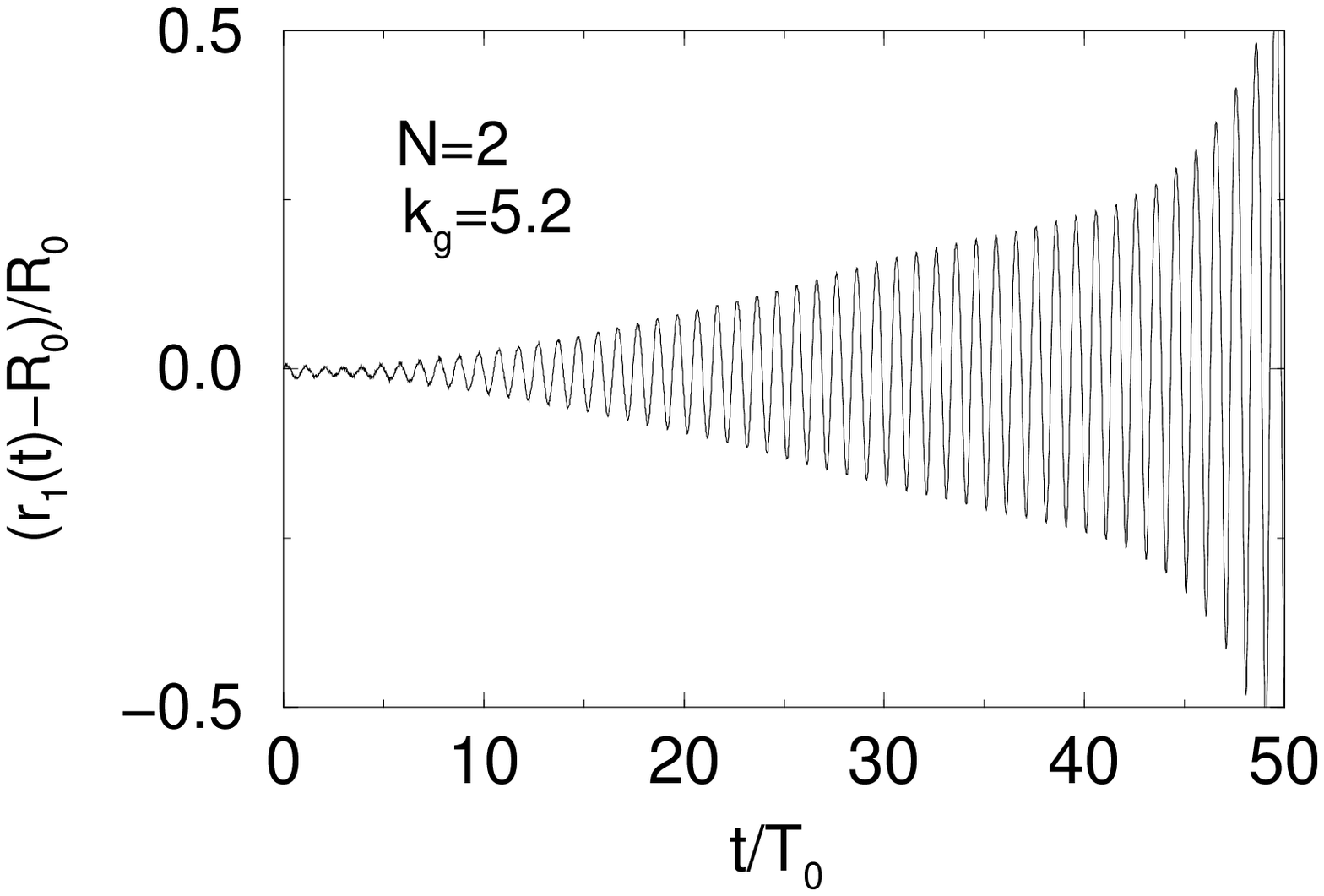,width=.85\textwidth}}
\caption{Plot of the radial displacement of one of
the spiral tip vs time showing the instability of
a two-arm spiral waves in the FN kinetics studied in
Ref. \protect\cite{vas}. The kinetics is defined by
the equations $\partial_t g=D\nabla g-k_rg(g-a)(g-1)-k_rg$
and $\partial_tr=(g-r)/\tau$. We used $k_g=5.2$ and the
other parameters as defined in Ref. \protect\cite{vas}:
$D=1$, $k_r=1.5$, $a=0.05$, and $\tau=5$.}
\label{2armvasiev}
\end{figure}


\begin{thebibliography}{999}

\bibitem{Win0}
A.T. Winfree, {\it When Time Breaks Down} (Princeton
U. Press, N. J. 1987).
\bibitem{Zyk}
V. S. Zykov, {\it Modelling of wave processes in excitable
media} (Manchester U. Press, Manchester, 1988).

\bibitem{cat} S. Jakubith, H.H. Rotermund, W. Engel, A. von Oertzen and
G.Ertl, Phys. Rev. Lett {\bf 65}, 3013 (1990).

\bibitem{Win1} A.T. Winfree, Science {\bf 181}, 937 (1973);
W. Jahnke, W. E. Skaggs and A. T. Winfree,
J. Phys. Chem {\bf 93},
740 (1989).

\bibitem{Plesetal} T. Plesser et al , J. Phys.
Chem {\bf 94}, 7501 (1990).

\bibitem{Skietal} G.S. Skinner and H.L. Swinney, Physica D
{\bf 48}, 1 (1991).

\bibitem{belfles} A. Belmonte, Q. Ouyang and J.M. Flesselles,
J. Phys. II (France) {\bf 7}, 1425 (1997).

\bibitem{dic} W. F. Loomis, {\em 
Dictyostelium Discoideum, A Developmental System}
(Academic Press, New York, 1975; F. Siegert and C. J. Weijer, Physica {\bf
49 D}, 224 (1991).

\bibitem{heart} 
See e.g. the several review articles
in the focus issue of
CHAOS {\bf 8}, 1 (1998).

\bibitem{fitzhugh} R. FitzHugh, Biophys. J. {\bf 1}, 445 (1961); J. Nagumo,
 S. Arimoto and S. Yoshizawa, Proc. I. R. E. {\bf 50}, 2061 (1962).

\bibitem{zykfg} V. S. Zykov, Biofizika {\bf 31}, 862 (1986).

\bibitem{Win2} A.T. Winfree, Chaos {\bf 1}, 303 (1991).

\bibitem{notefig1} Fig.~\ref{flowergarden}
 was constructed from our own simulations and 
agrees qualitatively
with the results of previous studies \cite{zykfg,Win2} regarding
the existence of various boundaries 
between different regimes.
It focuses, however, more specifically on the weakly excitable limit and
thus provides a more quantitative characterization of
these boundaries in this limit.
 
\bibitem{Hopf} E. Lugosi, Physica D {\bf 40}, 331 (1989);
D. Barkley et al,
Phys. Rev. A{\bf 42}, 2489 (1990);
Phys. Rev. Lett. {\bf 68}, 2090 (1992); I. Mitkov
et al,
Phys. Rev. E {\bf 54}, 6065 (1996).

\bibitem{hopfak} 
A. Karma, Phys. Rev. Lett {\bf 65}, 2824 (1990).

\bibitem{Cod2} D. Barkley, Phys. Rev. Lett {\bf 72}, 164 (1994);
D. Barkley and I. G. Kevrekidis Chaos, {\bf 4}, 453 (1994).
V. N. Biktashev et al, Int. J. Bif.  Chaos, (1996).

\bibitem{Lietal} G. Li et al, Phys. Rev. Lett.
{\bf 77}, 2105 (1996).

\bibitem{mikzyk} see A.S. Mikhailov, V.A. Davidov
and V.A. Zykov, Physica D {\bf 70}, 1 (1994) and references therein.

\bibitem{agl} K.I. Agladze, V. A. Davydov and A.S. Mikhailov,
Pisma Zh. Eksp. Teor. Fiz. {\bf 45}, 601 (1987)\
[JETP Lett. {\bf 45}, 767 (1987)]; O. Steinbock, V. Zykov and S, C. M \"{u}ller,
Nature {\bf 366}, 322 (1993).
\bibitem{ag} K.I. Agladze and P. de Kepper, J. Phys. Chem. {\bf 96}, 5239
(1992).
\bibitem{stein} O. Steinbock, J. Sch\"{u}tze and S.C. M\"{u}ller,
 Phys. Rev. Lett.
{\bf 68},248 (1992).
\bibitem{apmun} A. P. Mu\~{n}uzuri et al., Phys. Rev. E {\bf 50} 4258 (1994).
\bibitem{bel} A. Belmonte and J.M.  Flesselles, Europhys. Lett. {\bf 32},
267 (1995).
\bibitem{mit} I. Mitkov, I. Aranson and D. A. Kessler,
 Phys. Rev. {\bf E 52}, 5974  (1995).
\bibitem{krins} V. Krinsky, E. Hamm and V. Voignier,
Phys. Rev. Lett. {\bf 76}, 3854 (1996).

\bibitem{vas} B. Vasiev, F. Siegert and C. Weijer,
Phys. Rev. Lett. {\bf 78}, 2489 (1997).

\bibitem{fbp} J.J. Tyson and J.P. Keener, Physica D{\bf 32}, 327 (1988).

\bibitem{keenty} J.J. Tyson and J.P. Keener, Physica {\bf 29D},215 (1987).

\bibitem{klr}
D.A. Kessler, H. Levine, and W.N. Reynolds, Physica D{\bf 70}, 115 (1994).

\bibitem{ak92}
A. Karma, Phys. Rev. Lett. {\bf 68}, 397 (1992).

\bibitem{fife}
P.C. Fife, J. Stat. Phys. {\bf 39}, 687 (1985).

\bibitem{ps1} P. Pelce and J. Sun,
Physica D {\bf 48}, 353 (1991).

\bibitem{ak1} A. Karma, Phys. Rev. Lett. {\bf 66}, 2274 (1991).

\bibitem{ak2} A. Karma, in {\it Nonlinear Phenomena Related to Growth and
Form}, M. Ben Amar et al eds,
(Plenum Press, N. Y., 1991).

\bibitem{hk1} V. Hakim and A. Karma, Phys. Rev. Lett. {\bf 79}, 665 (1997).

\bibitem{note32}
The numerically
determined constant $\zeta\simeq 0.685$ of \cite{mikzyk}
is given in our notation by 
$\zeta=b'^{-3/2}=|a'_1|^{-3/2}/\sqrt{2}=0.6876\cdots$.
This value of $\zeta$
corresponds to replacing $b$ in Eq. \ref{e2} by
the constant $b'$ given by the location $a_1'$ of the
maximum of the Airy function as explained at the end
of section \ref{srspi}.
 
\bibitem{Biktension} V. N. Biktashev, A. V. Holden
and H. Zhang, Phil. Trans. R. Soc. Lond. A{\bf 347}, 611 (1994).

\bibitem{ps2} P. Pelce and J. Sun,
Physica D{\bf 63}, 273 (1992).

\bibitem{pier} This analysis was performed
in collaboration with B. Pier, and has been described in
B. Pier DEA report (ENS Lyon, 1995).

\bibitem{bcf} W.K. Burton, N. Cabrera and F. C. Franck, Phil. Trans. Roy.
 Soc. (London) {\bf A 243}, 299 (1951).

\bibitem{noteB} This definition coincides with Eq.~(\ref{Bdef}) since
$R_{tip}=\epsilon/c_0,\, W= 2\Delta c_0 \tau_e$
with the adimensionned units of (\ref{eqrd1},\ref{eqrd2}).
\bibitem{abr} M. Abramovicz and I. A. Stegun {\it Handbook of
Mathematical Functions} (Dover, N. Y., 1972) p. 478
\bibitem{pan} A.M. Pertsov, E. A. Ermakova and 
A.V. Panfilov, Physica {\bf 14D},
117 (1984).

\bibitem{noFm} Note that a factor of 3 has been included in the definition
of $F$ as compared to \cite{hk1}. So, the function $F$ here defined is three
times larger than in \cite{hk1}. Correspondingly, the parameter $m$ 
(\ref{mexp}) is three
times smaller. Note also that here, lengths have been measured from the start
on the scale of the tip radius $\epsilon/c_0$ so that the quantities
with tilde in \cite{hk1} appear here without tilde (e.g. $d$ here is denoted
$\tilde d$ in \cite{hk1}).

\bibitem{kesskupf}
D.A. Kessler and R. Kupferman, Physica D{\bf 105}, 207 (1997).
\bibitem{keen3d}
 J.P. Keener, Physica {\bf D31},269 (1988). 
\bibitem{win3d} C. Henze, E. Lugosi and A. T. Winfree, Can. J. Phys. {\bf 68}
683 (1989).
\bibitem{fenkar} F. Fenton and A. Karma, Phys. Rev. Lett. {\bf 81}, 481 (1998);
Chaos {\bf 8},20 (1998)
\end{thebibliography}
\end{document}